\documentclass[onecolumn]{elsarticle}

\journal{Physica Medica}
\usepackage{hyperref}
\usepackage{lineno,color}
\modulolinenumbers[5] %remove for submission

\usepackage{amsmath}
\usepackage{graphicx}
\usepackage{subcaption}
\usepackage[left=2.5cm,right=2.5cm,top=2.5cm,bottom=2.5cm]{geometry}
\biboptions{numbers}%\biboptions{authoryear}

\begin{document}

\begin{frontmatter}

\title{Large-Area SiPM Pixels (LASiPs): a cost-effective solution towards compact large SPECT cameras}

\author[a,b]{D.~Guberman}
\ead{daniel.guberman@pi.infn.it}
\author[a,b]{R.~Paoletti}
\author[a]{A.~Rugliancich}
\author[a,b]{C.~Wunderlich}
\author[c]{A.~Passeri}

\address[a]{Istituto Nazionale di Fisica Nucleare (INFN), Sezione di Pisa, I-56126 Pisa, Italy}
\address[b]{Dipartimento di Scienze Fisiche, della Terra e dell’Ambiente, Università di Siena, I-53100 Siena, Italy}
\address[c]{Dipartimento di Scienze Biomediche Sperimentali e Cliniche (SBSC) Università di Firenze, I-50134 Florence, Italy}

\begin{abstract}
Single Photon Emission Computed Tomography (SPECT) scanners based on photomultiplier tubes (PMTs) are still largely employed in the clinical environment. A standard camera for full-body SPECT employs $\sim50$-100 PMTs of 4-8~cm diameter and is shielded by a thick layer of lead, becoming a heavy and bulky system that can weight a few hundred kilograms. The volume, weight and cost of a camera can be significantly reduced if the PMTs are replaced by silicon photomultipliers (SiPMs). The main obstacle to use SiPMs in full-body SPECT is the limited size of their sensitive area. A few thousand channels would be needed to fill a camera if using the largest commercially-available SiPMs of 6$\times$6~mm$^2$. As a solution, we propose to use Large-Area SiPM Pixels (LASiPs), built by summing individual currents of several SiPMs into a single output. We developed a LASiP prototype that has a sensitive area 8 times larger than a 6$\times$6~mm$^2$ SiPM. We built a proof-of-concept micro-camera consisting of a 40$\times$40$\times$8~mm$^3$ NaI(Tl) crystal coupled to 4 LASiPs. We evaluated its performance in a central region of $15\times15$~mm$^2$, where we were able to reconstruct images of a $^{99m}$Tc capillary with an intrinsic spatial resolution of $\sim2$~mm and an energy resolution of $\sim11.6$\% at 140 keV. We used these measurements to validate Geant4 simulations of the system. This can be extended to simulate a larger camera with more and larger pixels, which could be used to optimize the implementation of LASiPs in large SPECT cameras. We provide some guidelines towards this implementation. %With a proper optimization of the pixel size and geometry, LASiPs could be suitable for large SPECT cameras.
\end{abstract}


\end{frontmatter}

%\linenumbers %remove for submission

\section{Introduction}
\label{sec:intro}

Single-photon emission computed tomography (SPECT) is a nuclear imaging technique that has been used in clinical environments since at least forty years.  It provides high efficiency in diagnosing several diseases, like Alzheimer's~\cite{Valotassiou_2018} and Parkinson~\cite{Son_2016}.

Most common clinical SPECT scanners are constituted by one or more gamma camera heads~\cite{Anger_1958}, roughly consisting of: (i) a lead collimator; (ii) a large ($\sim50\times40\times1$~cm$^3$) scintillating crystal; (iii) an array of 50--100 photomultiplier tubes (PMTs) of relatively large area (4--8~cm diameter). The whole camera (including at least a part of the PMT electronic readout) is shielded by a $\sim1$--3~cm thick layer of lead, which turns a SPECT camera into a heavy and bulky system that can weight a few hundred kilograms. A gantry with two cameras can weigh more than 2000~kg.

Standard clinical full-body SPECT systems exhibit an energy resolution of $\sim10$\% at 140 keV and an intrinsic spatial resolution better than 5~mm\footnote{https://www.siemens-healthineers.com}. The intrinsic spatial resolution is a property of the scintillation detector (scintillating crystal and photodetectors), that combined with the collimator resolution gives the extrinsic (total) spatial resolution of the system~\cite{Van_Audenhaege_2015}. In most SPECT scans the extrinsic spatial resolution is actually driven by the characteristics of the collimator~\cite{CHERRY_2012} (with typical values going from 6 to 15~mm at a source-collimator distance of 10~cm). The clinical potential of SPECT relies on the compensation for qualitative and quantitative image degradation due to non-stationary experimental data, such as Compton scattering, attenuation and geometrical system response, the latter representing the main source of blurring for the system spatial resolution. The ability of the reconstruction algorithm to manage those effects plays a key-role in the final impact SPECT can have in the clinical practice~\cite{Passeri_2004}. Nonetheless, there is another point of primary concern from the clinical point of view that is strictly related to the hardware: the possibility of reducing the size, weight and cost of a SPECT scanner. A lighter and more compact system would be safer to operate, would give more flexibility both to the patient and the physician, and could be installed in smaller rooms (and would be easier to fit in smaller hospitals).

Recently, SPECT scanners based on cadmium-zinc-telluride (CZT) technology have been introduced. With respect to traditional systems, they are characterized by a substantial improvement in both energy and (intrinsic) spatial resolution, while being lighter and more compact. The main limitation for a wider use of this technology in full-body SPECT is the camera price, which increases significantly with size~\cite{Agostini_2016}. The first multi-purpose scanners became available very recently~\cite{Goshen_2018, Morelle_2019, Desmonts_2020}. However, most CZT-SPECT systems developed in the last ten years were relatively small instruments, dedicated to image specific organs (e.g., cardiac imaging, see~\cite{NUDI_2017} for a recent review). In these applications the organs to be scanned are close to the collimator, thus making the improvement in intrinsic spatial resolution of primary value. 

A cost-effective approach to build lighter and more compact gamma cameras would be to replace the PMTs by silicon photomultipliers (SiPMs). Replacing PMTs by SiPMs would be beneficial for several reasons. They provide higher photodetection efficiency (PDE), do not require high-voltage operation and their cost is trending down. Moreover, SiPMs are not sensitive to magnetic fields, which is particularly interesting for combining SPECT and Magnetic Resonance Imaging~\cite{INSERT_2018}. Particularly, SiPMs are much more compact than PMTs. A typical SPECT PMT is $\sim15$~cm long (without electronics), which corresponds to more than 50~\% of the thickness of the camera (the millimeter thickness of a SiPM is negligible in comparison). A SPECT camera using compact photodetectors and electronics would be also much more compact. It would be lighter because the amount of lead needed for the shielding would be reduced. And it would simplify the construction of the gantry, which could allow to also reduce the weight and size of the whole scanner.

Probably the main obstacle for using SiPMs in full-body SPECT cameras is their limited pixel size: SiPMs are rarely commercially available in sizes larger than $6 \times 6$~mm$^2$. A few thousand channels would be needed to fill a $50 \times 40$~cm$^2$ camera using SiPMs, dramatically increasing the cost and complexity of the system. This is one of the reasons why up to now almost all research in SiPMs for SPECT has been limited to small cameras~\cite{INSERT_2018, Bouckaert_2014, Popovic_2014}.

Larger SiPMs are normally not built because their capacitance increases significantly with size. This is particularly critical for many high-energy physics and astrophysics experiments where fast timing and close to single-photoelectron resolution are needed. Several solutions have been developed in these fields to build large SiPM pixels that keep their capacitance at a reasonable level~\cite{SiPM-LST,HAHN_2018,Guberman_2019, Nagai_2019,Mallamaci_2019}. Inspired by these developments, we performed a feasibility study on the implementation of Large Area SiPM Pixels (LASiPs) in SPECT, as a solution to use SiPM technology in large gamma cameras. In section~\ref{sec:methods} we introduce the LASiP concept and present a prototype pixel we developed and characterized. We also describe the experiments performed with a NaI(Tl) crystal coupled to 4 of those LASiP prototypes and the Monte Carlo simulations we performed to further study the system. The results of such experiments and simulations are shown in section~\ref{sec:results}. The impact of these results and the feasibility of developing large SPECT cameras equipped with LASiPs are discussed in section~\ref{sec:discussion}.

\section{Materials and Methods}\label{sec:methods}

\subsection{The LASiP concept}\label{sec:LASiP}

One way of building large SiPM pixels is summing the analog currents of several SiPMs into a single output. SiPM signals are filtered and summed using custom-designed adder circuits based, for instance, on operational or common-base transistor amplifiers. This way the pixel size can be increased by a factor equal to the number of SiPMs that are being summed, while keeping the capacitance at a reasonable level. In this solution, which has been successfully applied in Very High Energy (VHE) astrophysics~\cite{SiPM-LST,HAHN_2018, Mallamaci_2019}, the adder circuits are typically built using discrete components. In our case, the sum is performed using an Application-Specific Integrated Circuit (ASIC) named MUSIC (8 channel Multiple Use IC for SiPM anode readout). The MUSIC was designed for multiple purposes, but mainly targeting the readout of SiPMs in Cherenkov telescopes. One of its main functionalities is the possibility to sum up to 8 SiPMs using a current switch consisting of two common base transistors and an operational transconductance amplifier (see~\cite{Gomez_2016} for the details). Using an ASIC offers several advantages, including compactness and ease of reproduction on a large scale, which is particularly relevant for SPECT applications. The MUSIC offers many other functionalities. Some of them will be described in section~\ref{sec:pixel}.

\subsection{LASiP prototype}\label{sec:pixel}

\begin{figure}[tpb]
    \begin{subfigure}{.33\textwidth}
    \centering
    \includegraphics[clip=true, trim=2cm 6cm 3cm 6cm, height=5cm]{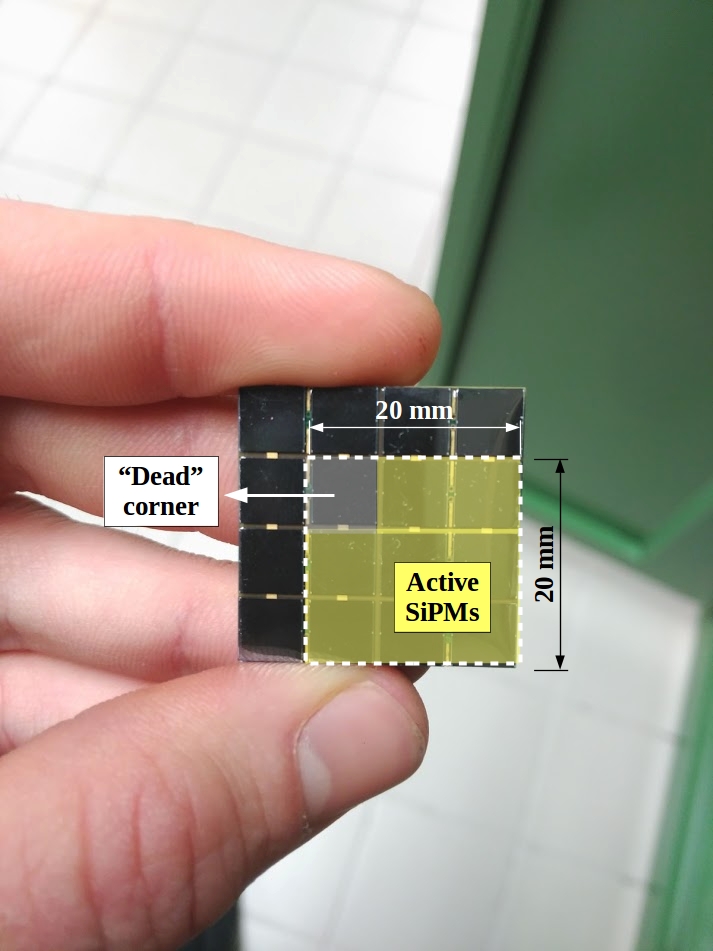}
    \subcaption{}\label{fig:LAPIS_a}
    \end{subfigure}
    \begin{subfigure}{.33\textwidth}
    \centering
    \includegraphics[clip=true, trim=0cm 0cm 0cm 0cm, height=5cm]{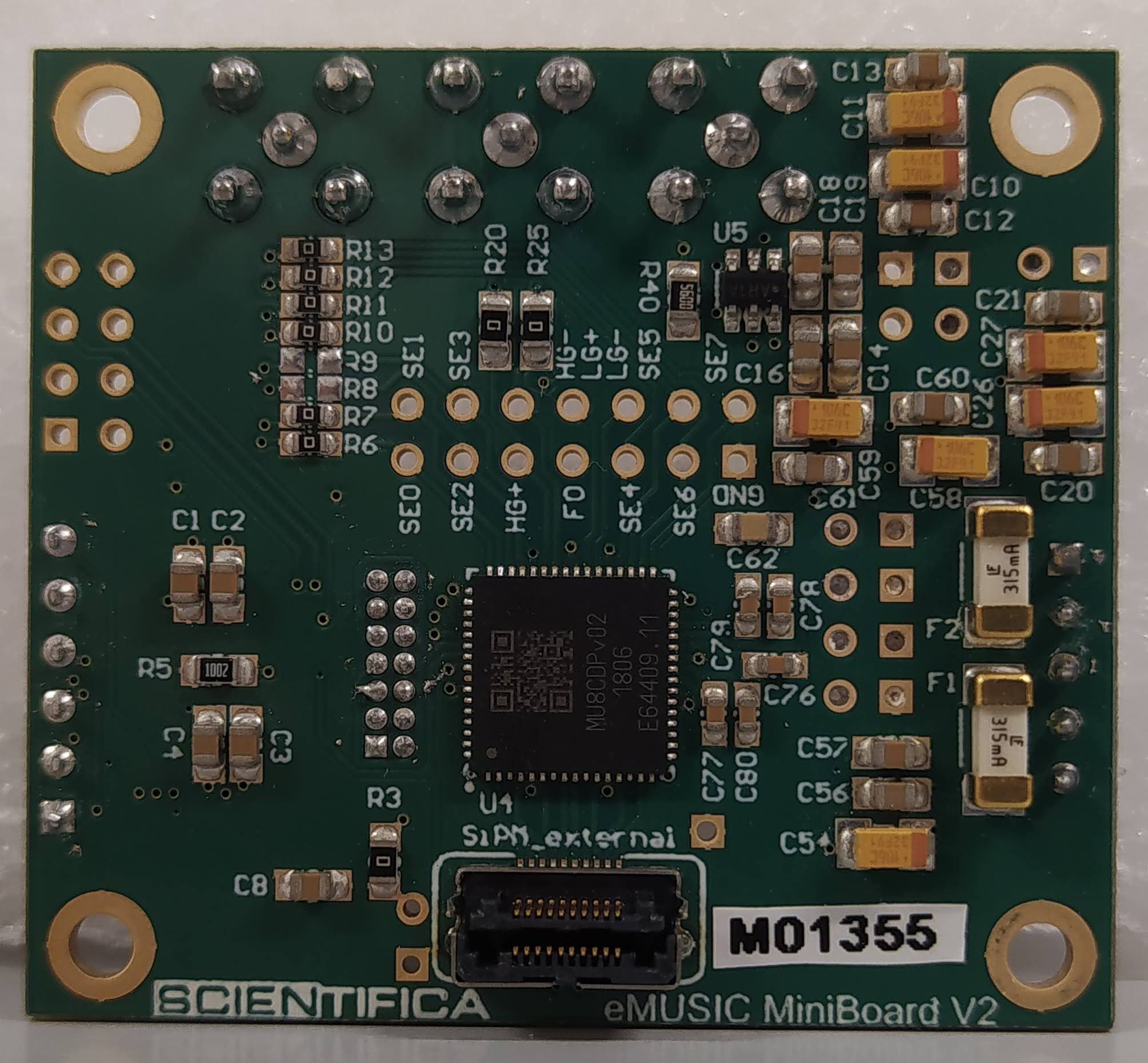}
    \subcaption{}\label{fig:LAPIS_b}
    \end{subfigure}
    \begin{subfigure}{.33\textwidth}
    \centering
    \includegraphics[clip=true, trim=3cm 7cm 2cm 8cm, height=5cm]{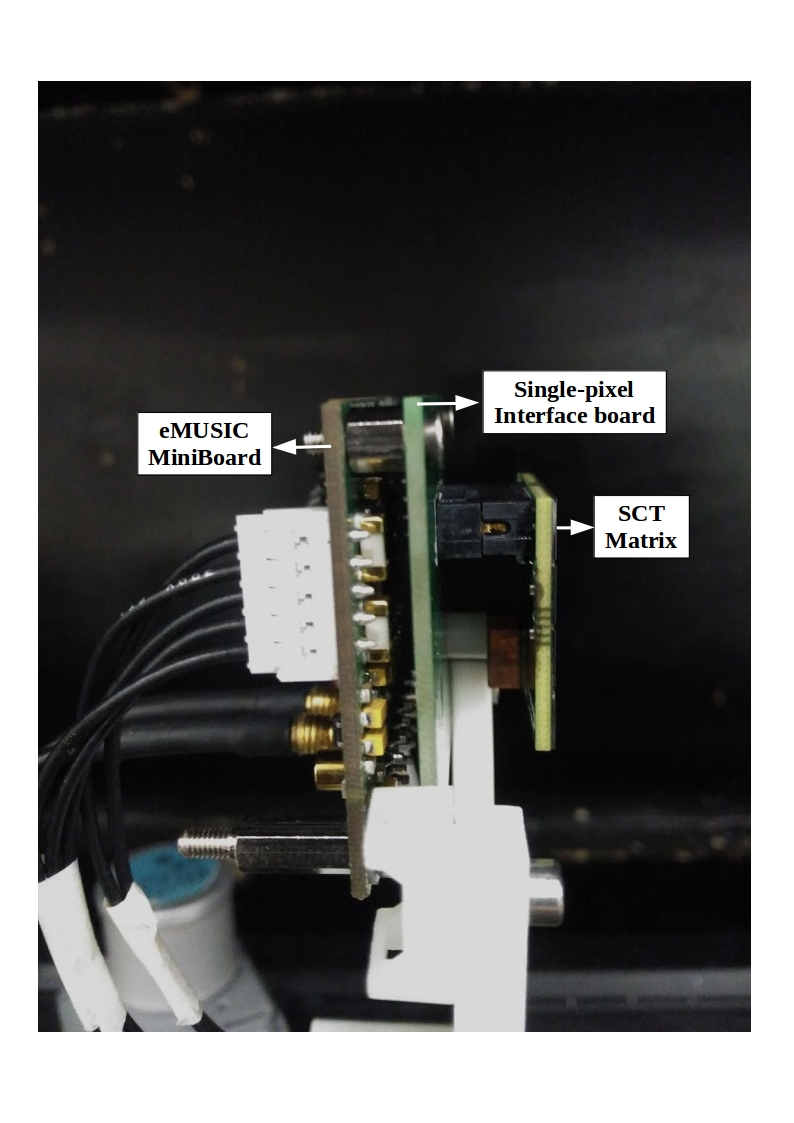}
    \subcaption{}\label{fig:LAPIS_c}
    \end{subfigure}
    \caption{Essential components of a LASiP prototype. \textbf{(a)} SCT Matrix: only the 8 SiPMs marked in yellow are summed by the MUSIC and are actually part of the pixel; \textbf{(b)} eMUSIC MiniBoard; \textbf{(c)} A fully-assembled LASiP seen from the side: the SCT matrix is connected to the eMUSIC MiniBoard by a custom-made interface board.}
    \label{fig:LAPIS}
\end{figure}

For the proof of concept we developed a LASiP prototype in which 8 SiPMs of $\sim6\times6$~mm$^2$ are summed, resulting in a single pixel of $\sim2.9$~cm$^2$ (with an active area of $\sim2.2$~cm$^2$). We used technology that was already available, that had been developed for VHE astrophysics applications: a single prototype employs one \textit{SCT matrix} and one \textit{eMUSIC MiniBoard}  (see Figure~\ref{fig:LAPIS}).

The SCT matrices were originally produced to equip the Schwarzschild-Couder medium-size Telescope~\footnote{After which the ``SCT matrix'' name is given.} proposed for the Cherenkov Telescope Array~\cite{leslie}. A single matrix holds 16 FBK NUV-HD3 SiPMs of $\sim6\times6$~mm$^2$ (nominal chip size: $6.24\times6.24$~mm$^2$), with a $\sim0.5$~mm gap between adjacent SiPMs (Figure~\ref{fig:LAPIS_a}). These SiPMs provide a peak PDE of $\sim60~\%$ at $\sim350$~nm~ and a dark count rate (DCR) of $\sim0.13$~MHz$/$mm$^2$ at 20$^{\circ}$C when operated at $\sim33$~V~\cite{leonardoSPIE_2019}. The LASiP prototype uses only 8 of the 16 SiPMs that the SCT matrix holds. The limitation was imposed by the MUSIC, which can sum up to 8 channels.

The eMUSIC MiniBoard (bought from SCIENTIFICA, S.L.U\footnote{https://www.scientifica.es/products/emusic-miniboard}) is an evaluation board for the MUSIC chip (Figure~\ref{fig:LAPIS_b}). It can be connected to up to 8 SiPMs and allows to program the MUSIC from the computer. The eMUSIC MiniBoard is a plug-and-play board that allows to exploit all the functionalities the MUSIC provides (see~\cite{Gomez_2016}). Those particularly relevant for our detector are:
\begin{itemize}
    \item An output with the sum of up to 8 SiPMs.
    \item Enabling/disabling channels. The output signal of each individual SiPM can be accessed during calibration by disabling the remaining seven.
    \item Applying individual offset values to the bias voltage of each SiPM (useful to equalize the gain).
    \item Configure a filter with pole-zero cancellation to shape the pulse.
\end{itemize}

\begin{figure}
    \centering
    \includegraphics[width=\textwidth]{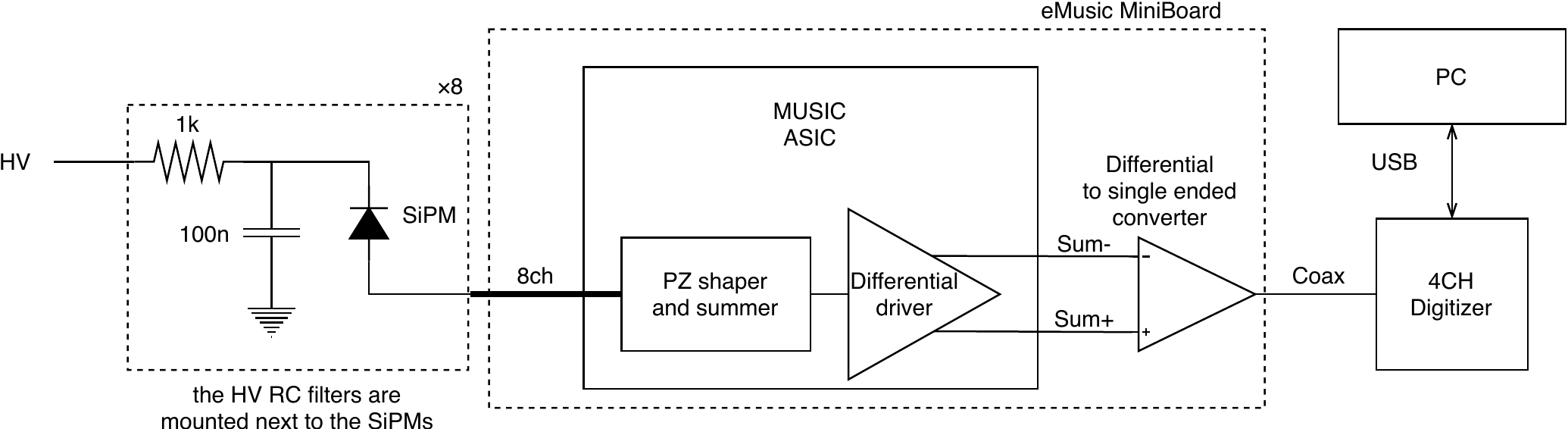}
    \caption{Scheme of the electronic readout including the summing stage.}
    \label{fig:sum_scheme}
\end{figure}

A scheme of the electronic readout of a LASiP prototype can be found in Figure~\ref{fig:sum_scheme}. The 8 individual SiPM signals are shaped and summed by the MUSIC. The MUSIC outputs the sum of the signals in differential mode. Then it is converted to single-ended and is sent to a digitizer or an oscilloscope for the acquisition.

Using the SCT matrix and the eMUSIC MiniBoard was useful for the proof-of-concept and especially efficient in terms of time and cost. However, this constrained the size and geometry of the pixel we could build. We decided to adopt the geometry shown in Figure~\ref{fig:LAPIS_a}: in our LASiP prototype the SiPMs are organized forming a square of $\sim2\times2$~cm$^2$ area, with a \textit{dead} corner of $\sim6\times6$~mm$^2$. We considered this was the best solution that would allow us to place four LASiPs near each other, building the \textit{micro-camera} that is described in the next section. The signal loss due to the pixel dead corner was studied with Monte Carlo (MC) simulations (see section~\ref{sec:met_sims}). The 8 active SiPMs of the SCT matrix are connected to the eMUSIC MiniBoard by means of a custom-made interface board (Figure~\ref{fig:LAPIS_c}).

\subsection{Proof-of-concept micro-camera}

To test the feasibility of using LASiPs in SPECT we built a \textit{micro-camera} consisting of a NaI(Tl) scintillation crystal coupled to an array of four LASiPs. This was the minimum number of LASiPs with which we could achieve our aims: (i) evaluate the energy resolution when charge is shared between pixels and compare it with that of standard SPECT scanners, (ii) prove that we can reconstruct simple images with a reasonable spatial resolution and (iii) have a system that we could use to validate MC simulations that let us study with further detail the impact of LASiPs in the performance of a gamma camera. We considered that these three goals were milestones that should be achieved before even considering to build a large SPECT camera based on LASiPs (which would be significantly more complex and expensive since it would involve much more channels). In this sense, the micro-camera was envisioned as a test bench for evaluating the developed LASiPs: we did not aim to perform a full characterization of its performance, especially in view of its limited number of channels.

\begin{figure}[tpb]
    \centering
    \includegraphics[height=6cm]{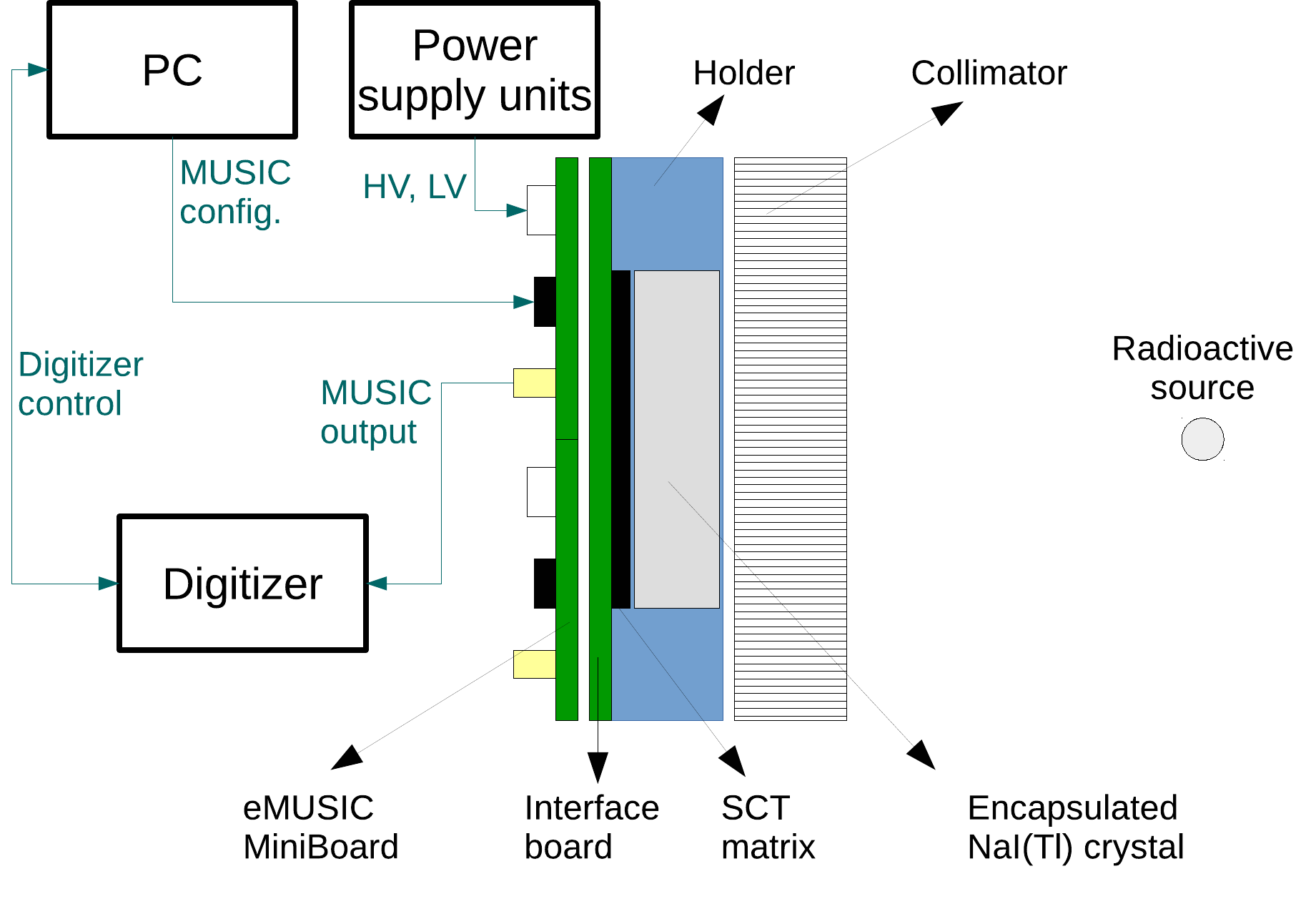}
    \caption{Scheme showing the different components of the micro-camera and setup employed.}
    \label{fig:setup}
\end{figure}

An overview of the micro-camera and the setup employed is shown in Figure~\ref{fig:setup}. The different components of the micro-camera are described in the next paragraphs.

\begin{figure}[htpb]
    \centering
    \includegraphics[height=5.5cm]{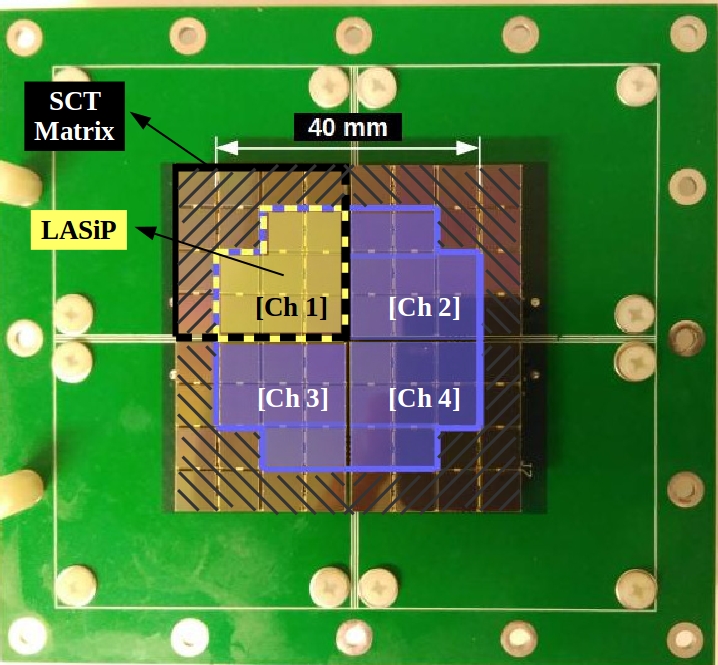}
    \includegraphics[height=5.5cm]{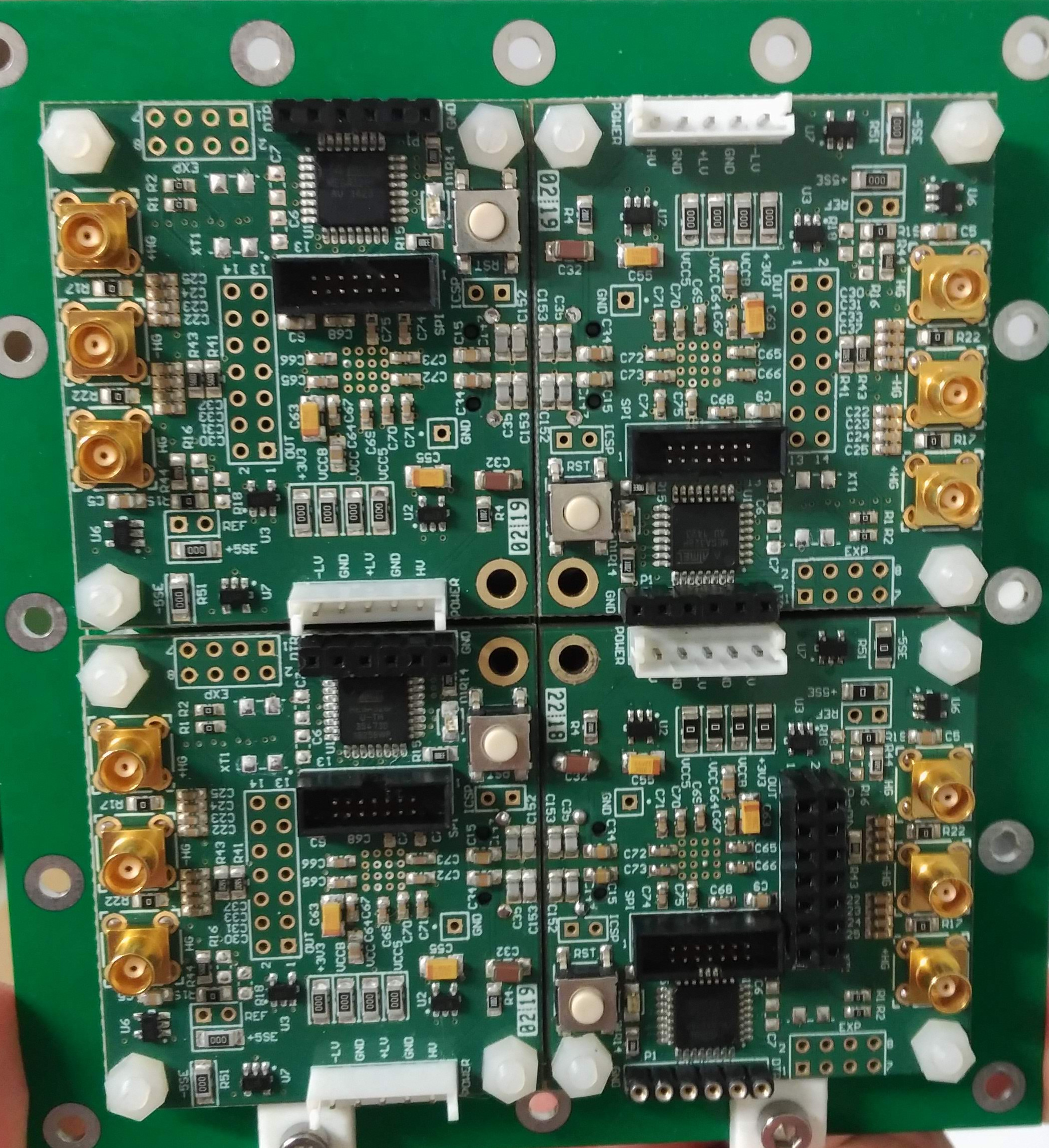}
    \includegraphics[clip=true, trim=6.5cm 8cm 4.2cm 4cm, height=5.5cm]{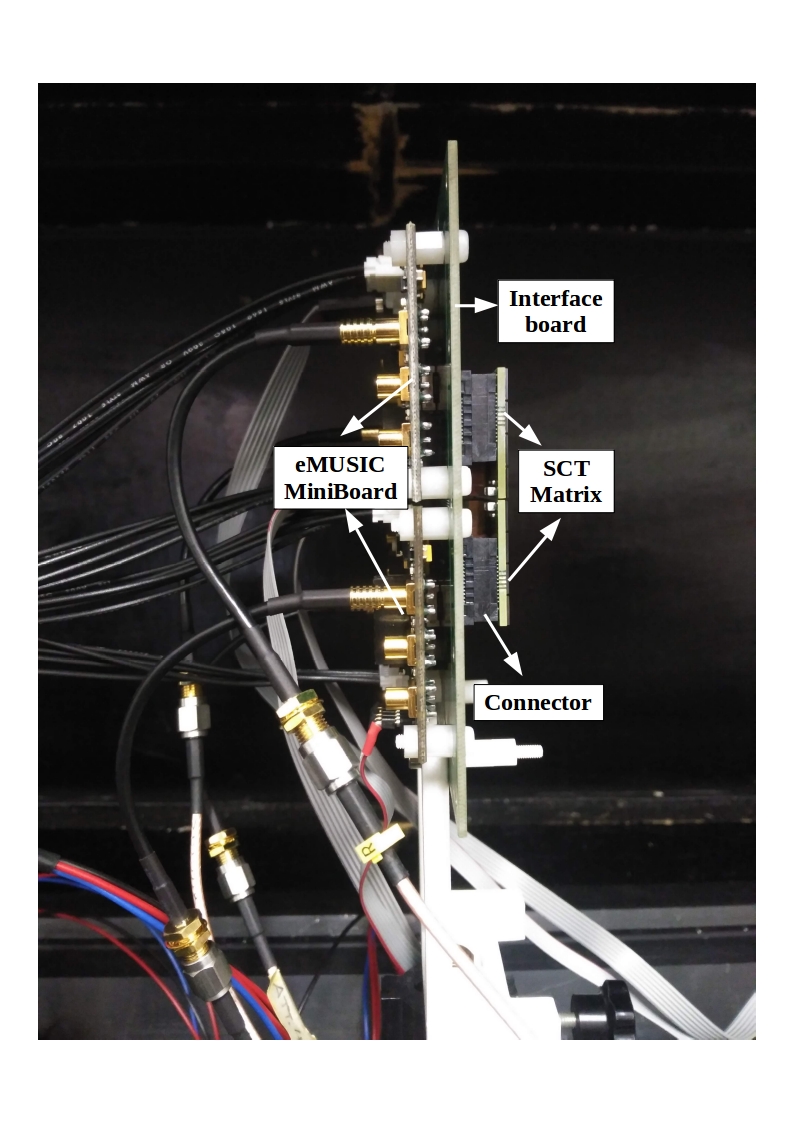}
    \caption{Electronic readout of the micro-camera. \textbf{Left:} Top-view of the micro-camera interface board holding 4 SCT matrices. As an example, the top-left matrix is delimited in black and the 8 SiPMs that are used to build a LASiP, in yellow. The micro-camera (in blue) uses 4 LASiPs. The 32 outermost SiPMs are not used. \textbf{Center:} The 4 eMUSIC MiniBoards seen from the back. \textbf{Right:} Side-view of the photodetectors and electronic boards.}
    \label{fig:LAPIS_4}
\end{figure}

\subsubsection*{Electronic readout}
The electronic readout of the micro-camera consists of 4 SCT matrices, 4 eMUSIC MiniBoards (one of each per LASiP) and an interface board (see Figure~\ref{fig:LAPIS_4}). The micro-camera interface board is designed to hold 4 LASiPs together and, in the frame of a single LASiP, connect the 8 SiPMs of each SCT matrix to their corresponding eMUSIC MiniBoard. The eMUSIC MiniBoard takes care of distributing the power supply to all SiPMs (applying individual offsets to each of them), shape each SiPM signal and sum all of them into a single output that is later sent to a digitizer. A common bias voltage ($V_b$) of 33~V was supplied to the 4 eMUSIC MiniBoards, which was chosen as a balance between PDE, DCR and the possibility to resolve the single-photoelectron (phe) pulse during the calibration. The pole-zero configuration was optimized to achieve a relatively high gain and short pulse tails that also facilitated the identification of the single-phe pulse. The individual offsets were adjusted until the conversion factor from charge to phes in all SiPMs were within 5\%.

\subsubsection*{Scintillation crystal, collimator and mechanics}
\begin{figure}
    \centering
    \includegraphics[clip=true, angle=90, trim=1cm 3cm 0cm 2cm, height=6cm]{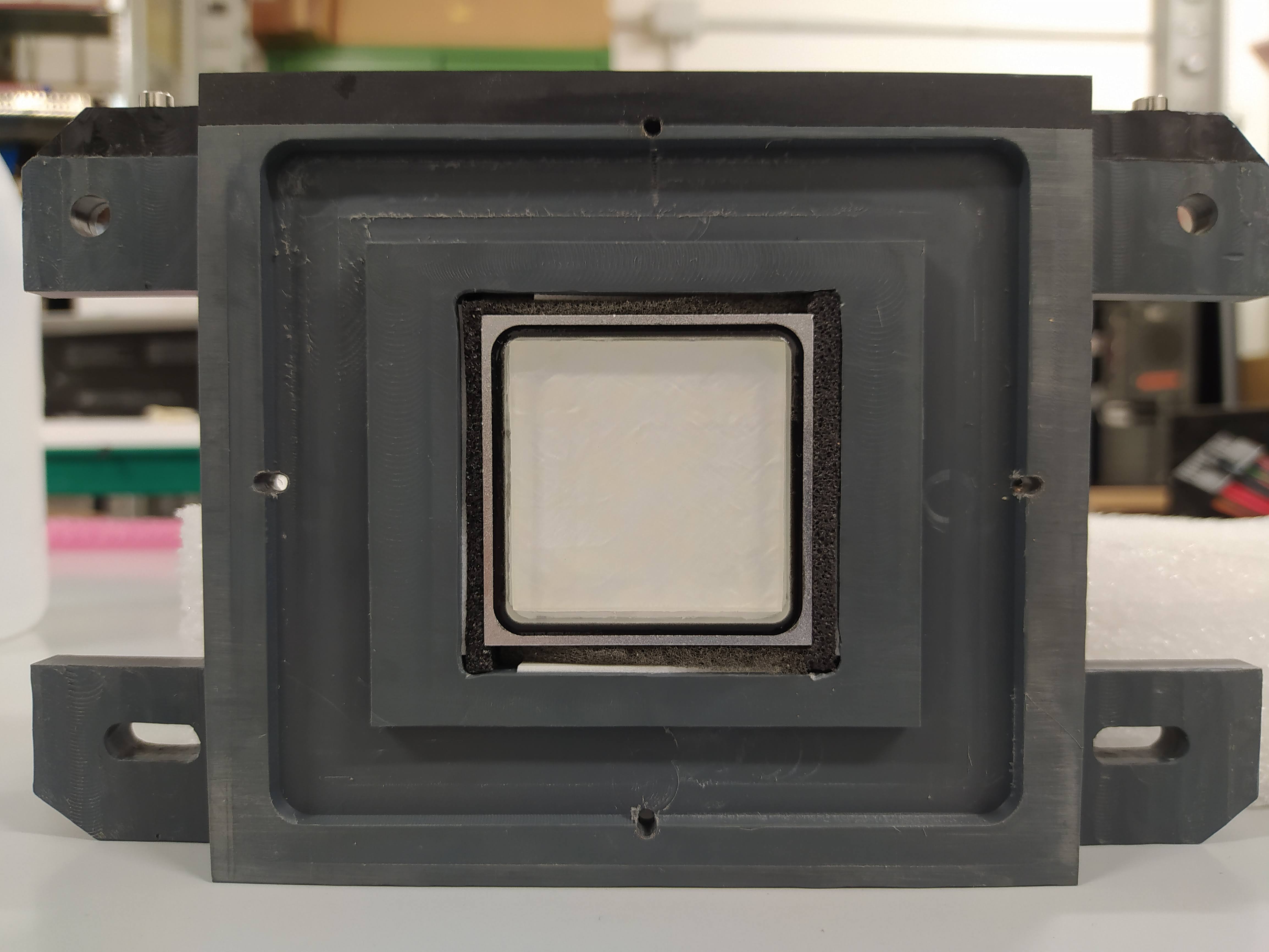}
    \includegraphics[clip=true, trim=4cm 2cm 1cm 2cm, height=6cm]{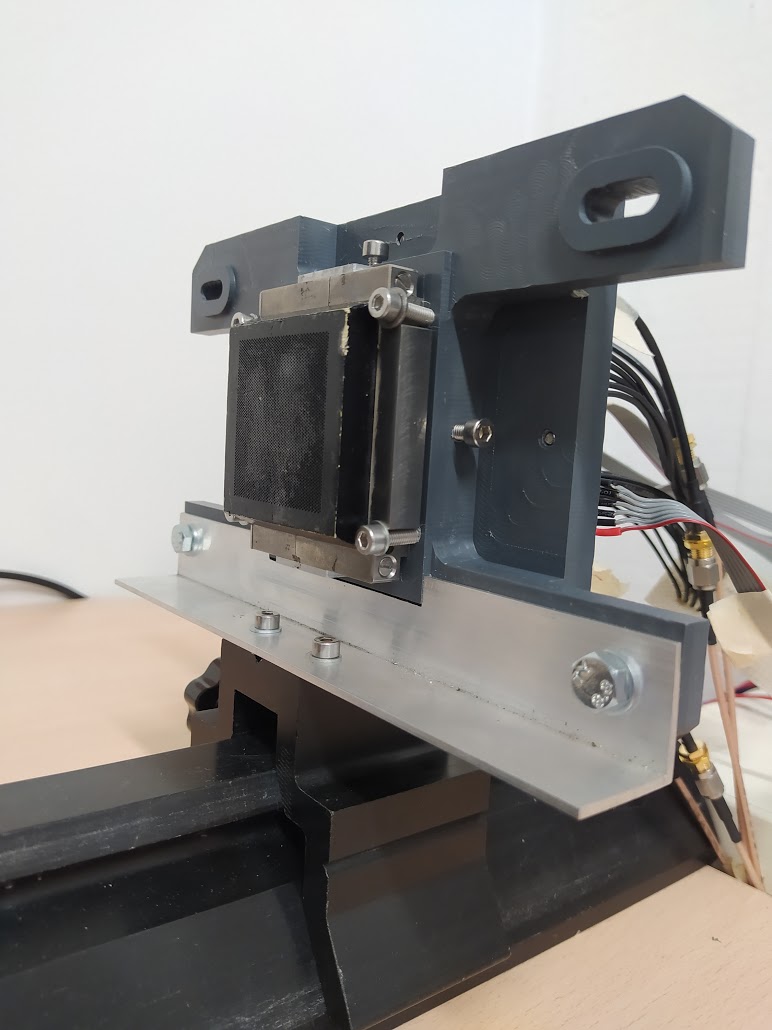}
    \includegraphics[clip=true, trim=6cm 2cm 4cm 7cm, height=6cm]{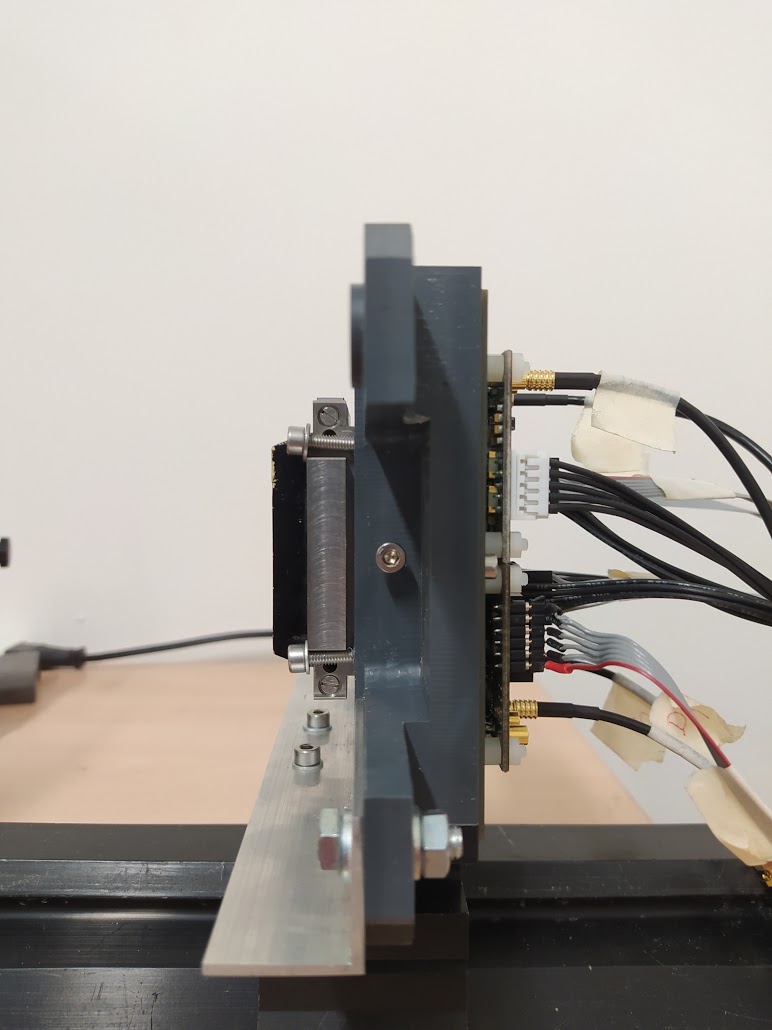}
    \includegraphics[clip=true, trim=2cm 0cm 10cm 5cm,height=6cm]{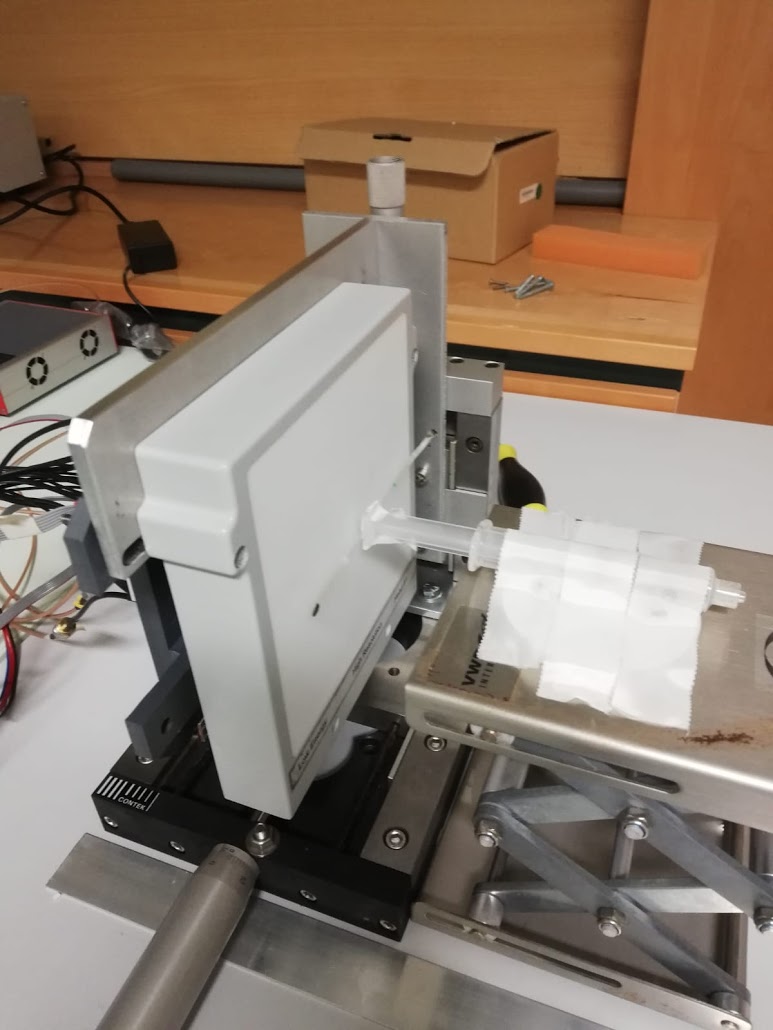}
    \caption{\textbf{Left:} The NaI(Tl) crystal and the custom-designed holder, seen from below, before mounting the electronic readout. \textbf{Center:} Two different views of the micro-camera and \textit{Coll 1} fixed in a rail during measurements in the lab. \textbf{Right:} Micro-camera and \textit{Coll 2} mounted in the positioning platform at Careggi Hospital (Firenze, Italy). In this image a capillary was filled with a solution containing $^{99m}$Tc and was placed close to the collimator.}
    \label{fig:microcam}
\end{figure}

A NaI(Tl) crystal of $40\times40\times8$~mm$^3$ bought from OST Photonics\footnote{{https://www.ost-photonics.com/product-category/scintillation-crystal-2/naitl-scintillator}} was coupled to the 4 LASiPs with SS-988 optical gel (refractive index 1.47, above $99\%$ transmission between 300 and 600~nm) from Silicone Solutions\footnote{{https://siliconesolutions.com/ss-988.html}}. According to the manufacturer, the crystal, which is sealed in an aluminum housing, is surrounded by an MgO diffuse reflector and has a 3~mm thick fused silica glass exit window (see Figure~\ref{fig:microcam}). Crystal, LASiPs and the electronic readout boards were mounted in a 3D-printed holder that was designed to be attached to two different lead collimators. The first one, \textit{Coll 1} had a hole diameter $d\simeq0.5$~mm and a thickness $a\simeq2$~cm. The second one, \textit{Coll 2} was a clinical LEUHR collimator ($d\simeq1.2$~mm and $a\simeq3$~cm). Holder and collimator could be mounted in a positioning platform, specifically designed to allow movements of the camera in the detector plane with sub-millimeter precision.

\subsubsection*{Data acquisition} A CAEN DT5720 digitizer was used for the acquisition (250 MS/s). Individual discriminator thresholds were set to each channel, optimized to minimize the triggering by dark count events. For each event, a 2~$\mu$s waveform was acquired on each channel. The charge was integrated offline in a 600~ns window. The width and position of the integration window were both optimized aiming to maximize energy resolution. The longer the integration window, the larger the number of collected photons, but also the number of integrated dark counts. The optimal integration time depends on the SiPM bias voltage, since it affects both PDE and DCR. The Pole-Zero shaper may also have an impact the optimal integration time, although we did not study in detail such eventual dependence.

\subsection{Energy reconstruction}\label{sec:enrec}
 
 The total charge $Q$ collected in an event is defined as the sum of the charges collected by each LASiP. With all the values of $Q$ obtained during a measurement we built a histogram where we could identify the photopeak corresponding to the energy of the gamma rays emitted by the radioactive source employed. Only events with an energy within $\pm15\%$ of the photopeak position were used for the image reconstruction (see section~\ref{sec:evtrec}).
 
\subsection{Event position reconstruction.}\label{sec:evtrec}
 
 The method used to reconstruct the position of an event begins with a simple centroid method. Spatial linearity and uniformity corrections are later applied to produce the final image~\cite{Simmons_1988}. More complex image-reconstruction algorithms exist (most of them also start from the centroid method) and have the potential to provide better spatial resolution (see for instance~\cite{Passeri_1993, Li_2010, Morozov_2015}). However, testing other methods was beyond the scope of this work.
 
 \subsubsection{Raw images reconstructed through the centroid method.}
 In the centroid method, the coordinates $(x_c,y_c)$ of an event are reconstructed as
 \begin{equation}
     x_c =  \frac{\sum^{4} _{i=1} x_i q_i}{\sum^{4} _{i=1} q_i} \;\;\; ; \;\;\; y_c =  \frac{\sum^{4} _{i=1} y_i q_i}{\sum^{4} _{i=1} q_i}
 \end{equation}
 where $q_i$ is the charge measured by the $i$-th pixel that has its center in $(x_i,y_i)$. 
This method, although fast and simple, has several limitations. Events cannot be reconstructed outside the region delimited by the $(x_i,y_i)$ of the four pixels. This is true even in the ideal case in which the crystal surfaces are perfectly polished and all scintillation photons are carrying information of their initial direction. In a more realistic scenario, events contain also a diffuse component. For instance, due to diffuse reflections in the crystal walls, as in our case. As a result, with the centroid method events are reconstructed within an area that is much smaller than the one delimited by the center of the four pixels.

\subsubsection{Spatial linearity and uniformity correction.}
To recover spatial linearity we followed a similar method to what was described in~\cite{Popovic_2014}. A radioactive source was fixed near the micro-camera and, using the positioning platform, the camera was moved to scan the field of view (FOV). In our case, the size of the FOV after corrections was limited by the range of the moving platform (15~mm).

We built a grid of measurements in which the relative position between source and detector was known. For each measurement we mapped the mean position reconstructed with the centroid method $(x^{rec} _c, y^{rec} _c)$ to its known position $(x^{true} _c,y^{true} _c)$. This map can be interpolated so that each event reconstructed with the centroid method can be corrected to recover spatial linearity.

A uniformity correction map was generated by taking a long-exposure flood-field irradiation: a $^{99m}$Tc source was placed far (more than 50~cm away) from the micro-camera. The image was reconstructed with the centroid method and corrected by spatial linearity. The inverse of this image was the uniformity correction map, which was applied to every reconstructed image.

\subsection{LASiP noise measurements}\label{sec:met_noise}
One of the main limitations to increase the number of SiPMs that are summed to build a pixel is the degradation of the single-phe resolution. The noise of all the SiPMs that build a LASiP are summed, which degrades the timing performance and single-phe resolution of a pixel. The noise (e.g., DCR, optical cross-talk) of a single FBK NUV-HD3 SiPM has been studied in~\cite{leonardoSPIE_2019}. Here we focused on the additional noise introduced by the summing stage.  We studied in particular two forms of noise that could have a significant impact in the performance of a SPECT system: the single-phe resolution (the ability to resolve differences of a few photons, which depends on the amplitude of the measured signal) and uncorrelated noise, which includes SiPM dark counts, electronic and digitizer noise and is independent of the measured signal.

\begin{figure}
    \centering
    \includegraphics[height=6cm]{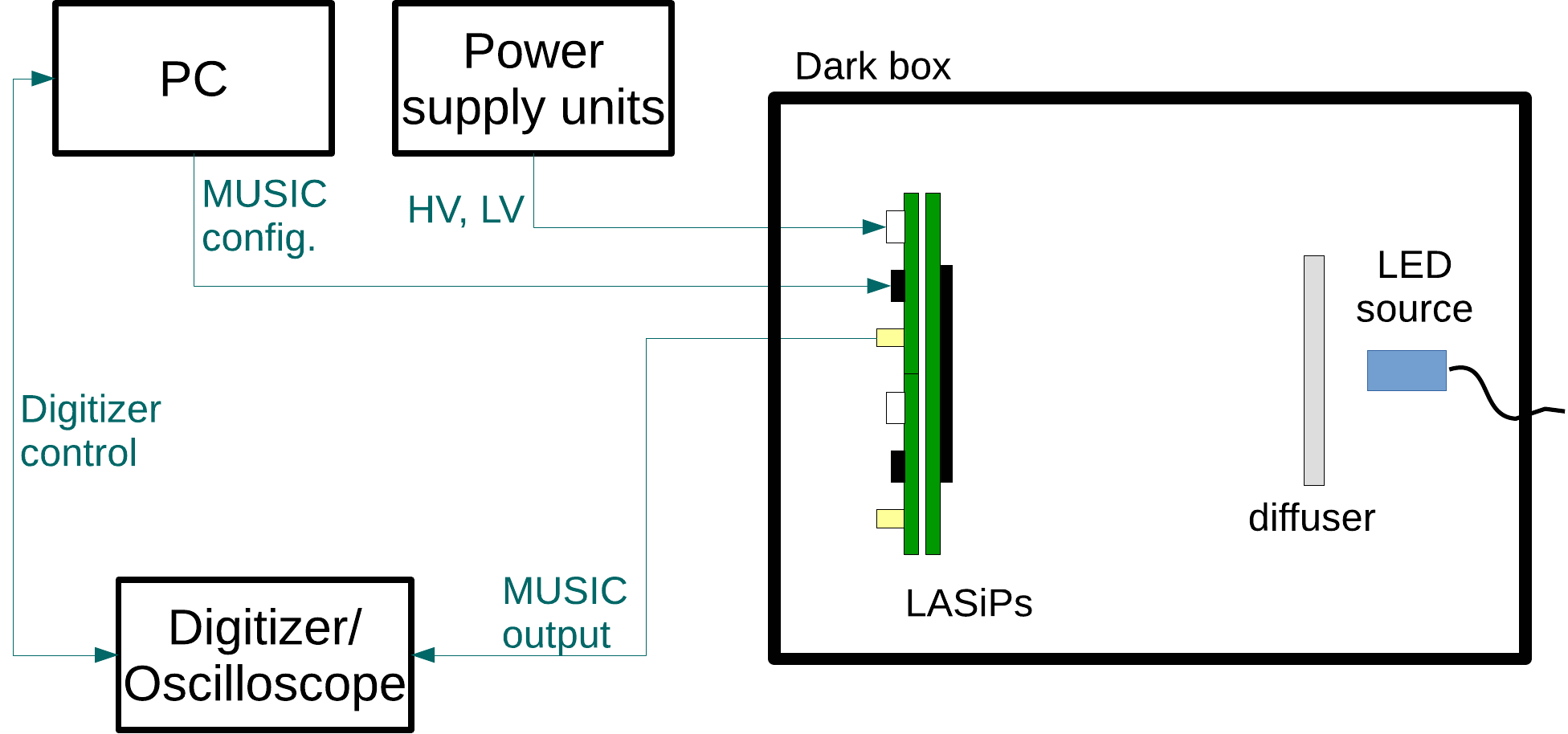}
    \caption{Setup employed to characterize LASiP noise.}
    \label{fig:setup_noise}
\end{figure}

The setup employed for noise measurements is shown in Figure~\ref{fig:setup_noise}. LASiPs were placed inside a dark box where they could be illuminated by $\sim0.5$~ns pulses of $\sim380$~nm generated by a PicoQuant PDL 800-B LED driver. To study the degradation of the single-phe resolution introduced by the summing stage, we turned on the LED driver, set a pulse frequency of 1~kHz, and recorded the waveforms with an oscilloscope. We repeated the measurement eight times, changing the number of SiPMs summed by the MUSIC from 1 to 8. The LED intensity was regulated in each measurement to keep the mean number of detected photons at the level of a few phes to facilitate the identification of the single-phe peak. We obtained the charge spectrum for each of these measurements and fitted them with eq.~2 of~\cite{Guberman_2019}, a multi-peak fit function that includes optical cross-talk and is standard for describing SiPM charge spectra:
\begin{equation}\label{eq:XTfit}
    f(x) \sim P(0\mid \mu) \; G(x-x_0, 0, \sigma_{0}) + \sum^N _{n=1} \sum^n _{m=1} p_{n,m}(p_{XT}) P(m\mid \mu) G(x-x_0, n, \sigma_{t})
\end{equation}
$P(m\mid \mu)$ is the Poisson probability of having $m$ cells fired given a mean number of interacting photons $\mu$ and $G(x, n, \sigma)$ is a Gaussian function of expected value $n$ and variance $\sigma^{2}$. The optical cross-talk probability $p_{XT}$ is modeled by a binomial function $p_{n,m}$ (see eq.~1 of~\cite{Guberman_2019}). $x$ is measured in phe and $x_0$ is the position of the pedestal peak. $\sigma_{t} = \sqrt{\sigma^{2}_{0}+n \sigma^{2}_{1}}$ gives the width of the $n$-th peak, where $\sigma_{0}$ is defined as the pedestal noise and $\sigma_{1}$ is typically associated to SiPM cell-to-cell gain fluctuations. The lower $\sigma_{t}(n)$ is, the better is the SiPM resolution of the $n$-th peak. Both $p_{XT}$ and $\sigma_{t}(n)$ introduce uncertainties in the measured photon flux that could have an impact on the energy resolution of the system. %With the measurements we performed and the MC simulations described in section~\ref{sec:met_sims} we were able to evaluate that impact.

To study the impact of uncorrelated noise we recorded events in the absence of visible-light or radioactive sources (with all eight channels enabled for the sum). Random-triggered events were acquired with the digitizer in the same conditions than during standard measurements with the micro-camera. Charge was integrated in the same 600~ns window employed for the charge extraction. We fitted the charge distribution with a Gaussian function of variance $\sigma^2_{UN}$, which was used as input for the simulations described in section~\ref{sec:met_sims}.

\subsection{Evaluation of the micro-camera performance}\label{sec:met_perf}

We evaluated the energy and spatial resolution of the micro-camera in a small region around the FOV center. We limited our measurements to this region aiming to be as far as possible from the dead corners.  The performance closer to the edges would be naturally worse because in the micro-camera all 4 pixels are \textit{outer-most pixels}, but especially because of the proximity to the pixel dead corners. Then it would not be very representative of what may occur in a large camera equipped with ``proper'' LASiPs (i.e., without dead corners). The central region was the only place in which we could obtain results that could eventually be used to undesrtand how LASiPs would perform in a larger camera. For the measurements we employed the setup of Figure~\ref{fig:setup}. We used a sealed source of $^{241}$Am (which we assumed to be point-like) and a liquid radioactive solution of $^{99m}$Tc, embedded into a glass capillary (0.5~mm inner diameter, 100~mm long).

The energy resolution was measured employing $^{99m}$Tc, using two different setups. In the first setup, the capillary was completely filled with the radioactive solution (total activity of $\sim50$~$\mu$C) and placed in different positions, near the collimator. In the second setup the collimator was removed and the capillary was filled for a length not exceeding 2~mm. Its active volume was positioned at a distance of $\sim500$~mm from the camera, with its longitudinal axis perpendicular to the camera plane. This way we produced a flood-field irradiation of the detector.

Spatial resolution was evaluated with both $^{241}$Am and $^{99m}$Tc sources. First, the $^{241}$Am source was imaged at a distance $h\simeq15$~mm from the camera using \textit{Coll~1}. We fitted the reconstructed image by a 2D-gaussian and defined the extension $R$ of the image as the FWHM of the function that resulted from the fit. Then the capillary was fully-filled with the $^{99m}$Tc solution and was imaged at a distance $h\simeq20$~mm from the camera using \textit{Coll~2}. In this case we define $R$ as the FWHM of the projection of the reconstructed image in the axis perpendicular to the capillary orientation. In both cases, with the source fixed, the micro-camera was moved in the detector plane to acquire data from different regions of the FOV.

The measured $R$ results from the contribution of the source diameter $R_{src}$ (negligible for the point-like source), the collimator resolution $R_{c}$ and the intrinsic detector resolution $R_{d}$ as:
\begin{equation}\label{eq:ResExt}
    R = \sqrt{R^2 _{d}+R^2 _{c}+R^2 _{src}}
\end{equation}
The collimator resolution depends on the hole diameter $d$, the distance $h$ between source and collimator and the effective collimator thickness $a_{eff}$ as 
\begin{equation}\label{eq:Rcoll}
    R_{c} (h) = d \frac{a_{eff}+h}{a_{eff}}
\end{equation}
where $a_{eff}=a-2/\mu$, with $a$ the collimator thickness and $\mu$ the linear attenuation coefficient ($\mu^{-1}= 0.37$~mm at 140 keV in lead).

\subsection{Simulations}\label{sec:met_sims}

To better understand the system behavior we performed Geant4 simulations~\cite{Allison_2006} of a system with similar characteristics to the micro-camera. We did not aim to perfectly match the micro-camera response, but to model a system in which we could study the impact that LASiP noise and the pixel dead corner had on the overall performance.
 
\begin{figure}[t]
    \begin{subfigure}{.33\textwidth}
    \centering
    \includegraphics[width=0.9\textwidth]{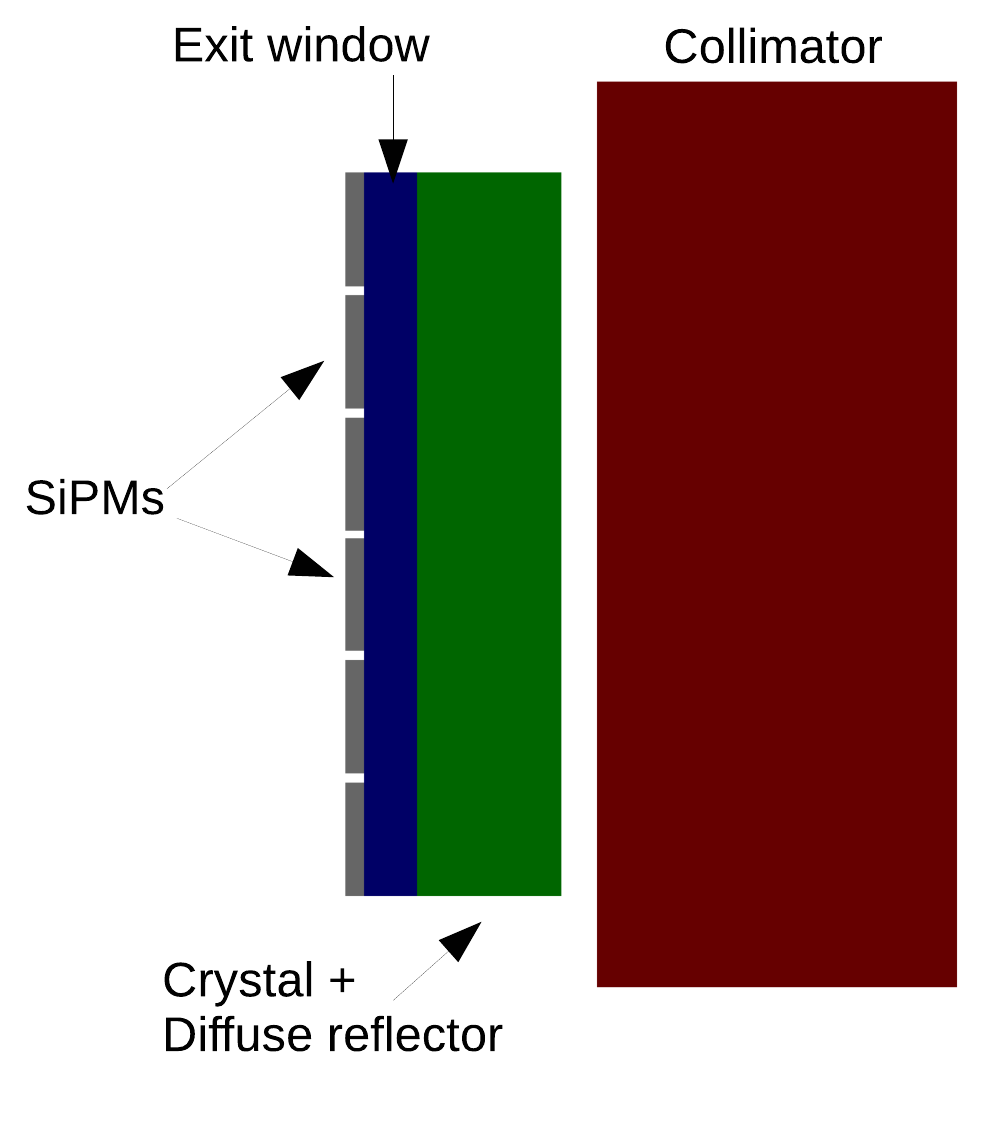}
    %\subcaption{}\label{fig:sims_side}
    \end{subfigure}
    \begin{subfigure}{.24\textwidth}
    \centering
    \includegraphics[width=0.9\textwidth]{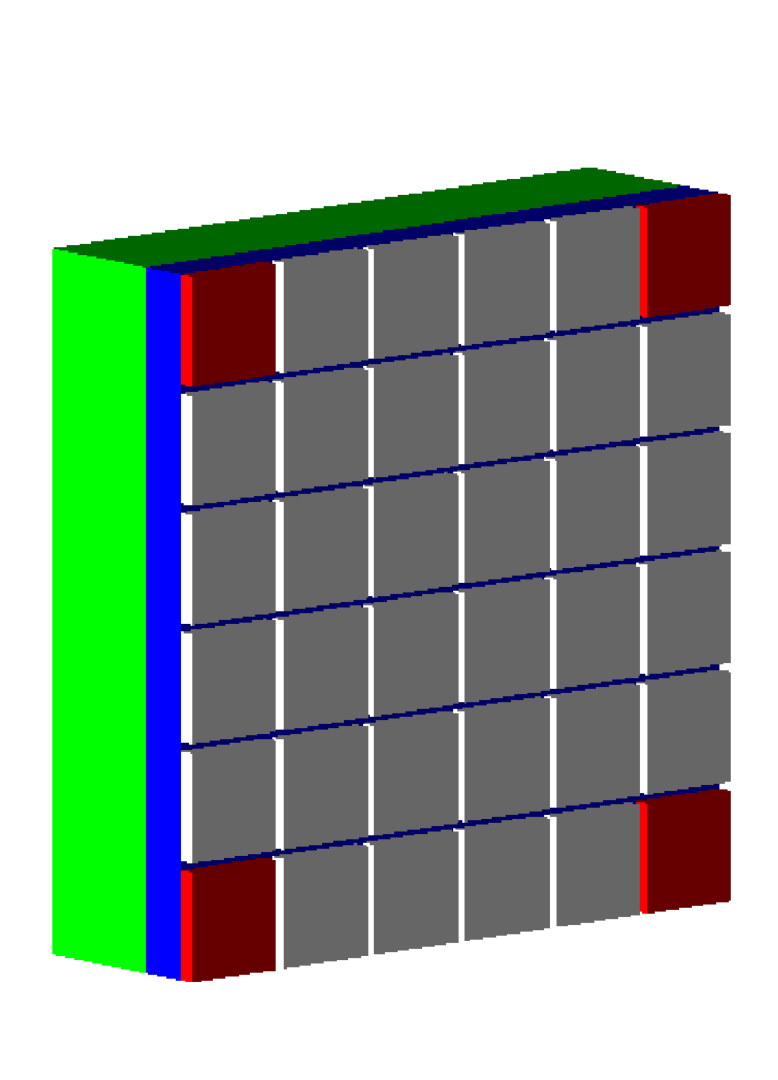}
    %\subcaption{}\label{fig:sims_sipms}
    \end{subfigure}
    \begin{subfigure}{.33\textwidth}
    \centering
    \includegraphics[width=0.9\textwidth]{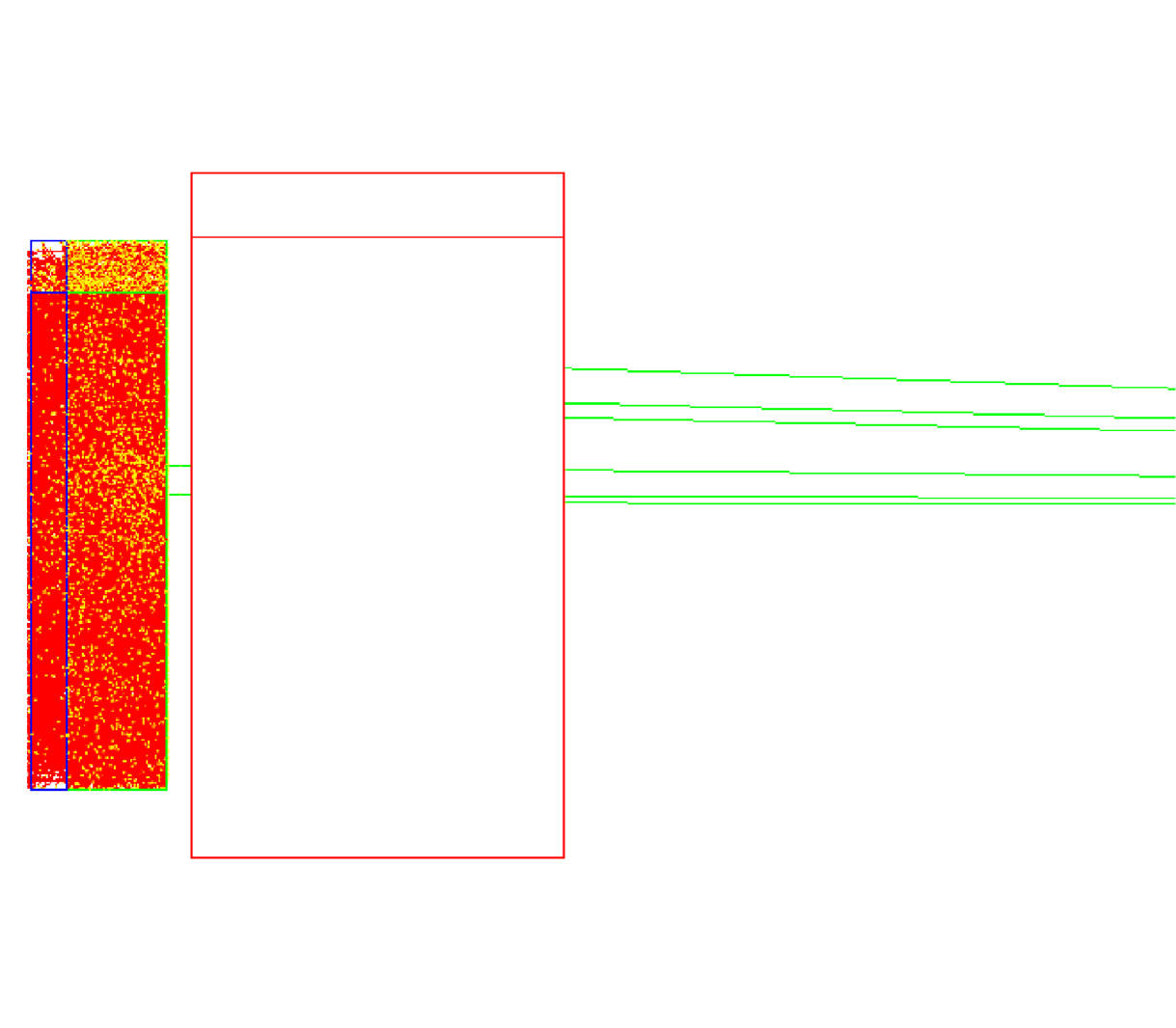}
    %\subcaption{}\label{fig:sims_ex}
    \end{subfigure}
    \caption{Overview of the simulated system. \textbf{Left:} The different components of the simulated system (side view). \textbf{Center:} The back part of 36 the simulated SiPMs are shown. The output from the SiPMs in the corner (in red) can be enabled/disabled to study the impact of the LASiP dead corner. \textbf{Right:} A representation showing different gamma rays approaching the detector. A few of them are able to go through the collimator and produce scintillation photons inside the crystal.}
    \label{fig:setup_sims}
\end{figure}

The simulated system features a $40\times40\times8$~mm$^3$ NaI(Tl) crystal coupled to 36 SiPMs of $\sim6\times6$~mm$^2$ (Figure~\ref{fig:setup_sims}). A 140~keV capillary source of 0.5~mm diameter and 40~mm long was also simulated. The optical photons generated by scintillation inside the crystal were tracked until they were absorbed, escaped or detected by one of the SiPMs. An MgO reflector surrounding the crystal was also included in the simulations. We used the Geant4 \textit{RoughTeflon LUT Davis} model~\citep{Stockhoff_2017} for describing the interaction of optical photons in the crystal-reflector interface.  It was the one that was better reproducing the relative charge distribution between pixels observed in the micro-camera, at least compared to the other models we tested: \textit{Glisur} (with several surface roughness) and \textit{Unified} (with different surface finish like \textit{groundteflonair} and \textit{etchedteflonair}). A 3~mm fused silica glass exit window was placed between SiPMs and crystal. Lead collimators with the same geometrical characteristics of \textit{Coll1} and \textit{Coll2} could be placed in front of the camera.

The 36 SiPMs were distributed in the same way as in the micro-camera. The number of scintillation photons detected by each SiPM were recorded independently. Then they were summed in groups of 8 to mimic a micro-camera LASiP or in groups of 9 to study the performance degradation introduced by the dead corner (see Figure~\ref{fig:setup_sims}). The PDE of the SiPMs was not simulated. Instead, we scaled the scintillation light output so that the total number of photons was comparable to what we estimated from the measurements with the micro-camera. According to the manufacturer, the NaI(Tl) crystal employed in the micro-camera coupled to a PMT exhibits an energy resolution of $\sim8\%$ at 662~keV. In the simulations we scaled (and fixed) the scintillation light-yield to achieve an energy resolution of $\sim9\%$ at 140~keV when all 36 SiPMs are enabled and no noise is added.
 
We also did not simulate the LASiP pulse shape or the waveforms acquired with the digitizer. Instead, we injected noise in three steps. In a simulated event in which $N$ scintillation photons hit a single LASiP we:
 \begin{enumerate}
     \item Simulate cross-talk events: we add $\Delta N$ artificial counts, that are randomly generated with a Poisson distribution with mean $\mu(N,p_{XT})$, where $p_{XT}$ is the cross-talk probability. At the end of this step $N$ is replaced by $N'=N+\Delta N$
     \item Simulate the finite resolution of the detector: we build a Gaussian distribution with expected value $N'$ and variance $\sigma_{t}(\sigma_0,\sigma_1,N')$ (see definition in section~\ref{sec:met_noise}). At the end of this step $N'$ is replaced by a random number $N''$ generated with this Gaussian distribution.
     \item Simulate uncorrelated noise: $\Delta N''$ random counts are generated with a Gaussian distribution of variance $\sigma^2 _{UN}$. $\sigma_{UN}$ is the standard deviation of the charge distribution collected in the absence of signal (see definition in section~\ref{sec:met_noise}). At the end of this step $N''$ is replaced by $N'''=N''+\Delta N''$.
 \end{enumerate}
Note that four parameters should be input to simulate the noise: $p_{XT}$, $\sigma_{0}$, $\sigma_{1}$ and $\sigma_{UN}$. The reference values for these parameters were estimated from the LASiP noise measurements described in section~\ref{sec:met_noise}. Then we could vary them to study the individual impact of the different forms of noise in the micro-camera performance.
 
A charge histogram is finally built with the sum of the noise-corrected charge measured in each pixel. This charge histogram is later used to evaluate the energy resolution and to extract the acceptance window for the image reconstruction, the same way it was done with the micro-camera (section~\ref{sec:enrec}). Since we wanted to study the specific impact of the LASiP noise in the performance, the rest of the detector components (crystal, reflector, coupling material, exit window) were treated as ideal objects. Simulations do not include background effects that are present in the data like Iodine escape peak or scattering in other materials that are not those belonging to the micro-camera.

\section{Results}\label{sec:results}

\subsection{LASiP Noise}\label{sec:res_noise}
\begin{figure}[htb]
    \centering
    \includegraphics[width=\textwidth]{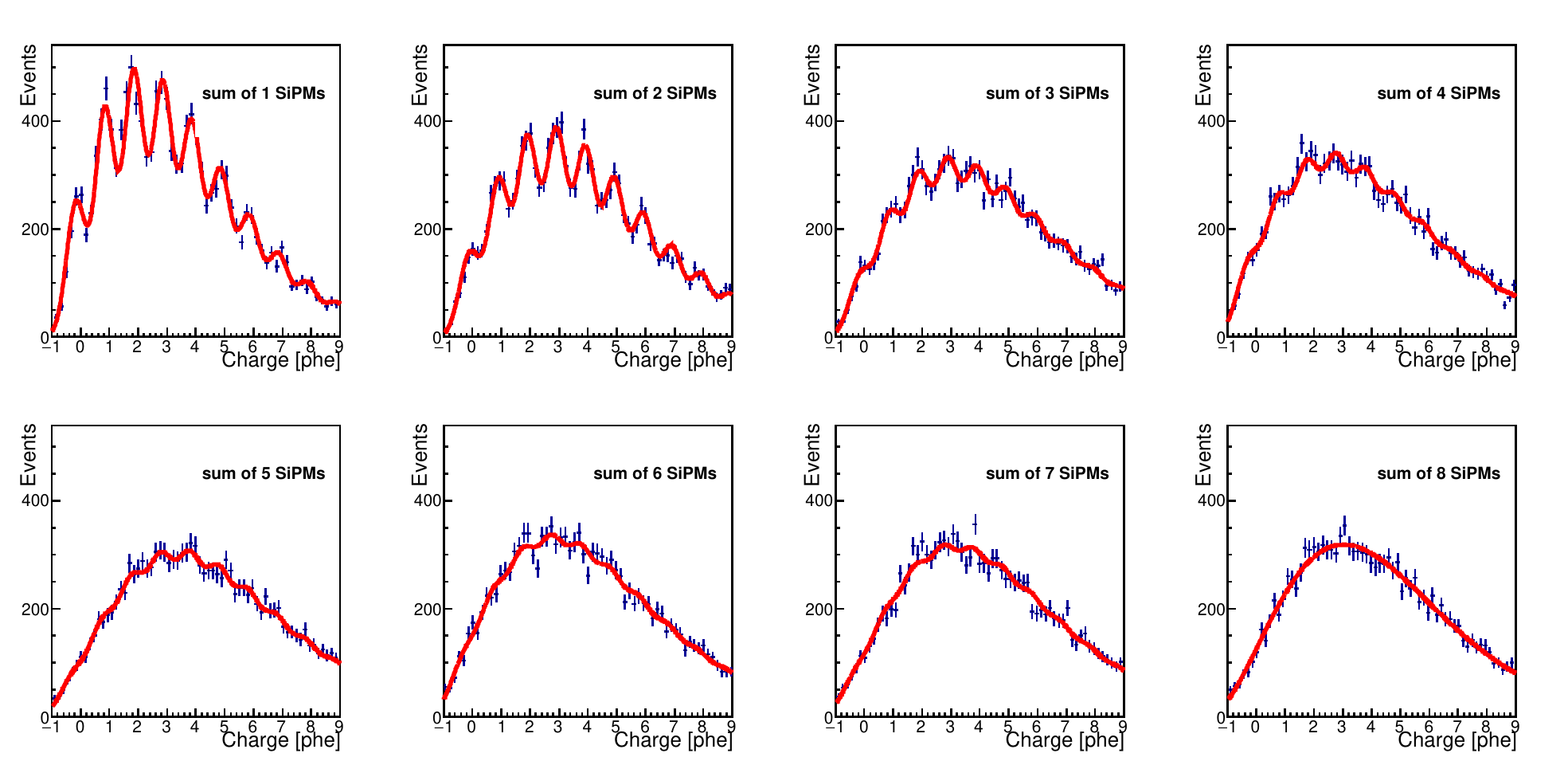}
    \caption{From left to right and top to bottom: evolution of the single-phe spectrum as the number of summed SiPMs increases. In red the fit performed using eq.~\ref{eq:XTfit}. The data were acquired with an oscilloscope.}
    \label{fig:singlephe}
\end{figure}

\begin{figure}[t]
    \begin{subfigure}{.48\textwidth}
    \centering
    \includegraphics[width=0.9\textwidth]{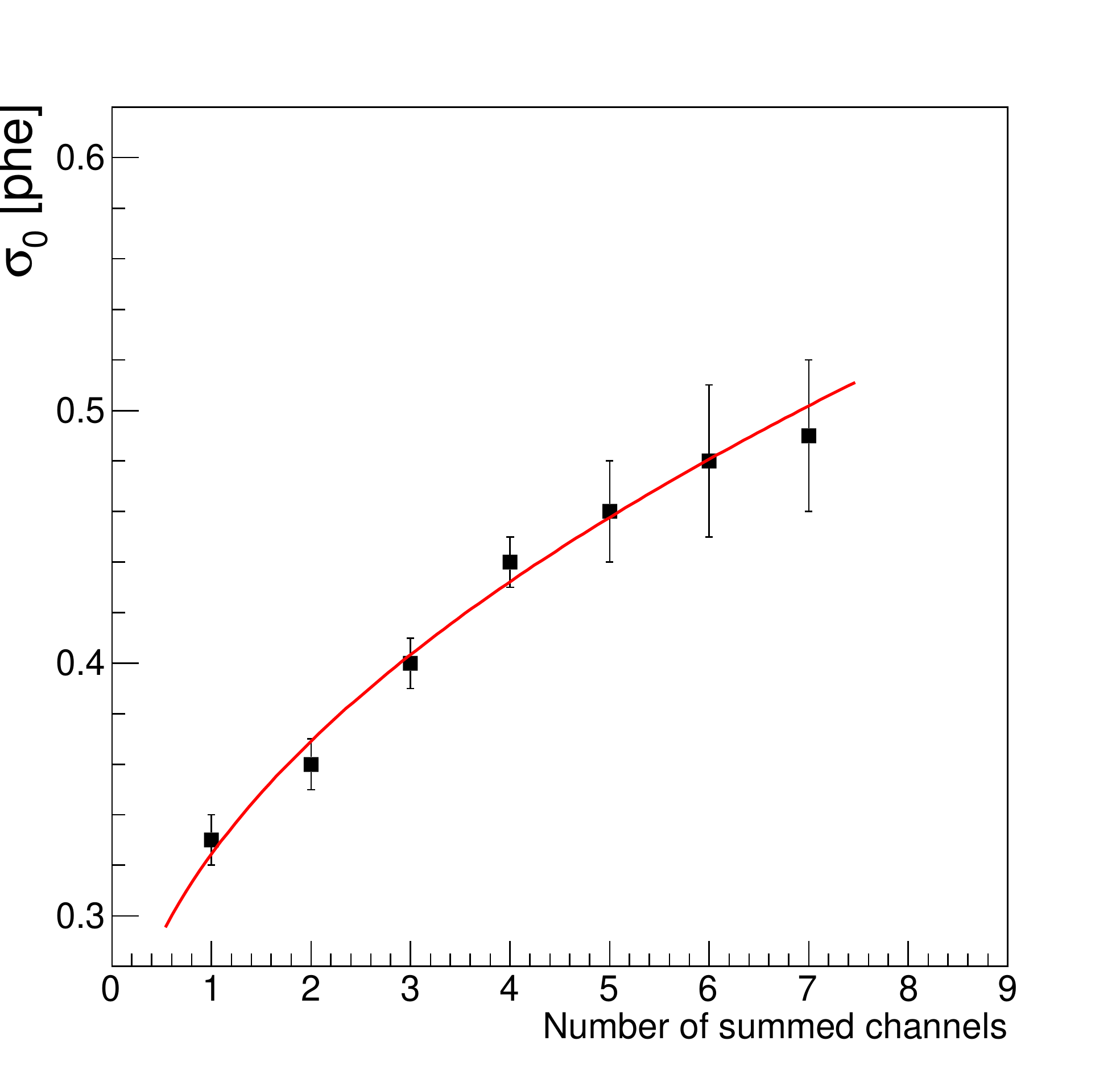}
    \subcaption{}\label{fig:sigma0}
    \end{subfigure}
    \begin{subfigure}{.48\textwidth}
    \centering
    \includegraphics[width=0.9\textwidth]{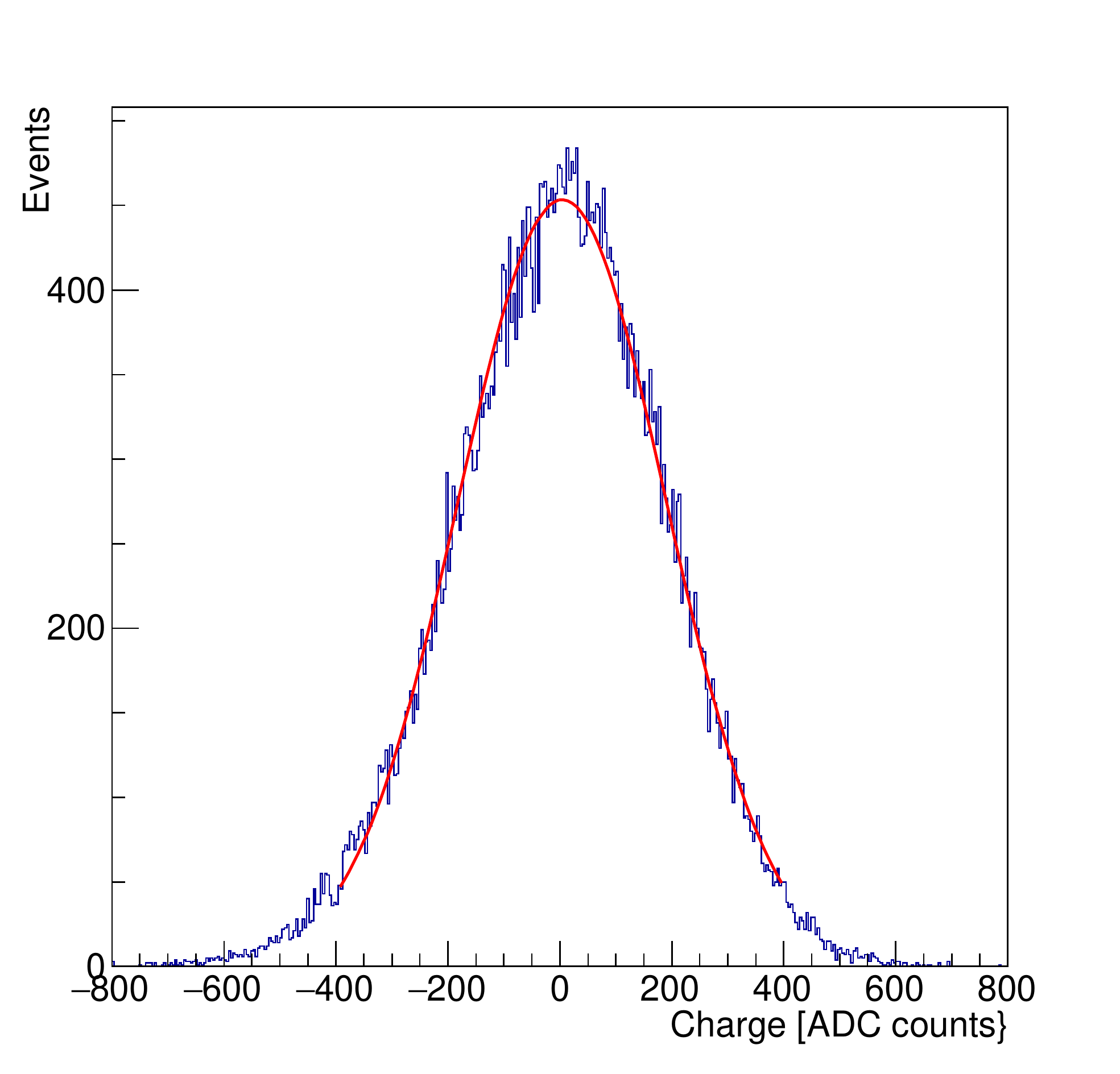}
    \subcaption{}\label{fig:diginoise}
    \end{subfigure}
    \caption{\textbf{Left:} Values of $\sigma_0$ obtained from the fits performed in Figure~\ref{fig:singlephe} as a function of the number of summed SiPMs. In red the fit described in the text. \textbf{Right:} Charge distribution recorded with the digitizer for a single LASiP summing 8 SiPMs in the absence of visible-light or radioactive sources. In red the Gaussian fit performed from which we obtained $\sigma_{UN}\simeq186$~ADC counts. }
    \label{fig:my_label}
\end{figure}

Figure~\ref{fig:singlephe} shows the single-phe spectra obtained when flashing a LASiP with a pulsed LED as described in section~\ref{sec:met_noise}. The single-phe resolution degrades as the number of summed SiPMs increases. We were able to fit all spectra using eq.~\ref{eq:XTfit}, except for the case in which we summed 8 SiPMs in which it was impossible to identify peaks. The fit parameters $p_{XT}$ and $\sigma_1$, were reasonably constant in all seven cases, with values of $\sim25\%$ and $\sim0.07$~phe respectively. As expected, $\sigma_0$ increased as $\sqrt{N_{SiPM}}$, with $N_{SiPM}$ the number of summed SiPMs. This can be seen in Figure~\ref{fig:sigma0}, where we fitted the points in the range between 1 and 7 SiPMs as $\sigma_0 = p_0 + p_1  \sqrt{N_{SiPM}}$, with $p_0 = (0.22 \pm 0.02)$~phe and $p_1 = (0.11 \pm 0.01)$~phe. The fitting function can be evaluated to estimate $\sigma_0 \left(N_{SiPM}=8\right)\simeq 0.53$~phe.

Figure~\ref{fig:diginoise} shows the charge distribution obtained using the digitizer for the acquisition, for a LASiP summing 8 SiPMs in the absence of sources (uncorrelated noise measurements). The distribution was fitted by a Gaussian of $\sigma_{UN}=186$~ADC counts, which corresponds to $\sim0.6\%$ of the mean position of the photopeak (see section~\ref{sec:eres_res}). 

From the results reported in this section we defined the reference values of the input parameters for simulating the pixel noise (see section~\ref{sec:met_sims}). These values are shown in Table~\ref{tab:ref_noise}. The MC images shown in sections~\ref{sec:res_evtrec} to~\ref{sec:spres_res} contain noise simulated with those values.

\begin{table}[ht]
    \centering
    \begin{tabular}{ c c c c}
          $p_{XT} [\%]$ & $\sigma_0$ [phe] & $\sigma_1$ [phe] & $\sigma_{UN}$ [Q] \\
         \hline \hline
         25 & 0.53 & 0.07 & 6.e-3  \\
         \hline
    \end{tabular}
    \caption{Reference input noise parameters of the simulations. Note that $\sigma_{UN}$ is given in units of the total charge $Q$ collected in the four pixels at the photopeak.}
    \label{tab:ref_noise}
\end{table}

\subsection{Event reconstruction}\label{sec:res_evtrec}

\begin{figure}[htpb!]
\begin{center}
    \begin{subfigure}{.62\textwidth}
    \centering
    \begin{subfigure}{.45\textwidth}
    \centering
    \includegraphics[width=\textwidth]{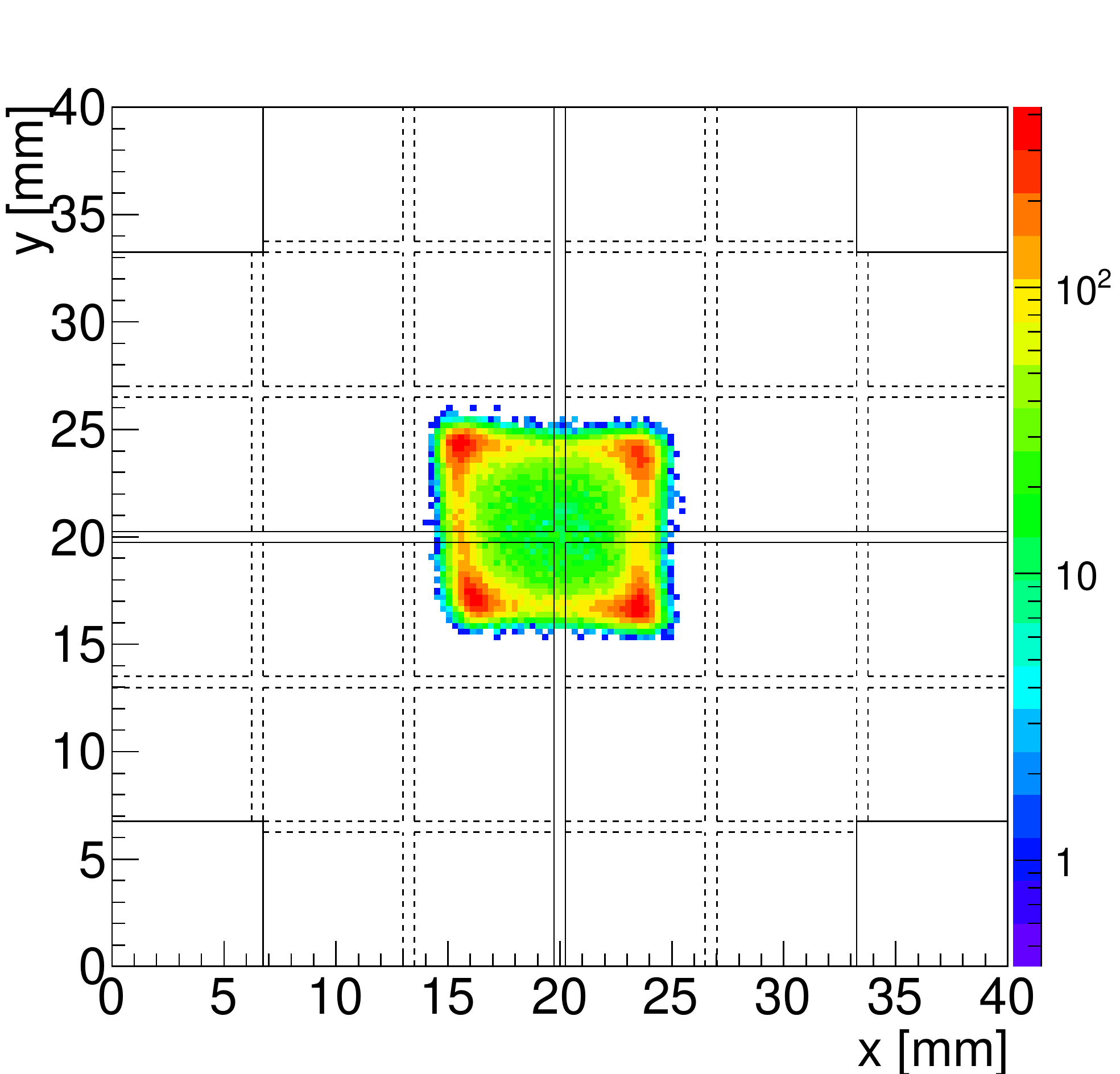}
    \subcaption{Data: raw reconstruction}\label{fig:CorrectionsData_a}
    \end{subfigure}
    \begin{subfigure}{.45\textwidth}
    \centering
    \includegraphics[width=\textwidth]{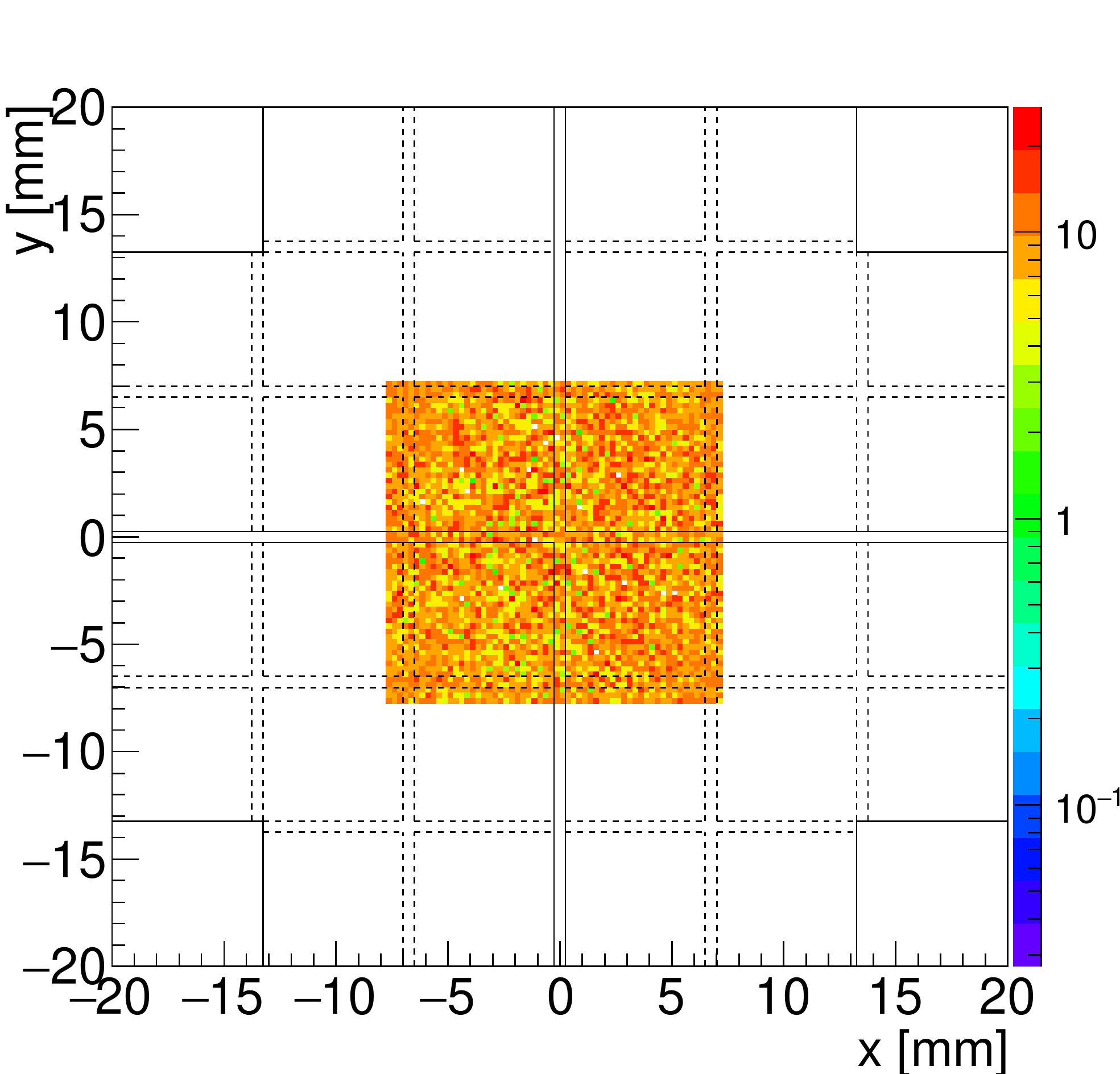}
    \subcaption{MC: raw reconstruction}\label{fig:CorrectionsMC_c}
    \end{subfigure}
    \begin{subfigure}{.45\textwidth}
    \centering
    \includegraphics[width=\textwidth]{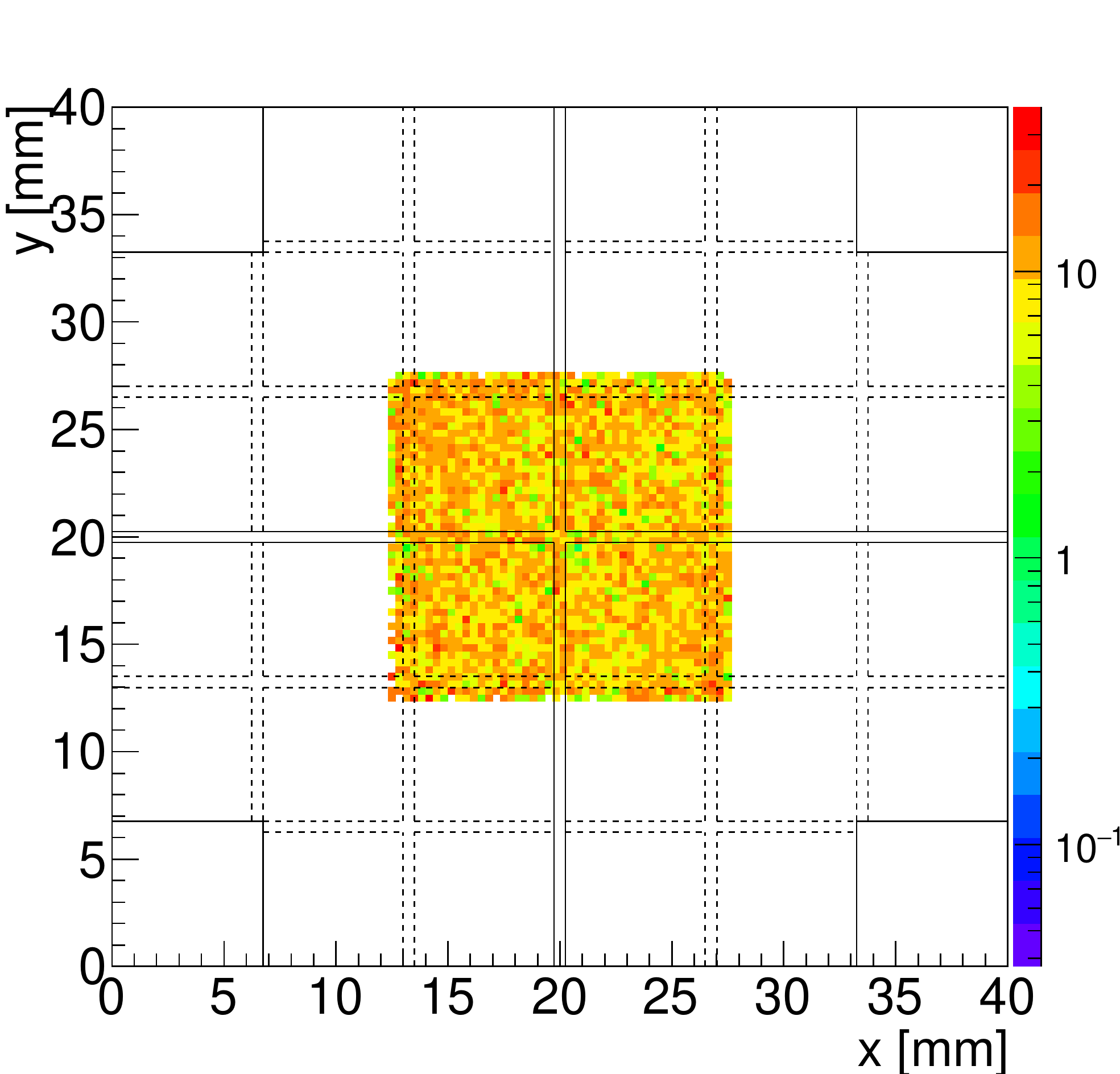}
    \subcaption{Data: after corrections}\label{fig:CorrectionsData_b}
    \end{subfigure}
    \begin{subfigure}{.45\textwidth}
    \centering
    \includegraphics[width=\textwidth]{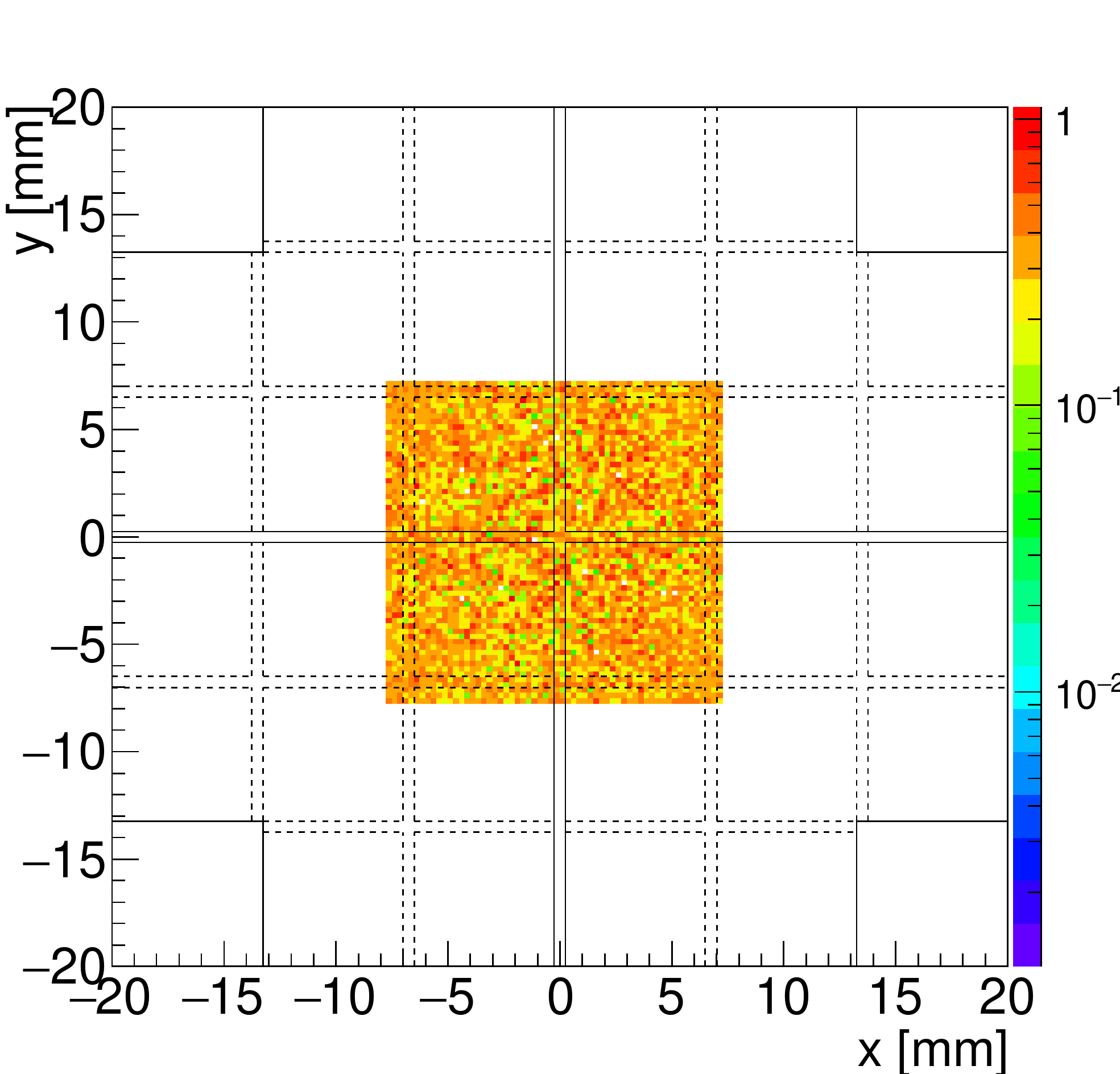}
    \subcaption{MC: after corrections}\label{fig:CorrectionsMC_d}
    \end{subfigure}
    \end{subfigure}
    \begin{subfigure}{.36\textwidth}
    \centering
    \includegraphics[width=\textwidth]{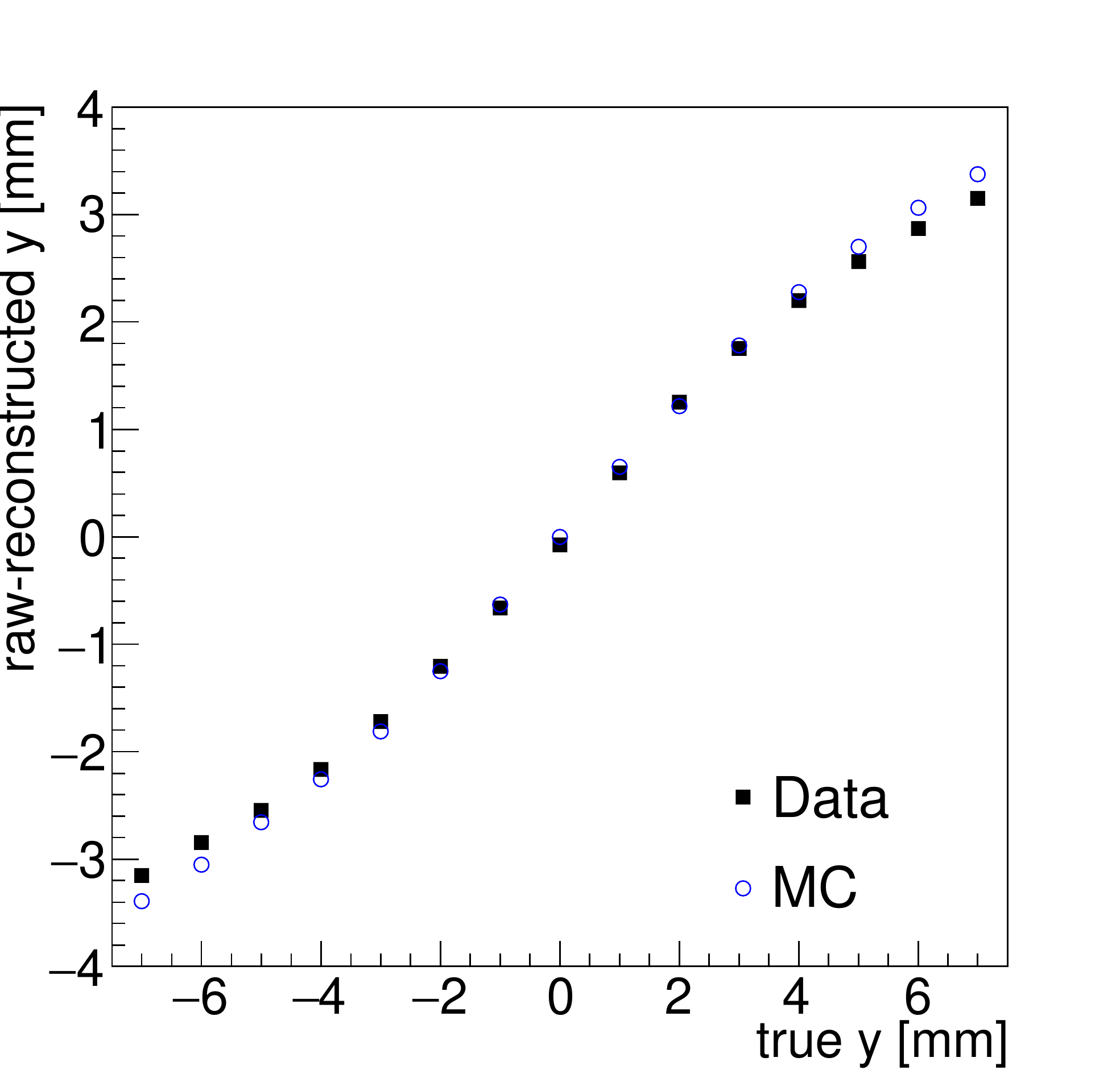}
    \subcaption{}\label{fig:Corrections_DataVsMC}
    \end{subfigure}
    \end{center}
    \caption{\textbf{(a)--(d)} Flood-field images obtained in the laboratory with the micro-camera and Monte Carlo simulations, before (raw-reconstructed with the centroid method) and after spatial linearity and uniformity corrections. \textbf{(e)} raw-reconstructed position in the \textit{y} axis for a capillary oriented parallel to the \textit{x} axis as a function of its true position, both for data and MC.}
    \label{fig:CorrectionsData}
\end{figure}

Figure~\ref{fig:CorrectionsData} shows the image of a flood-field image before and after spatial linearity and uniformity corrections, both for the experimental measurements and the MC simulations. As mentioned in section~\ref{sec:evtrec} the raw images reconstructed with the centroid method appear ``collapsed'' in the center of the FOV (Figures~\ref{fig:CorrectionsData_a} and~\ref{fig:CorrectionsMC_c}). The response of the real micro-camera is not uniform all across the FOV, probably caused by a combination of several factors including a non-homogeneous crystal response due to the presence of impurities, a non-uniform crystal-LASiP coupling or a non-perfect SiPM gain equalization. These inhomogeneities were not simulated, which explains why Figure~\ref{fig:CorrectionsMC_c} seems to have a rotation symmetry that Figure~\ref{fig:CorrectionsData_a} does not have. However, the bulk of this effect is corrected after spatial linearity and uniformity corrections, as can be seen in Figure~\ref{fig:CorrectionsData_b} and as it will be shown in the reconstructed images of section~\ref{sec:spres_res}. After corrections the FOV is also enlarged. Note that in the simulations we limited the FOV to an area of $15\times15$~mm$^2$ aiming to emulate the conditions of the laboratory measurements.

Figure~\ref{fig:Corrections_DataVsMC} shows the raw-reconstructed position (before corrections) of a capillary source that was oriented parallel to the \textit{x} axis as a function of its true (measured) position in the \textit{y} axis, both for experimental data and MC simulations. The agreement between the experimental and the simulated curves is good enough for the scope of this work, which illustrates that the \textit{RoughTeflon LUT\_DAVIS} model employed in the simulations reproduces accurately enough the light distribution inside the crystal. There is an acceptable disagreement between the two curves near the edges that should not affect the conclusions derived in the next sections.

\subsection{Energy resolution}\label{sec:eres_res}

\begin{figure}[tpb]
    \begin{subfigure}{.48\textwidth}
    \centering
    \includegraphics[clip=true, trim=0cm 0cm 0cm 0cm, width=0.9\textwidth]{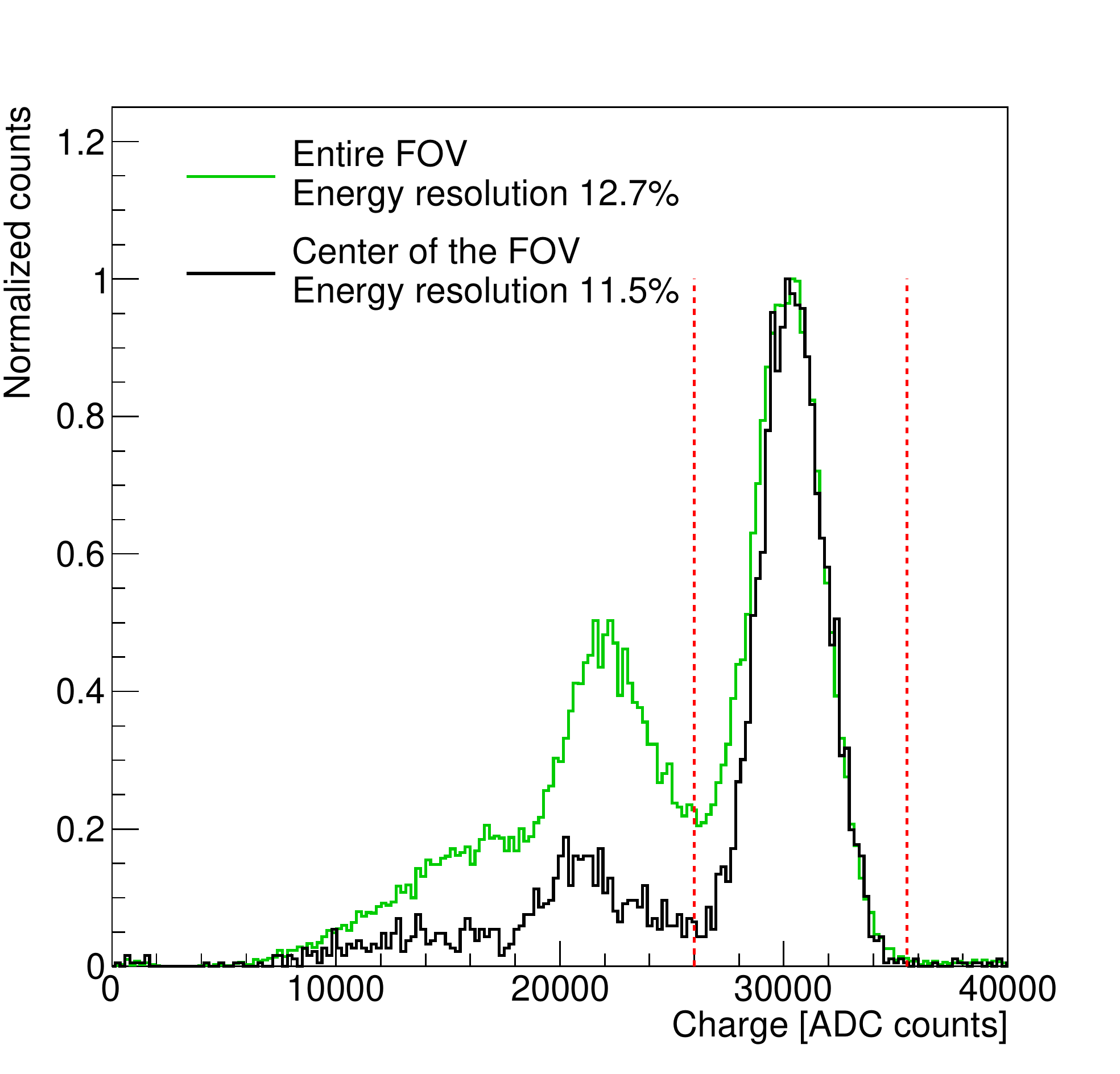}
    \subcaption{Data: flood-field}\label{fig:EnergyRes_flood}
    \end{subfigure}
    \begin{subfigure}{.48\textwidth}
    \centering
    \includegraphics[clip=true, trim=0cm 0cm 0cm 0cm, width=0.9\textwidth]{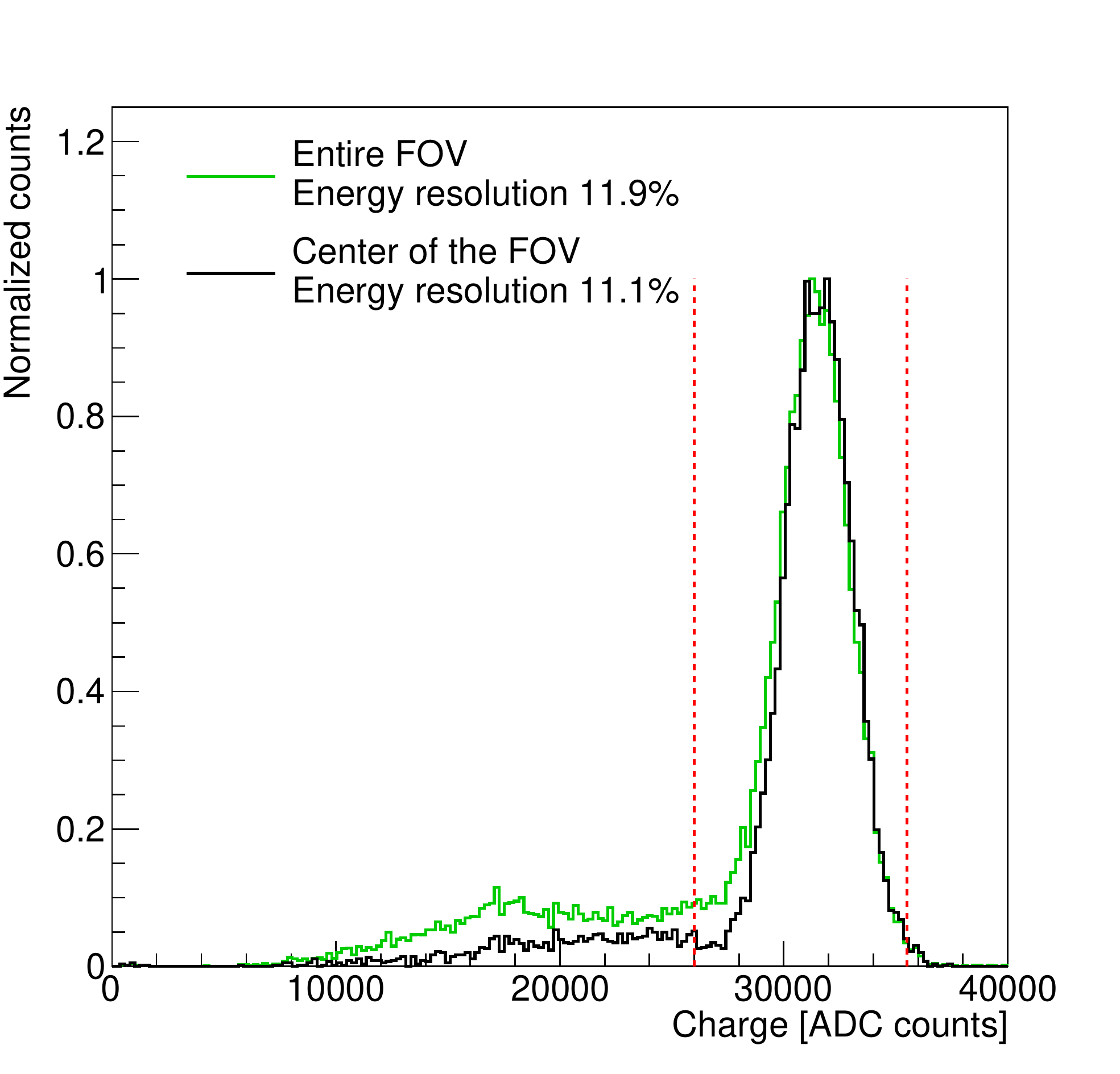}
    \subcaption{Data: capillary}\label{fig:EnergyRes_cap}
    \end{subfigure}
    \caption{Charge histograms obtained during: \textbf{(a)} a flood-field irradiation with $^{99m}$Tc (corresponding image in Figure~\ref{fig:CorrectionsData}); \textbf{(b)} a capillary fully-filled with $^{99m}$Tc, placed at 20~cm from \textit{Coll2} (corresponding image in Figure~\ref{fig:CapSrc}). Green histograms contain all the events in the UFOV. Black histograms only those reconstructed in a $6\times6$~mm$^2$ region around the camera center. The peak corresponds to 140~keV. Vertical red lines show the limits of the acceptance window used for image reconstruction.}
    \label{fig:EnergyRes}
\end{figure}

The measured energy resolution at 140 keV slightly depends on the measuring position. The average value was 11.6\%, ranging from a minimum value of 10.9\% to a maximum of 12.7\%. The variations of the photopeak position measured in different parts of the FOV were below 5\%. Figure~\ref{fig:EnergyRes_flood} shows the charge histograms obtained during a flood-field irradiation with $^{99m}$Tc. Figure~\ref{fig:EnergyRes_cap}, the charge histograms obtained when imaging the capillary using \textit{Coll~2}. The green histograms contain all the events reconstructed inside a $13\times13$~mm$^2$ region around the camera center (i.e., the full FOV excluding 1~mm at the edges). The black histograms include only those events that were reconstructed in a $6\times6$~mm$^2$ region around the camera center. The green histograms exhibit wider peaks, which was expected since they are more sensitive to a non-uniform charge collection across the crystal area, especially close to the crystal corners where light is not detected. In fact, as it will be shown in section~\ref{sec:res_sims}, simulations suggest that the LASiP dead corners could degrade significantly the energy resolution and be the main responsible for the second peak at 22000~ADC~counts that appears in Figure~\ref{fig:EnergyRes_flood}. The non-uniform light collection across the whole crystal area also explains why the photopeaks are broader in Figure~\ref{fig:EnergyRes_flood} than in Figure~\ref{fig:EnergyRes_cap}. While in the first case all parts of the FOV equally contribute to the charge histogram, in the second one most of the histogram counts come from the specific region in which the capillary was imaged, which was close to the camera center and far from the corners. For the same reason the mean of the photopeak in Figure~\ref{fig:EnergyRes_cap} is slightly higher than in Figure~\ref{fig:EnergyRes_flood}. As it will be shown in section~\ref{sec:res_sims}, the dead corner does not only impact the width of the photopeak, but also the mean collected charge.

\subsection{Intrinsic spatial resolution}\label{sec:spres_res}

\begin{figure}[htpb]
    \begin{subfigure}{.33\textwidth}
    \centering
    \includegraphics[width=\textwidth]{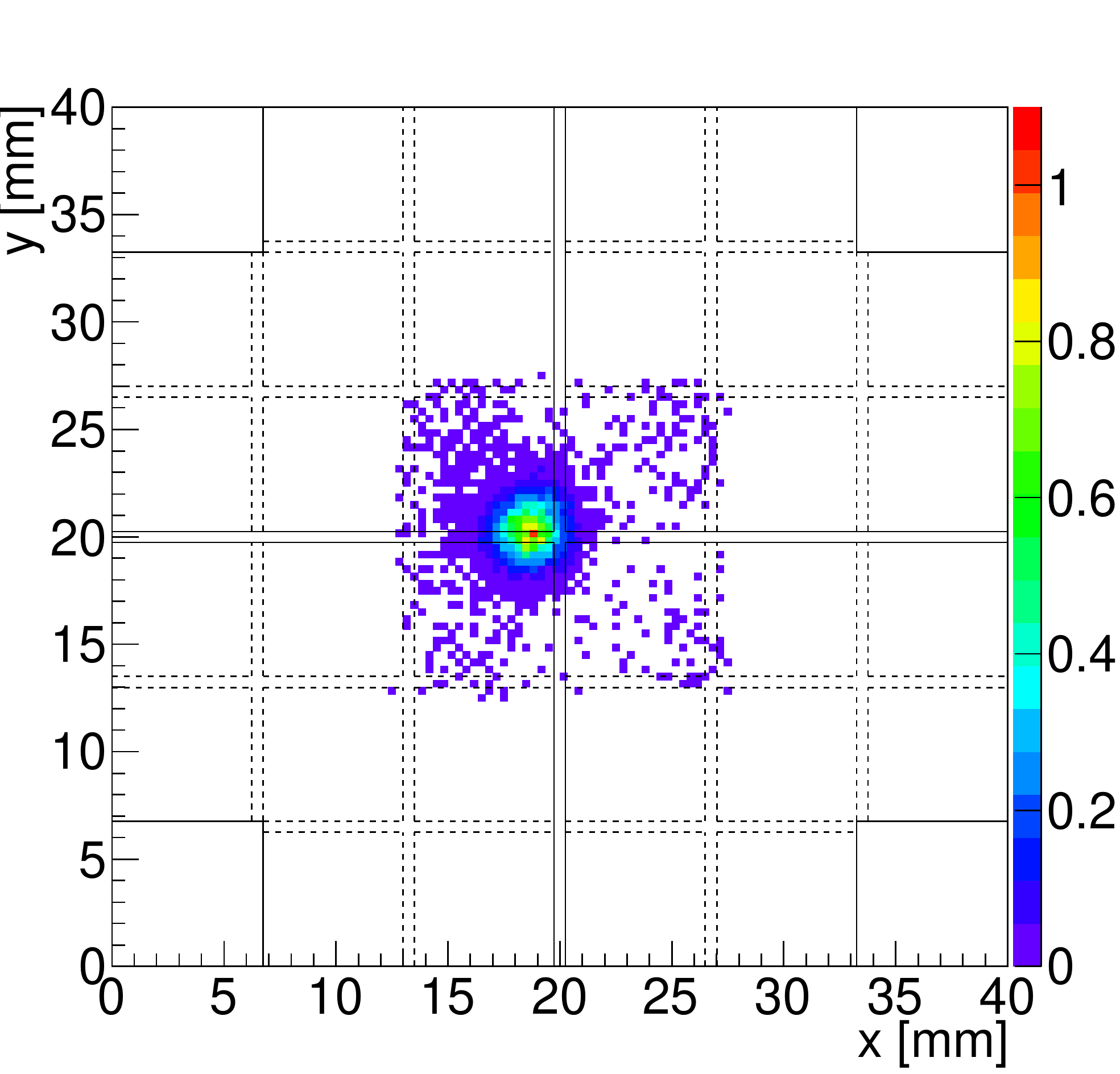}
     \end{subfigure}
     \begin{subfigure}{.33\textwidth}
    \centering
    \includegraphics[width=\textwidth]{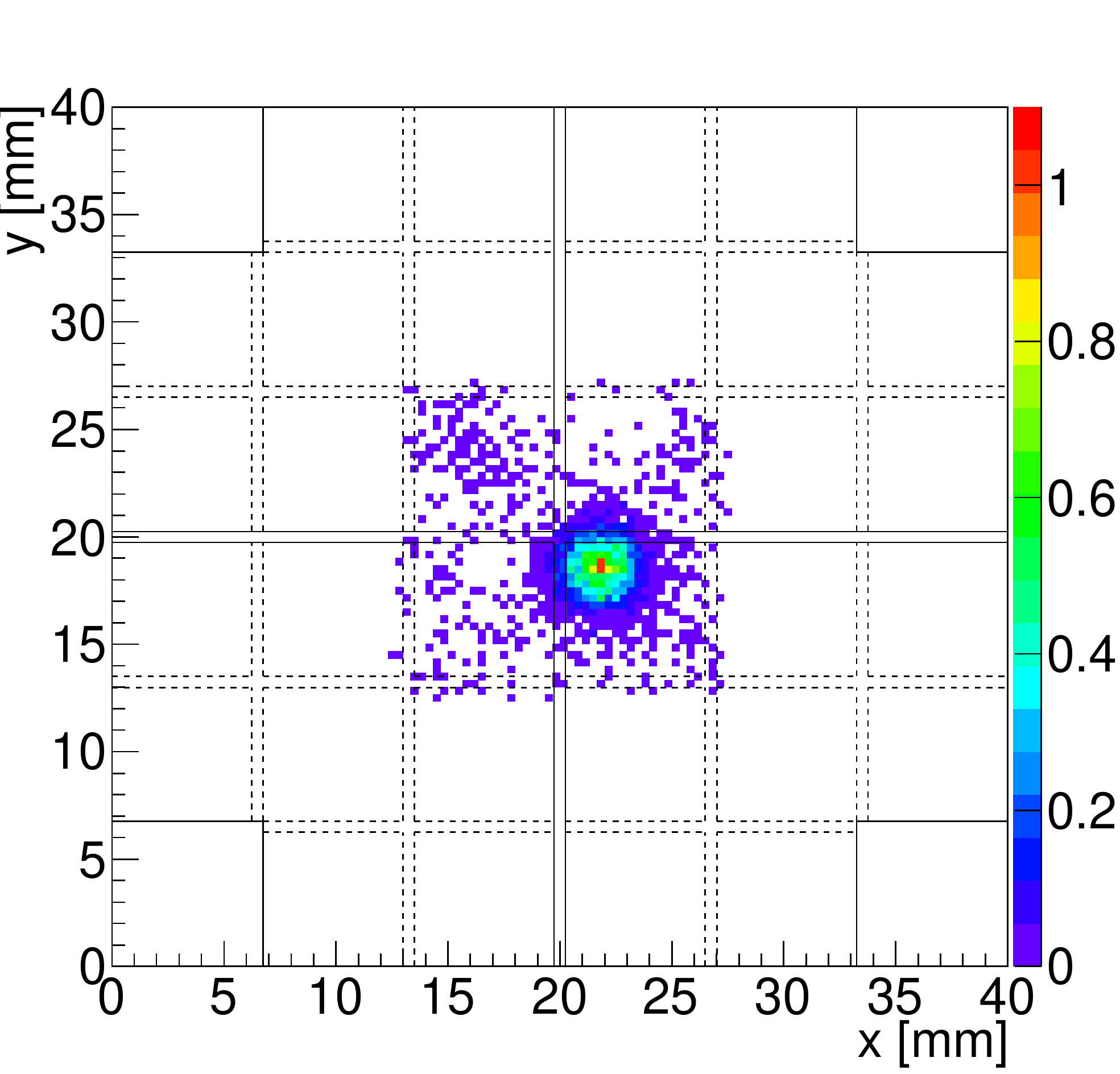}
     \end{subfigure}
     \begin{subfigure}{.33\textwidth}
    \centering
    \includegraphics[width=\textwidth]{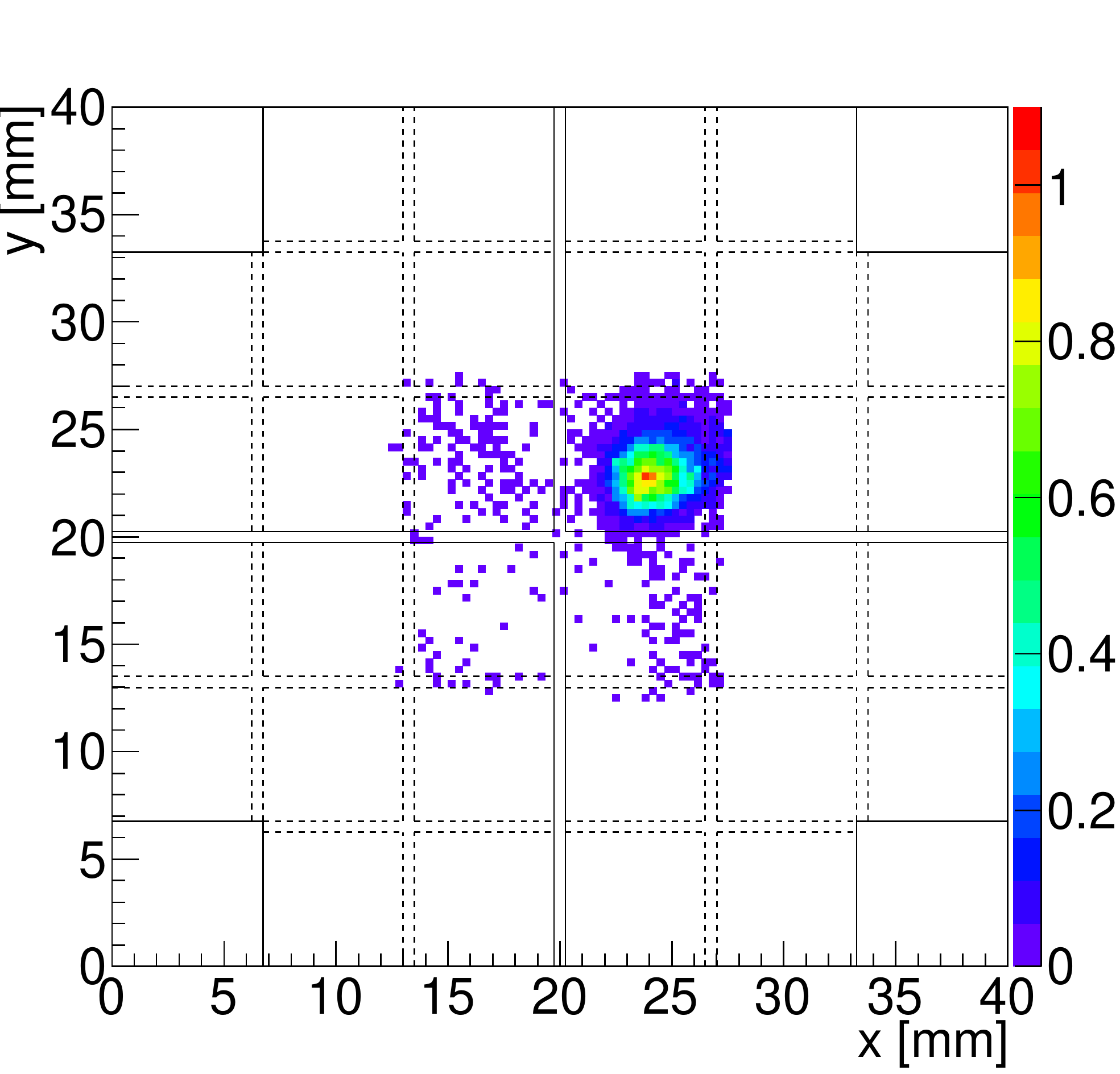}
    \end{subfigure}
    \begin{subfigure}{.33\textwidth}
    \centering
    \includegraphics[width=\textwidth]{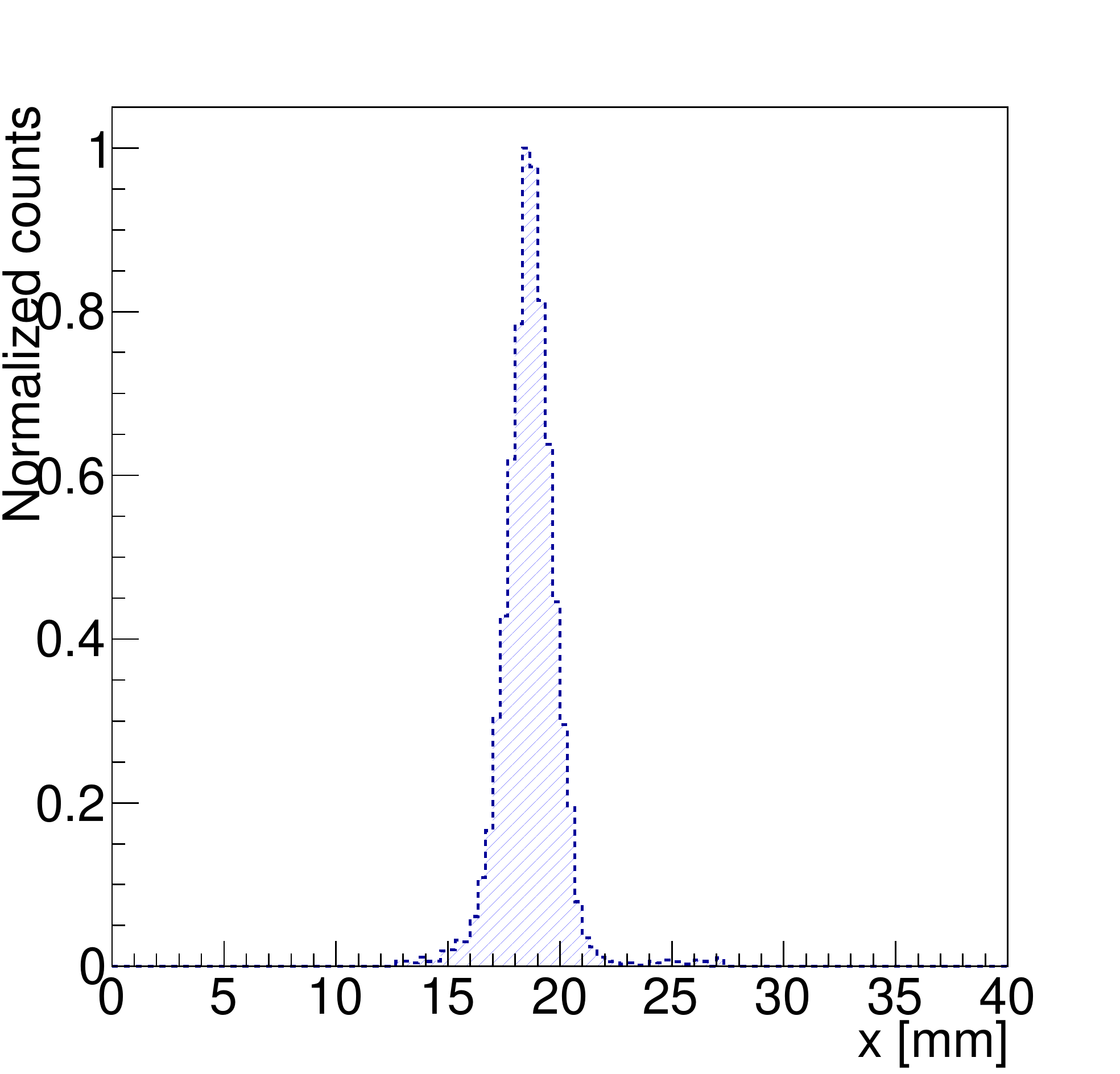}
    \end{subfigure}
    \begin{subfigure}{.33\textwidth}
    \centering
    \includegraphics[width=\textwidth]{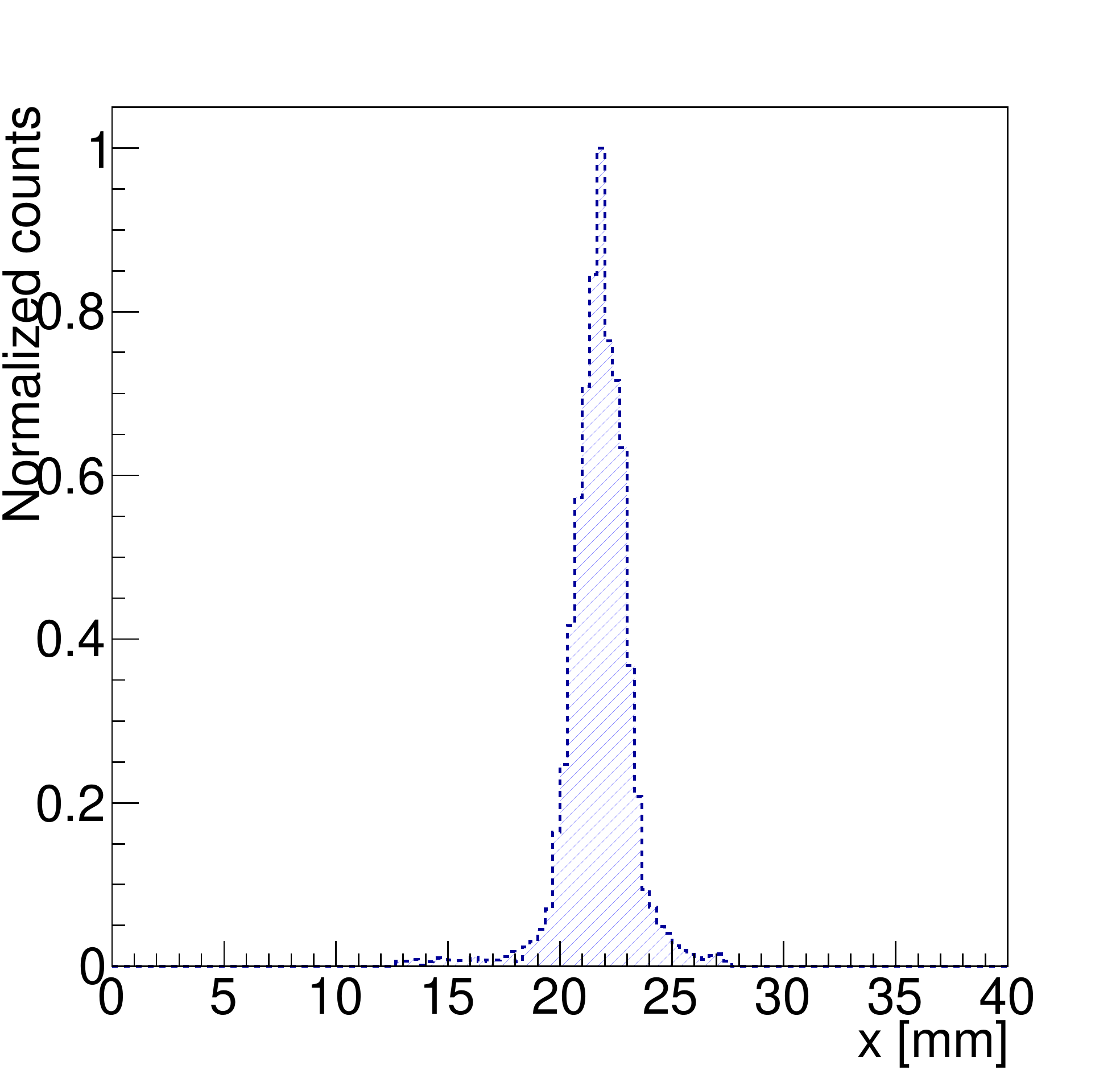}
    \end{subfigure}
    \begin{subfigure}{.33\textwidth}
    \centering
    \includegraphics[width=\textwidth]{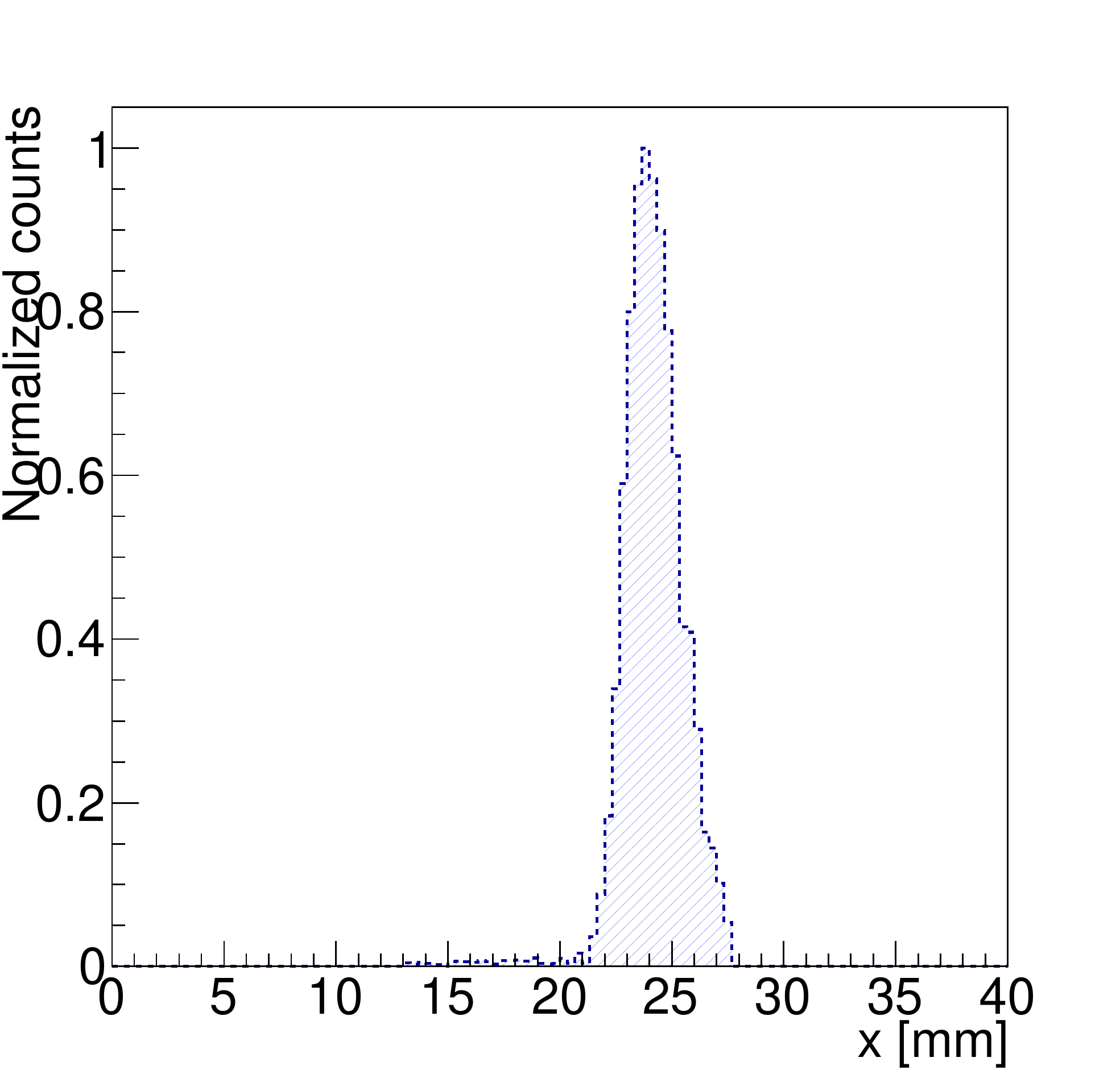}
    \end{subfigure}
    \caption{\textbf{Top:} Images of a $^{241}$Am point source, taken with the micro-camera and \textit{Coll 1}. \textbf{Bottom:} Projection of the images in the \textit{x} axis.}
    \label{fig:PointSrc}
\end{figure}

Images of the $^{241}$Am source were taken with the micro-camera in different test positions. Their positions were reconstructed with an accuracy better than 0.3~mm. Three of those images are shown in Figure~\ref{fig:PointSrc}. The extension $R_{point}$ of the reconstructed images obtained from the 2D-Gaussian fit was on average $(2.4\pm0.1)$~mm. Removing the collimator resolution (see~\ref{eq:ResExt}) we obtained an intrinsic spatial resolution of $R_d = (2.2\pm0.2)$~mm.

\begin{figure}[tpb]
    \begin{subfigure}{.33\textwidth}
    \centering
    \includegraphics[width=\textwidth]{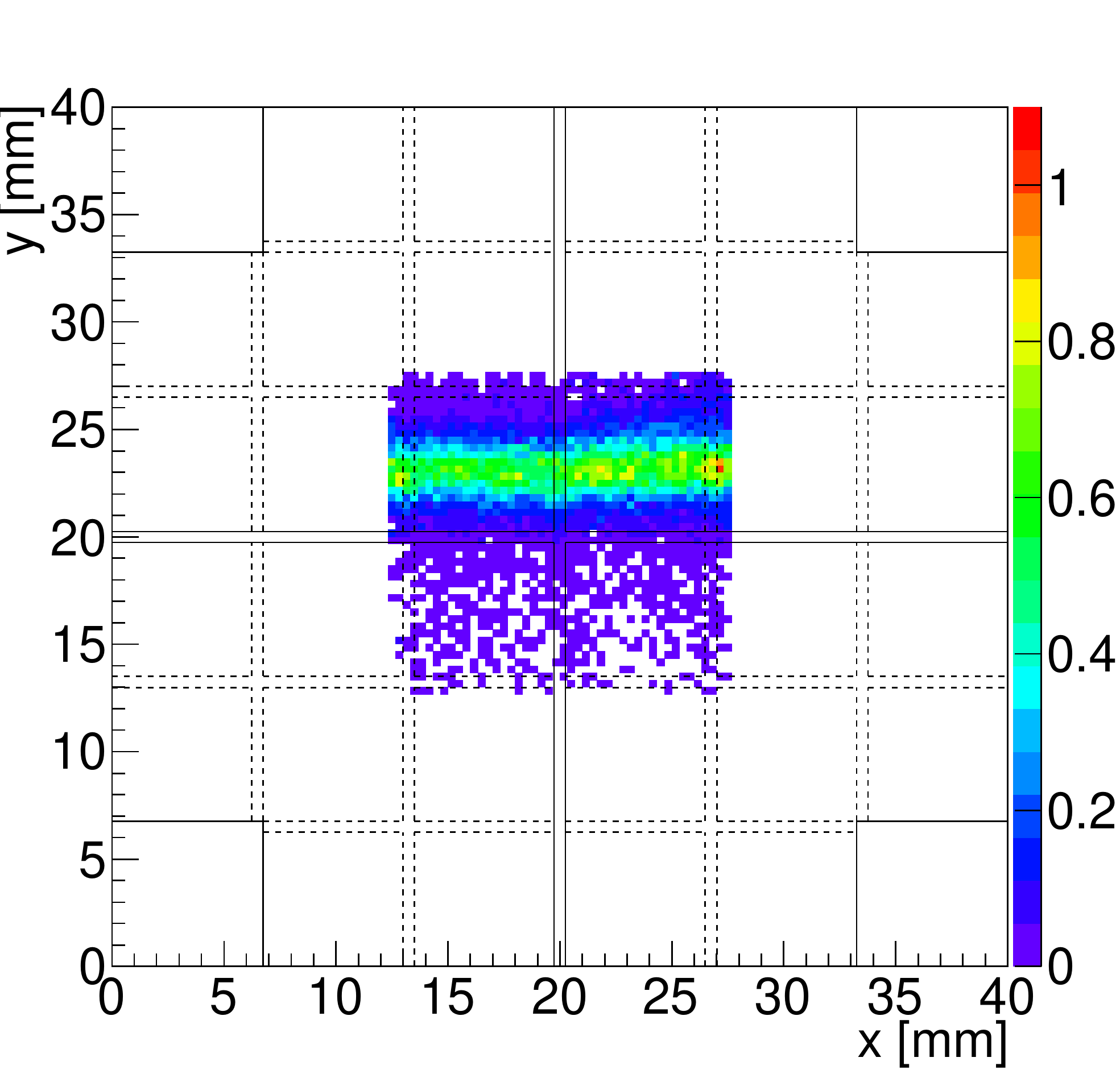}
    \end{subfigure}
     \begin{subfigure}{.33\textwidth}
    \centering
    \includegraphics[width=\textwidth]{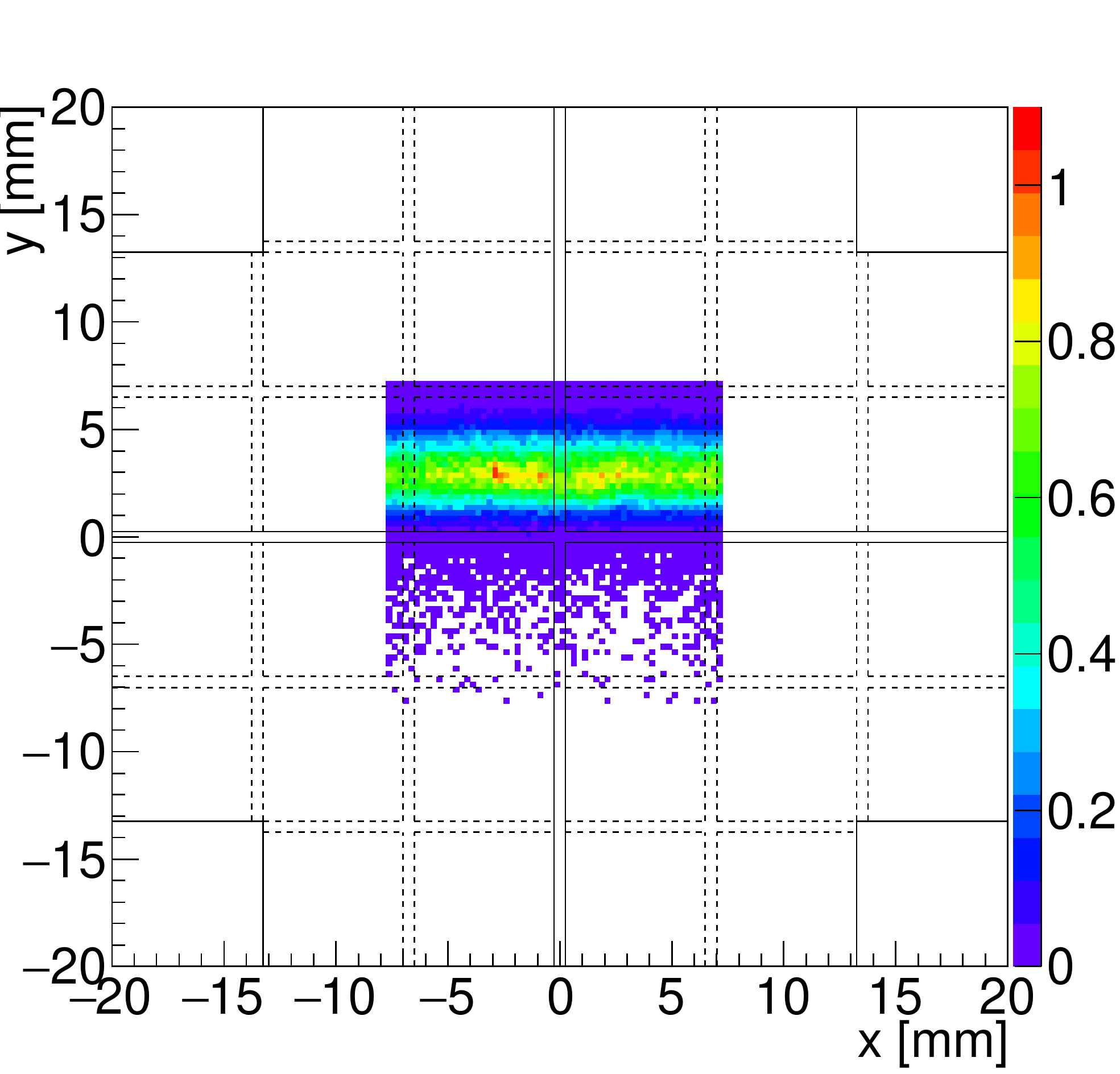}
    \end{subfigure}
     \begin{subfigure}{.33\textwidth}
    \centering
    \includegraphics[width=\textwidth]{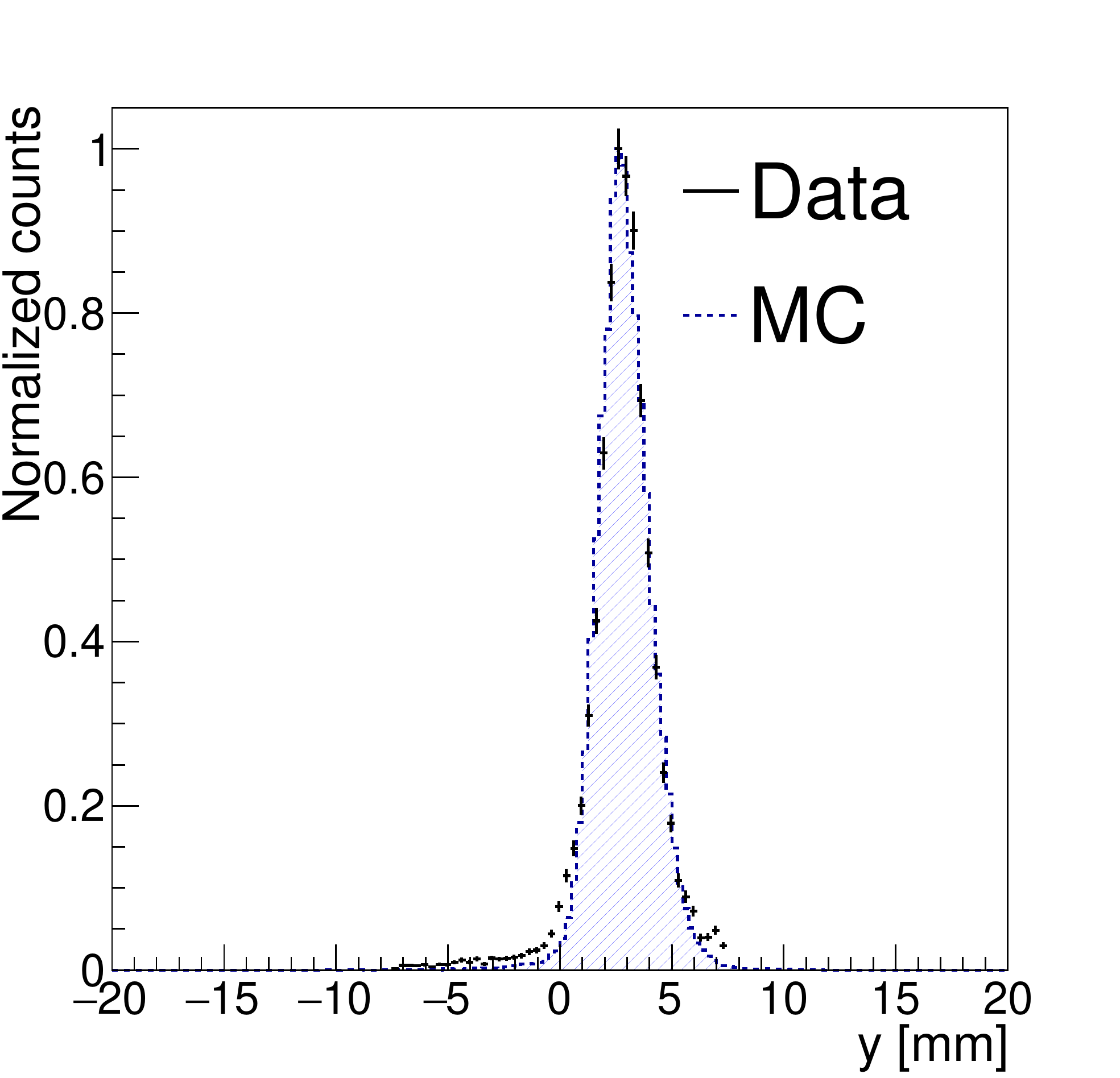}
    \end{subfigure}
     \begin{subfigure}{.33\textwidth}
    \centering
    \includegraphics[width=\textwidth]{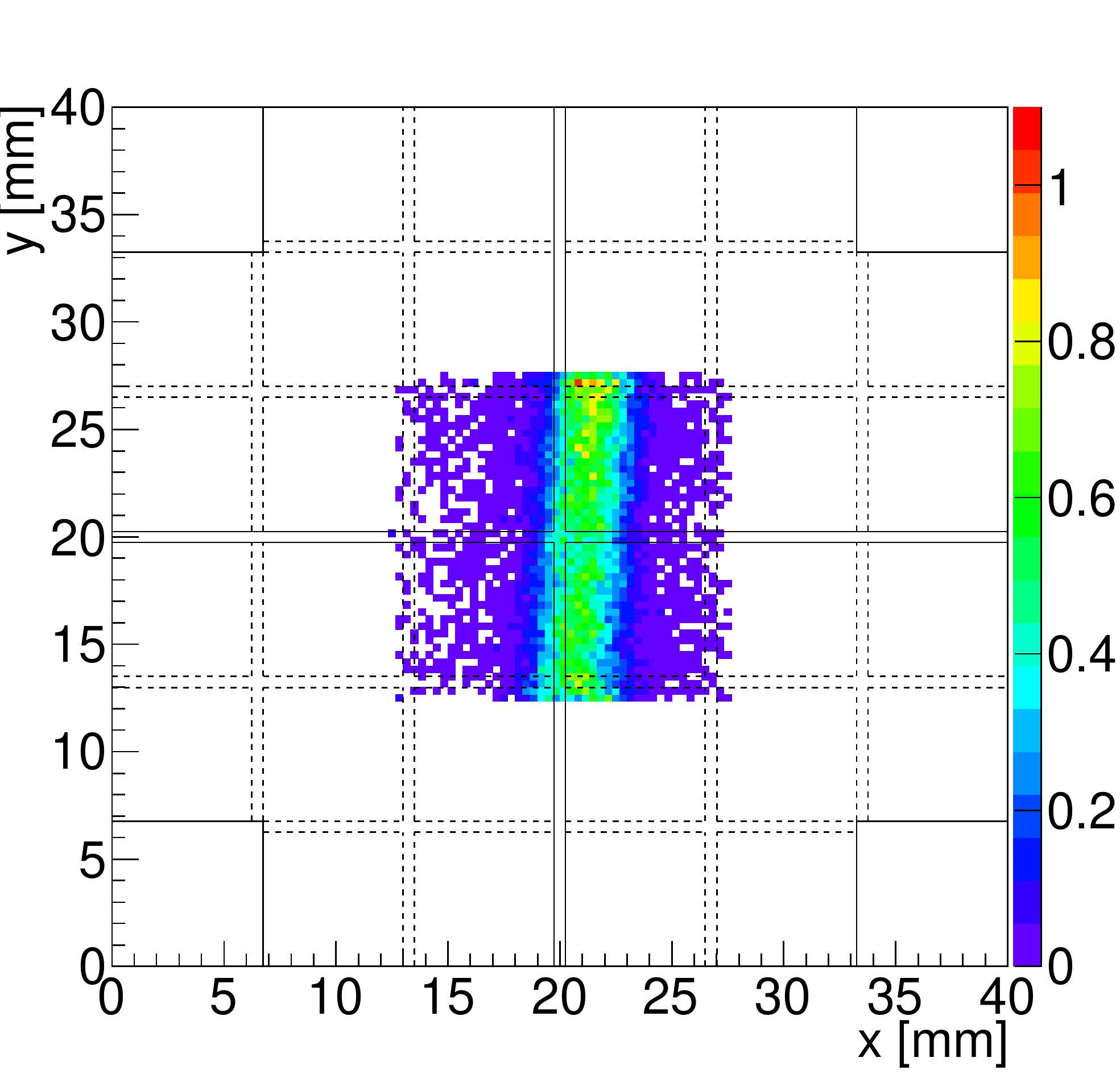}
    \subcaption{Data}
    \end{subfigure}
     \begin{subfigure}{.33\textwidth}
    \centering
    \includegraphics[width=\textwidth]{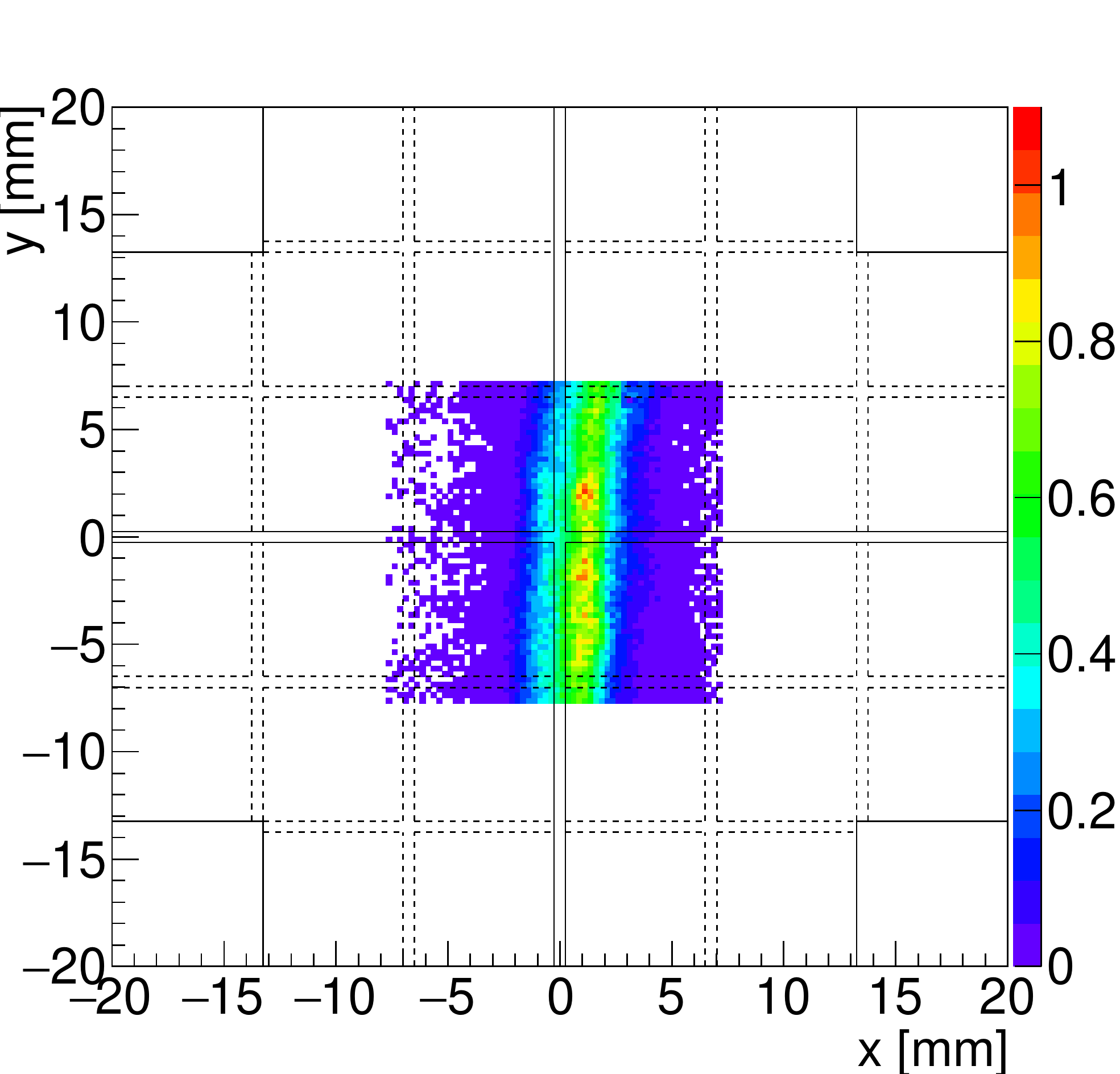}
    \subcaption{MC}
    \end{subfigure}
     \begin{subfigure}{.33\textwidth}
    \centering
    \includegraphics[width=\textwidth]{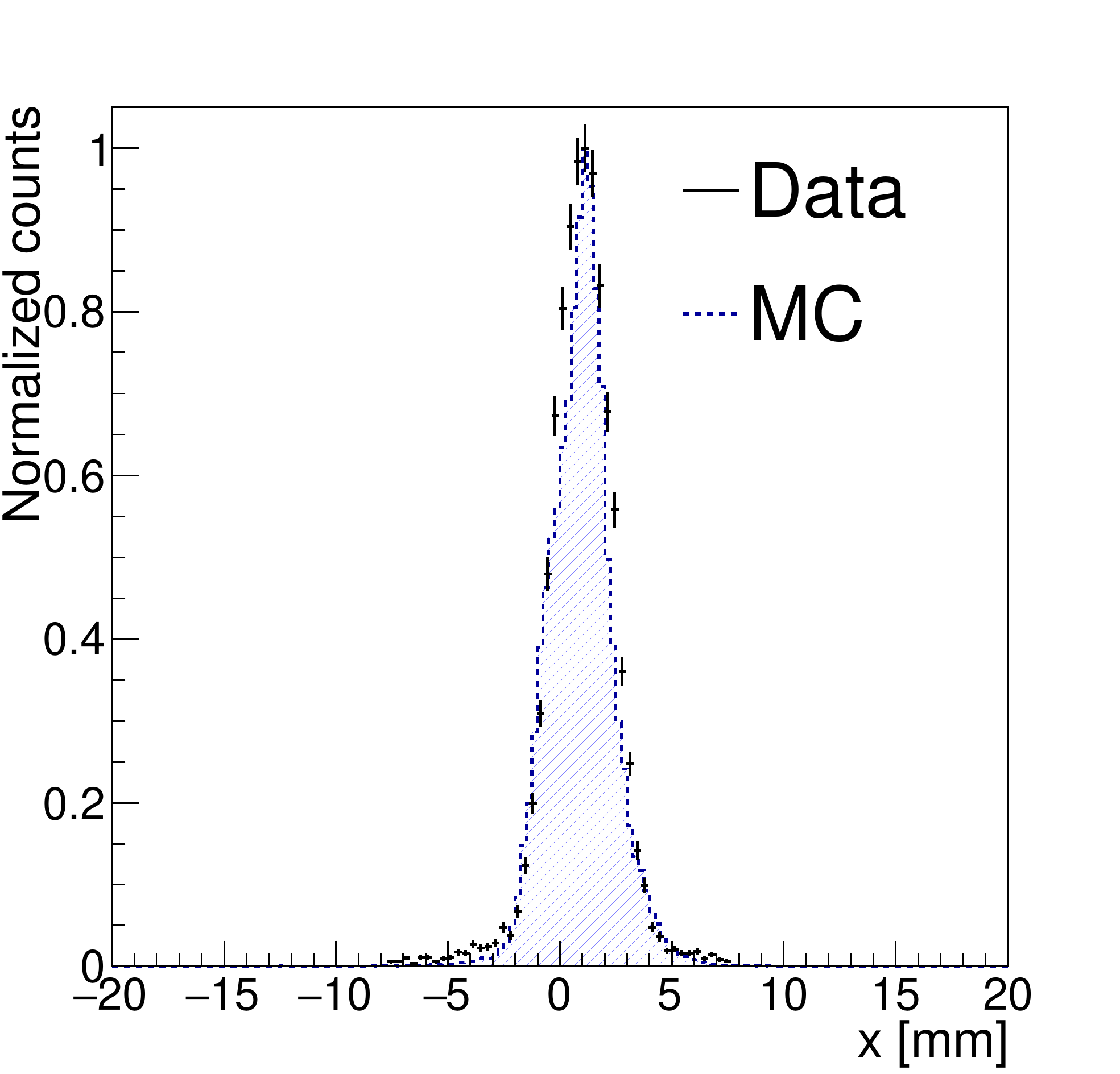}
    \subcaption{Projection}
    \end{subfigure}
    \caption{\textbf{Left:} Images of a capillary fully-field with a $^{99m}$Tc solution, taken with the micro-camera and \textit{Coll 2}. \textbf{Center:} MC images obtained when the same experimental conditions were simulated. \textbf{Right:} Projection of the left and center images in the axis \textit{y} (top) and \textit{x} axis (bottom).}
    \label{fig:CapSrc}
\end{figure}

Figure~\ref{fig:CapSrc} compares the reconstruction of two images of the $^{99m}$Tc capillary from laboratory measurements and simulations. To obtain the MC images we simulated the same conditions of the experiments: same orientation of the capillary, geometrical characteristics of the collimator and the capillary, and source-to-detector distance. As a reference the projection in one of the main axis of the detector plane are also shown. We consider that the agreement between data and simulations is good enough to use the simulations as a test-probe to study with more detail the impact of the LASiP characteristics in the system performance (see section~\ref{sec:res_sims}).

The capillary was imaged in different positions in the two orientations of Figure~\ref{fig:CapSrc}. The mean width of the capillary measured in the experiments was $R_{cyl}=(2.8\pm0.2)$~mm. Removing the source diameter and the collimator resolution we obtained $R_d = (1.9\pm0.4)$~mm.

\subsection{Simulations: impact of LASiP noise in the micro-camera energy resolution}\label{sec:res_sims}

\begin{table}[t]
    \centering
    \begin{tabular}{ c || c c c c || c | c }
         Nr & $p_{XT}$ & $\sigma_0$ & $\sigma_1$ & $\sigma_{UN}$ & $\epsilon$(8-SiPM LASiP) & $\epsilon$(9-SiPM LASiP) \\
          & [\%] & [$\times$ n.r.v.] &  [$\times$ n.r.v.] & [$\times$ n.r.v.] &  [\%] &  [\%] \\
         % &  & [m.c.u] & [m.c.u] & [m.c.u.] & [\%] & [\%] \\
         \hline \hline
         1 & - & - & - & - & 9.7 & 9.1 \\
         \hline
         2 & \textbf{5} & - & - & - & 9.8 & 9.1 \\
         3 & \textbf{10} & - & - & - & 10.0 & 9.3 \\
         4 & \textbf{25} & - & - & - & 10.1 & 9.4 \\
         5 & \textbf{40} & - & - & - & 10.4 & 9.7 \\
         \hline
         6 & - & \textbf{1} & \textbf{1} & - & 9.8 & 9.1 \\
         7 & - & \textbf{10} & \textbf{1} & - & 10.2 & 9.5 \\
         8 & - & \textbf{1} & \textbf{5} & - & 10.2 & 9.5 \\
         %9 & - & \textbf{5} & \textbf{5} & - & 10.3 & 9.6 \\
         \hline
         9 & - & - & - & \textbf{1} & 10.1 & 9.4 \\
         10 & - & - & - & \textbf{2} & 11.6 & 10.8 \\
         11 & - & - & - & \textbf{5} & 19.5 & 18.6 \\
         \hline
         12 & 25 & 1 & 1 & \textbf{1} & 10.7 & 9.9 \\
         \hline
    \end{tabular}
    \caption{Simulations performed with Geant4 to study the impact of LASiP noise in the micro-camera energy resolution. $p_{XT}$, $\sigma_0$, $\sigma_1$ and $\sigma_{UN}$ were defined in section~\ref{sec:met_noise} and their noise reference values (n.r.v) were listed in Table~\ref{tab:ref_noise}. The energy resolution $\epsilon$ was calculated assuming LASiPs built by summing 8 SiPMs (same geometry of the LASiP prototype) and 9 SiPMs (no dead corners).}
    \label{tab:overall}
\end{table}

Table~\ref{tab:overall} summarizes the impact of the different input noise parameters in the detector energy resolution measured in the center of the FOV ($6\times6$~mm$^2$ region around the camera center, as defined in section~\ref{sec:eres_res}) for a flood-field irradiation. Except for $p_{XT}$, the input noise parameters are given in units of the noise reference values (n.r.v) of Table~\ref{tab:ref_noise} and were varied to better understand their individual contribution. The obtained energy resolution is shown both for the case in which all 36 SiPMs are enabled (each pixel summing 9 SiPMs) and for the case in which each LASiP sums 8 SiPMs, as in the laboratory measurements. The first entry in the table shows the energy resolution when no noise is simulated. Entry nr. 12 represents the closest situation to our laboratory measurements (when the input noise parameters are exactly those of Table~\ref{tab:ref_noise}). Even if the energy resolution that was obtained with the 8-SiPM simulated LASiPs is slightly better than what was obtained in section~\ref{sec:eres_res} (expected due to the simplification of the simulations), we considered it to be close enough to the values measured with the micro-camera, at least for the scope of studying how energy resolution changes when we modify the input noise parameters.

\begin{figure}[t]
    \begin{subfigure}{.48\textwidth}
    \centering
    \includegraphics[clip=true, trim=0cm 0cm 0cm 0cm, width=0.9\textwidth]{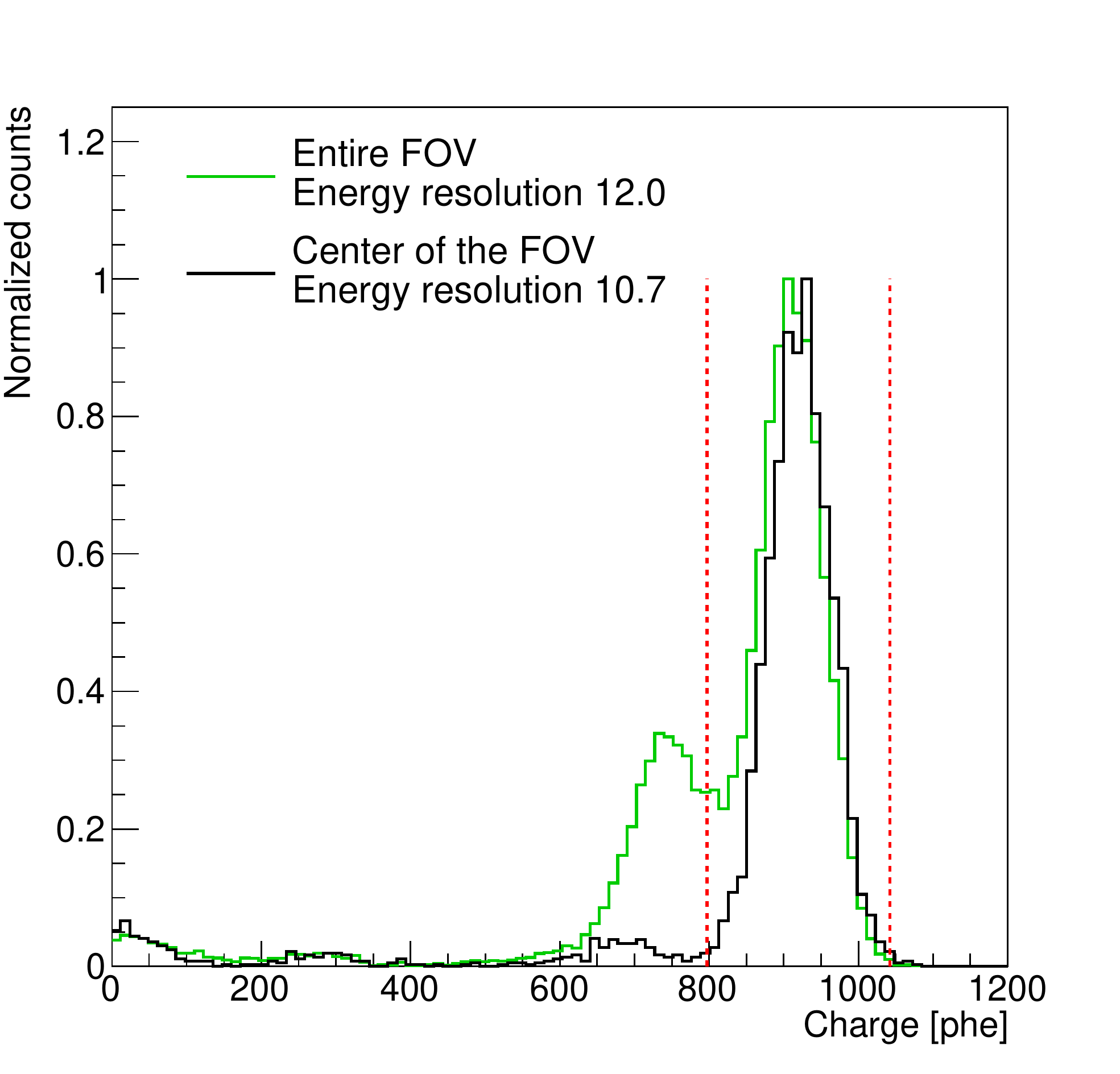}
    \subcaption{MC: 8-SiPM LASiPs}\label{fig:Sims_EnergyRes_corners0}
    \end{subfigure}
    \begin{subfigure}{.48\textwidth}
    \centering
    \includegraphics[clip=true, trim=0cm 0cm 0cm 0cm, width=0.9\textwidth]{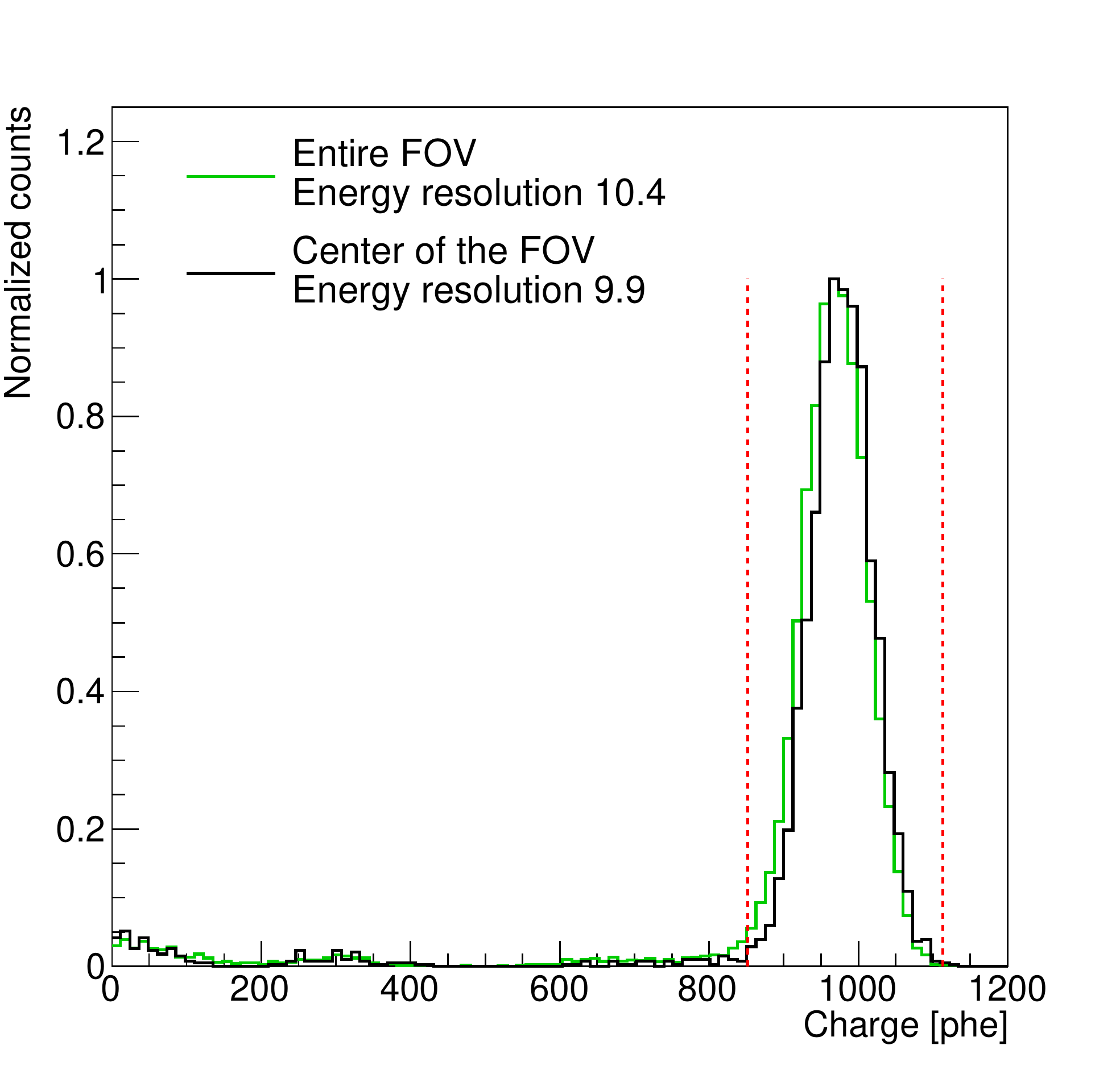}
    \subcaption{MC: 9-SiPM LASiPs}\label{fig:Sims_EnergyRes_corners1}
    \end{subfigure}
    \caption{Charge histograms obtained with Monte Carlo simulations (entry nr. 12 in Table~\ref{tab:overall}) during a flood-field irradiation with $^{99m}$Tc when: \textbf{(a)} the SiPMs in the corners are switched-off (each LASiP is the sum of 8 SiPMs, as in the micro-camera); \textbf{(b)} all 36 SiPMs are enabled (each LASiP is the sum of 9 SiPMs).}
    \label{fig:EnergyRes_MC}
\end{figure}

As anticipated in section~\ref{sec:eres_res}, the LASiP dead corners affect significantly the detector performance, as can be seen in all entries of Table~\ref{tab:overall}. This suggests that we expect to achieve a significantly better energy resolution in a camera in which LASiPs are built without dead corners, fully covering the crystal surface. The effect of the dead corner can also be seen in Figure~\ref{fig:EnergyRes_MC}, that shows the obtained charge spectra for entry nr. 12. The green histograms contain all the events reconstructed inside a $13\times13$~mm$^2$ region around the camera center. The black histograms include only those events that were reconstructed in the center of the FOV. In Figure~\ref{fig:Sims_EnergyRes_corners0} a second peak left to the main peak can be seen that does not appear in Figure~\ref{fig:Sims_EnergyRes_corners1}. It is more pronounced in the green histogram that includes events that were reconstructed closer to the camera corners. The dead corner impacts both the mean position and width of the photopeak.

It would seem that optical cross-talk has a minor impact on the energy resolution of the system  (entries nr. 2--5 in Table~\ref{tab:overall}), although not very critical: the results suggest that reducing $p_{XT}$ to $\sim10\%$ (typically achievable at the expense of a lower PDE) would not provide a significant improvement. The parameters $\sigma_0$ and $\sigma_1$ that describe the finite single-phe resolution of the SiPMs must be increased by a factor of 5 or 10 (which is rather unlikely to occur) with respect to their reference measured values to give a non-negligible contribution (entries nr. 6--8). Energy resolution seems to be much more sensitive to an increase in uncorrelated noise, likely dominated by dark counts (entries nr. 9--11). The noise level measured in the micro-camera LASiPs (relative to their mean signals) seems to be adequate if we compare entries 9 and 1. However, an increase in $\sigma_{UN}$ by a factor 2 (which is not so unlikely, for instance if using noisier SiPMs) already degrades significantly the energy resolution (entry nr. 10). This must be taken into account when designing a large camera with several and larger LASiPs.

%We focused on energy resolution when studying the impact of LASiP noise. Keeping energy resolution under control is extremely important in SPECT, since Compton scattering in tissue is the major factor that affects the quality of the images~\citep{NRC_1996}.
Concerning spatial resolution, we found that it was 10\% worse in the images of the capillary simulated with the n.r.v. than in the case in which no noise was added. This difference would not be very significant in the context of full-body SPECT, where the collimator contribution typically dominates spatial resolution. However, we note that the impact of noise on the detector spatial resolution should be studied in cameras holding more pixels, where the relative weight of noise will be high in pixels showing low (or no) signal. Such case should also require an optimization of the trigger settings, which is far beyond the scope of this work.

\section{Discussion}\label{sec:discussion}

In the previous sections we introduced the concept of LASiP and performed a series of experiments and simulations to study the feasibility of employing these pixels in large gamma cameras for SPECT. We were able to reconstruct simple images with a simple system like the proof-of-concept micro-camera, which supports the idea that LASiPs can be used in SPECT.

 The measured energy resolution of the micro-camera was $\sim11.6$\% at 140 keV, which is equivalent to typical values from both clinical SPECT systems based on PMTs and small SPECT cameras using SiPMs~\cite{INSERT_2018}. The measured value was affected by the LASiP prototype dead corner (as it was shown in the simulations). Hence we expect an improvement in an optimized design in which photodetectors cover the entire crystal exit window. Energy resolution is affected by all the detector components: scintillation crystal, reflective surface surrounding the crystal, photodetectors, coupling between crystal and photodetectors. From the LASiP side, the energy resolution could be improved by using SiPMs with higher PDE, reducing the pixel noise and increasing the photodetector active area. A better performance should be achieved using modern SiPMs with peak sensitivity at the 420~nm where NaI(Tl) scintillation light peaks. Some of them provide a PDE higher than 50\%, with cross-talk probability of $\sim10\%$ and a DCR of $\sim$70~kHz (half the DCR of the SiPMs used in the micro-camera)~\cite{Gundacker_2019}. The photodetector active area could be increased if reducing the dead space between the SiPMs that build the LASiP.
 
In our simulations we found that uncorrelated noise (DCR and electronic noise) is the dominant noise component affecting energy resolution. The impact of dark counts in the performance of gamma cameras employing silicon-based photodetectors had been studied in different works. In~\cite{Occhipinti_2015} it was shown that a relatively high dark count rate (400 kHz/mm$^2$ at 20$^\circ$) could significantly degrade energy resolution of a camera using SiPMs with a $\sim$30\% PDE in the wavelength of interest. In fact, the detector module that was under study (later developed in~\cite{INSERT_2018}) was cooled down to $\sim0^{\circ}$C to reduce the DCR. For the same reason, a cooling system was also employed in the camera of~\cite{Bouckaert_2014}, which was equipped with digital photon counters. With the micro-camera we were able to reconstruct images and achieve a reasonable energy resolution operating the LASiPs at room temperature. However, it should be noted that in the micro-camera all four pixels always exhibit a relatively large signal and hence achieve a high signal-to-noise (SNR) ratio. In a larger camera the situation could be different, as it will be discussed afterwards.

We were able to reconstruct the images produced by a $^{99}$Tc capillary and by a $^{241}$Am point-like source. Even if using only four pixels, we could produce simple images that support the idea that LASiPs could be an alternative for building compact SPECT cameras. The intrinsic spatial resolution measured close to the micro-camera center was $\sim2$~mm. The experiment with the capillary source was also simulated with Geant4. We were able to obtain a reasonable agreement between experiments and simulations in terms of spatial resolution and image reconstruction.

\subsection{Towards a large LASiP-based SPECT camera}

The micro-camera was useful to prove that LASiPs could be used to reconstruct simple images in the center of a gamma-camera and to validate the MC simulations that we used to understand the contribution of the LASiP noise to the overall performance. This was a necessary step towards the ultimate goal, which is to apply the proposed solution in a large camera (e.g., a camera of a full-body SPECT scanner). Extrapolating the performance evaluated with the micro-camera to a larger camera is not straightforward and it requires a dedicated study that is left for a future work. We would like to highlight that we have developed two key components for such a study: validated MC simulations that can be extended to a larger camera and a characterization of the LASiP noise as a function of the number of summed SiPMs, which let us model the noise in larger pixels. Nevertheless, in the next paragraphs we briefly discuss a few aspects that will be particularly relevant in a large camera in which a few thousand SiPMs may be employed.

\subsubsection*{LASiP size and geometry}

The 8-SiPM LASiP prototype was a cost-effective solution that was adopted as the first step to study the feasibility of using LASiP in SPECT. We do not intend to install pixels with such geometry in a camera hosting several tens to a few hundred pixels. As a minimum requirement, a more symmetric LASiP with no dead corners should be employed (e.g., building LASiPs of 9 SiPMs).

Using LASiPs of $\sim2\times2$~cm$^2$ like the prototype we developed, about 500 pixels would be needed to cover a $50\times40$~cm$^{2}$ SPECT camera. This is a significant improvement with respect to the thousand of pixels that would be needed if commercial SiPMs were used, although still high if compared to the 50--100 PMTs of a standard clinical camera. The proposed solution of building large SiPM pixels by summing SiPM signals, and in particular the MUSIC chip, were thought for VHE astrophysics. In these applications the photosensors must be sensitive to fluxes close to the single-phe level and provide a time resolution of $\sim1$~ns. As single-phe and time resolution degrade as the number of summed SiPMs increase, the developed pixels typically sum from 7 to 9 SiPMs. Since the timing and photon resolution requirements for SPECT are more relaxed, in principle it would be possible to build larger pixels by summing more SiPMs. Moreover, the idea of making large pixels by tiling SiPMs allows to build different geometries, especially when the number of SiPMs increases. SiPMs could be distributed forming circular or hexagonal pixels, which have the advantage of having radial symmetry. However, the number of SiPMs to be summed and their distribution inside a pixel should be carefully optimized, since pixel size and noise will affect the system performance.

A rather simple and low-cost solution to build larger LASiPs could be to connect, in parallel, pairs of SiPMs and send each output into one input channel of the MUSIC. One of the problems of this approach would be that the capacitance will significantly increase, resulting in a large amount of noise at the input of the preamplifier. This might not only impact the performance of the detector, but it would also make the calibration of the LASiPs more difficult. Another option that would probably produce less noise and be easier to calibrate would be to perform a sum in two (or more) stages, i.e., to sum the output of several MUSICs into a single channel. In this way it would still be possible to access individual SiPMs by switching off the rest of the MUSIC channels. The optimal solution would be to have an ASIC with as many input channels as SiPMs are summed inside a pixel. That hypothetical ASIC could even have individual configurable thresholds that would allow to perform the sum only over those SiPMs recording a signal above the single-SiPM noise level. A few ASICs have been developed to read the output of 16--64 SiPMs in time-of-flight Positron Emission Tomography (TOF PET) applications, which do not perform the sum of the signals but output a number of digital/binary signals equal to the number of inputs (see for instance~\cite{TOFPET2_2016,HRFlexToT}). However, reducing the number of readout channels is often also desirable in PET, as can be seen in \cite{CALVACORAZA_2017, PARK_2019}. The requirements for the electronics of TOF PET are in principle more demanding than in PET, since high-precision timing and some information on the single-SiPM charge to estimate the depth of interaction of gamma rays inside the crystal are needed. However, future developments in this field may also be adopted by LASiP-based solutions for SPECT.

\subsubsection*{Dark counts and their impact on the trigger settings}

\begin{figure}
    \begin{subfigure}{.53\textwidth}
    \centering
    \includegraphics[width=0.99\textwidth]{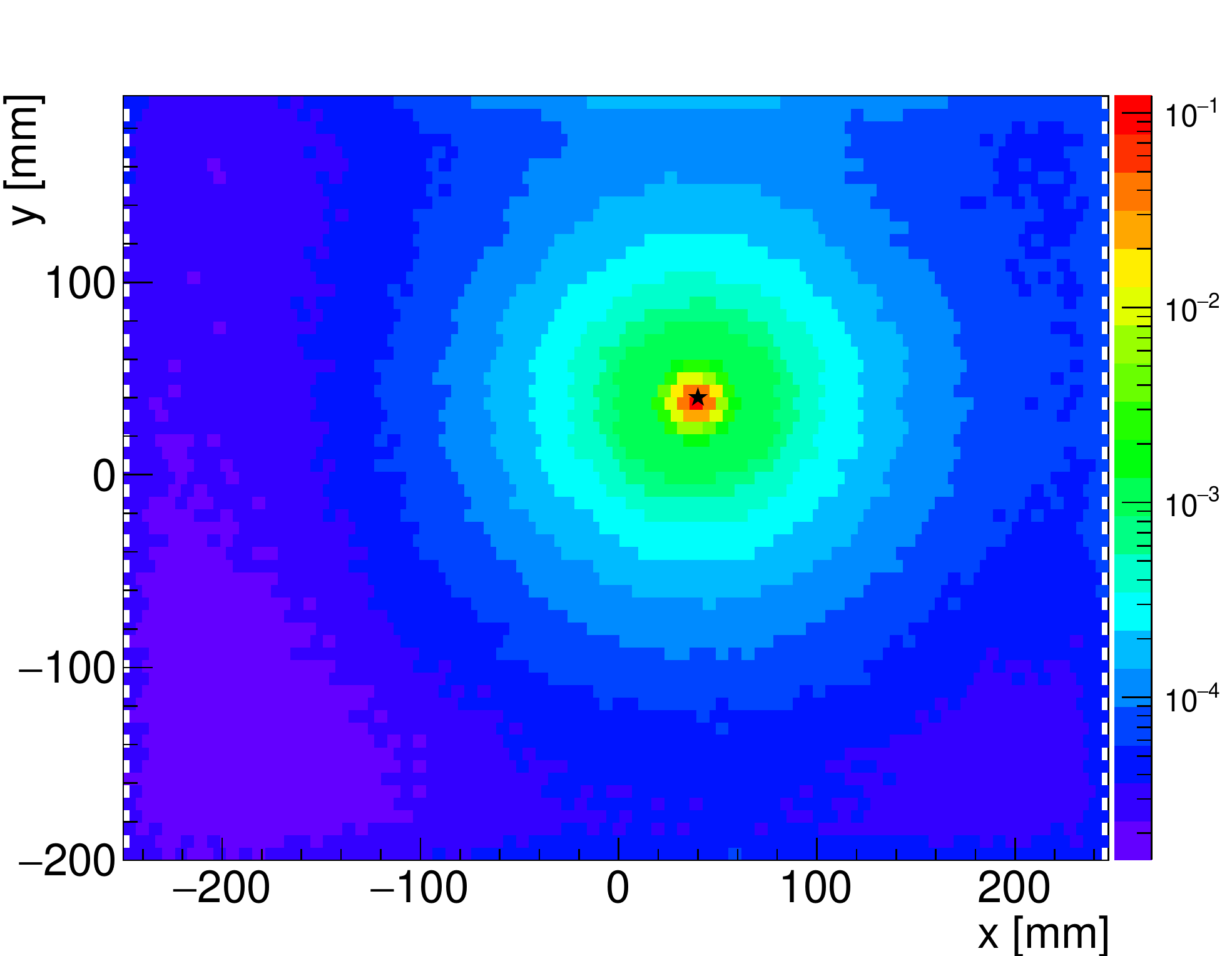}
    \subcaption{}\label{fig:trigger_a}
    \end{subfigure}
    \begin{subfigure}{.43\textwidth}
    \centering
    \includegraphics[width=0.99\textwidth]{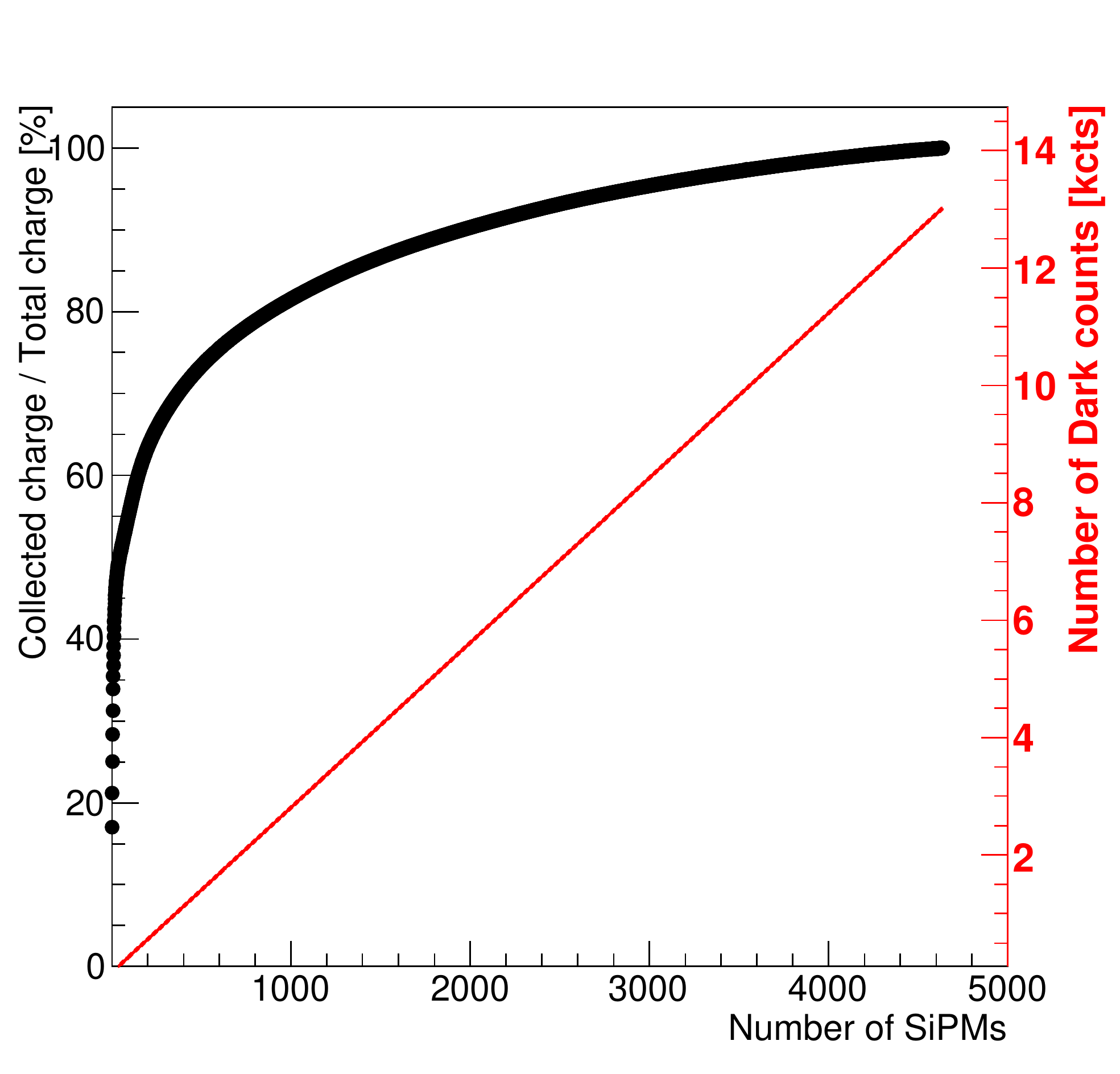}
    \subcaption{}\label{fig:trigger_b}
    \end{subfigure}
    \caption{\textbf{(a)} 2-D map with the mean charge collected by 4636~SiPMs of $6\times6$~mm$^2$ for a gamma-ray collimated beam of 140 keV directed into a NaI(Tl) crystal of $500\times400\times9.5$~mm$^3$. The black star marks the position of the beam in the \textit{(x,y)} plane. \textbf{(b)} Percentage of the total charge collected as a function of the number of SiPMs employed for the trigger. The expected number of integrated dark counts (DCR~$=0.13$~MHz$/$mm$^2$, integration time $=0.6$~$\mu$s) is shown in red.}
    \label{fig:trigger_area}
\end{figure}

In a large SPECT camera the scintillation photons will be distributed over a larger number of SiPMs than in the micro-camera. To illustrate this situation, we simulated a perfectly collimated beam of 140 keV directed into a NaI(Tl) crystal with similar characteristics to the one described in Section~\ref{sec:met_sims}, but with a size of $500\times400\times9.5$~mm$^3$. The crystal was equipped with 4636~SiPMs of $6\times6$~mm$^2$ arranged in a honeycomb geometry and the charge of each SiPM was readout individually. The spatial distribution of the mean charge collected by the SiPMs is shown in Figure~\ref{fig:trigger_a}. If more SiPMs are used to collect the charge, more dark counts will be integrated and this will degrade the performance of the system. One could limit the trigger area (i.e., the number of SiPMs used to measure the charge of an event) or set individual SiPM/pixel thresholds to reduce the impact of dark counts. However, if the trigger area is too small or the threshold too high many scintillation photons will be lost. Figure~\ref{fig:trigger_b} shows the evolution of the relative charge collected by the $N$ SiPMs with the highest signal as a function of $N$ (for the events simulated to build Figure~\ref{fig:trigger_a}). In the example $\sim200$ ($\sim800$) SiPMs are needed to collect $\sim60\%$ ($\sim80\%$) of the scintillation photons. Figure~\ref{fig:trigger_b} also shows the expected mean number of integrated dark counts for a DCR of $0.13$~MHz$/$mm$^2$ and an integration time of $0.6$~$\mu$s. Since both signal and dark counts increase with the number of SiPMs employed to measure the charge, there should be an optimal trigger area that maximizes the SNR.

With a similar reasoning, the integration time will also have to be optimized to maximize the SNR. The temporal distribution of the detected photons is dominated by the scintillation emission, which follows an exponential decay with a characteristic time $\tau=240$~ns. Dark counts, on the other hand, increase linearly with time. As in the micro-camera, there will be an optimal integration time that maximizes energy resolution. However, this effect will be more relevant in a large camera employing more SiPMs to collect the charge. It is then clear that in a larger LASiP-based camera not only the pixel size and geometry must be optimized, but also the trigger conditions. The optimal trigger conditions (trigger area and integration time) will differ depending on the crystal thickness and the pixel size, but also on the SiPM PDE and the DCR.

\subsubsection*{Final considerations}

Following the arguments exposed in the previous paragraphs, it is hard to estimate how LASiPs would perform in a full-body SPECT camera without a dedicated study. We may expect some degradation of the energy resolution due to a higher number of integrated dark counts. Depending on how large that degradation is, it may justify the use of a cooling system to lower the SiPM temperature and thus reduce the DCR. It is also reasonable to expect that the spatial resolution would degrade if the pixel size increases, although at the same time using more than 4 pixels to reconstruct the events would probably help. Also applying a better resolving algorithm during the image reconstruction should provide an improvement in this sense. The results of this work can be used as input for studying in detail the performance of a large SPECT camera based on LASiP through Monte Carlo simulations.

Undoubtedly the main advantage of replacing PMTs by SiPM-based pixels in SPECT is the reduction of the size and weight of the camera. The photodetectors would no longer be the main contributors to the camera volume, which could be reduced by~$\sim50$\%. With smaller and lighter cameras, the cost and the complexity of the whole SPECT scanner could be reduced. Besides this and the general advantages offered by SiPMs described in section~\ref{sec:intro}, LASiPs can be calibrated without the need of any external light-source using SiPM dark counts. This is relatively easy to do while single SiPMs can be accessed (as the MUSIC allows), although much more challenging when pixel size increases. Large PMTs also show a non-uniform spatial response along the photocathode, which can affect the performance of scintillation detectors~\cite{Mottaghian_2010}. The spatial uniformity on the LASiP response would be mostly determined by the differences in the manufacturing of individual SiPMs and the accuracy in the calibration of the electronic channels.

\section{Conclusions}

We studied for the first time the feasibility of using large SiPM pixels in SPECT, as a novel approach to build compact large-area SPECT cameras. We developed a LASiP prototype of $\sim2.9$~cm$^2$ area ($\sim2.2$~cm$^2$ active area) and evaluated its impact on the performance of a gamma camera through laboratory measurements and MC simulations. With a proof-of-concept micro-camera using 4 LASiP prototypes we were able to reconstruct simple images in a central region of $15\times15$~mm$^2$, with an energy resolution of $\sim11.6\%$ at 140 keV and an intrinsic spatial resolution of $\sim2$~mm. We performed Geant4 simulations of the micro-camera, that were validated with experimental data. In the future, these simulations can be extended to a larger camera, holding several LASiPs. The way in which we modeled LASiP noise can also be easily extended to larger pixels.

The results obtained in this work suggest that using LASiPs in large SPECT cameras could be feasible, providing that the trigger settings and the pixel size and geometry are optimized for the dimensions of the camera and that the impact of dark counts are properly mitigated. The employed LASiPs could be comparable in size and geometry to a SPECT PMT and could represent an affordable solution to use SiPMs in SPECT. In this work we provided two essential results that can be seen as the staring points for such optimization: we proved that with a very simple system we can reconstruct images with an energy resolution comparable to standard SPECT systems and validated MC simulations that can be easily adapted to study how LASiPs would perform in a larger camera.

\section*{Acknowledgments}
This work would not have been possible without the support from L. Stiaccini, who designed and built the mechanics of the holder. We would also like to acknowledge E. Fiandrini and V. Vagelli for kindly providing the SCT matrices we used for building the LASiPs. We would like to thank the ICCUB team (D. Gascón, S. Gómez, D. Sánchez) for the fruitful discussions and M. G. Bisogni for taking her time to go through the draft. This research has been carried on in the framework of the CompTo-NM project led by Imaginalis s.r.l. and financed by Regione Toscana - POR CREO FESR 2014-2020. The work of A. Rugliancich has been partially supported by the Tuscany Government, POR FSE 2014-2020, through the INFN-RT2 172800 Project.

%\section*{References}
\bibliography{mybibfile}

\begin{thebibliography}{35}
\expandafter\ifx\csname natexlab\endcsname\relax\def\natexlab#1{#1}\fi
\providecommand{\url}[1]{\texttt{#1}}
\providecommand{\href}[2]{#2}
\providecommand{\path}[1]{#1}
\providecommand{\DOIprefix}{doi:}
\providecommand{\ArXivprefix}{arXiv:}
\providecommand{\URLprefix}{URL: }
\providecommand{\Pubmedprefix}{pmid:}
\providecommand{\doi}[1]{\href{http://dx.doi.org/#1}{\path{#1}}}
\providecommand{\Pubmed}[1]{\href{pmid:#1}{\path{#1}}}
\providecommand{\bibinfo}[2]{#2}
\ifx\xfnm\relax \def\xfnm[#1]{\unskip,\space#1}\fi
%Type = Article
\bibitem[{Valotassiou et~al.(2018)Valotassiou, Malamitsi, Papatriantafyllou,
  Dardiotis, Tsougos, Psimadas, Alexiou, Hadjigeorgiou \&
  Georgoulias}]{Valotassiou_2018}
\bibinfo{author}{Valotassiou V}, \bibinfo{author}{Malamitsi J},
  \bibinfo{author}{Papatriantafyllou J}, \bibinfo{author}{Dardiotis E},
  \bibinfo{author}{Tsougos I}, \bibinfo{author}{Psimadas D} et~al.
\newblock SPECT and PET imaging in Alzheimer’s disease.
\newblock {Ann Nucl Med}  \bibinfo{year}{2018};
  \bibinfo{volume}{32}:\bibinfo{pages}{583--93}.
\newblock https://doi.org/10.1007/s12149-018-1292-6.
%Type = Article
\bibitem[{Son et~al.(2016)Son, Kim \& Park}]{Son_2016}
\bibinfo{author}{Son S}, \bibinfo{author}{Kim M}, \bibinfo{author}{Park H}.
\newblock Imaging analysis of Parkinson’s disease patients using SPECT and
  tractography.
\newblock {Sci Rep}  \bibinfo{year}{2016};
  \bibinfo{volume}{6}:\bibinfo{pages}{38070}.
\newblock https://doi.org/10.1038/srep38070.
%Type = Article
\bibitem[{Anger(1958)}]{Anger_1958}
\bibinfo{author}{Anger H}.
\newblock Scintillation camera.
\newblock {Rev Sci Instrum}  \bibinfo{year}{1958};
  \bibinfo{volume}{29}:\bibinfo{pages}{27}.
%Type = Article
\bibitem[{Van~Audenhaege et~al.(2015)Van~Audenhaege, Van~Holen, Vandenberghe,
  Vanhove, Metzler \& Moore}]{Van_Audenhaege_2015}
\bibinfo{author}{Van~Audenhaege K}, \bibinfo{author}{Van~Holen R},
  \bibinfo{author}{Vandenberghe S}, \bibinfo{author}{Vanhove C},
  \bibinfo{author}{Metzler S.~D}, \bibinfo{author}{Moore S.~C}.
\newblock Review of SPECT collimator selection, optimization, and fabrication
  for clinical and preclinical imaging.
\newblock Medical Physics  \bibinfo{year}{2015};
  \bibinfo{volume}{42}:\bibinfo{pages}{4796--813}.
\newblock https://doi.org/10.1118/1.4927061.
%Type = Incollection
\bibitem[{Cherry et~al.(2012)Cherry, Sorenson \& Phelps}]{CHERRY_2012}
\bibinfo{author}{Cherry S.~R}, \bibinfo{author}{Sorenson J.~A},
  \bibinfo{author}{Phelps M.~E}.
\newblock The Gamma Camera: Performance Characteristics.
\newblock In: \bibinfo{editor}{Cherry S.~R}, \bibinfo{editor}{Sorenson J.~A},
  \& \bibinfo{editor}{Phelps M.~E}, editors. \bibinfo{booktitle}{Physics in
  Nuclear Medicine}. \bibinfo{edition}{4th} ed.
\newblock  \bibinfo{address}{Philadelphia}: \bibinfo{publisher}{W.B. Saunders};
  \bibinfo{year}{2012}.
\newblock p. \bibinfo{pages}{209 -- 31}.
%Type = Incollection
\bibitem[{Passeri \& Formiconi(2004)}]{Passeri_2004}
\bibinfo{author}{Passeri A}, \bibinfo{author}{Formiconi A}.
\newblock SPECT and Planar Imaging in Nuclear Medicine.
\newblock In: \bibinfo{editor}{Del~Guerra A}, editor.
  \bibinfo{booktitle}{Ionizing Radiation Detectors for Medical Imaging}.
\newblock  \bibinfo{address}{Singapore}: \bibinfo{publisher}{World Scientific};
  \bibinfo{year}{2004}.
\newblock p. \bibinfo{pages}{235--85}.
%Type = Article
\bibitem[{Agostini et~al.(2016)Agostini, Ben-Haim, Rouzet, Songy, Giordano,
  Gimelli, Hyafil, Sciagrà, Bucerius, Verberne, Slart \&
  Lindner}]{Agostini_2016}
\bibinfo{author}{Agostini D}, \bibinfo{author}{Ben-Haim S},
  \bibinfo{author}{Rouzet F}, \bibinfo{author}{Songy B},
  \bibinfo{author}{Giordano A}, \bibinfo{author}{Gimelli A} et~al.
\newblock Performance of cardiac cadmium-zinc-telluride gamma camera imaging in
  coronary artery disease: a review from the cardiovascular committee of the
  European Association of Nuclear Medicine (EANM).
\newblock \EJNMMI  \bibinfo{year}{2016}; \bibinfo{volume}{43}.
\newblock https://doi.org/10.1007/s00259-016-3467-5.
%Type = Article
\bibitem[{Goshen et~al.(2018)Goshen, Beilin, Stern, Kenig, Goldkorn \&
  Ben-Haim}]{Goshen_2018}
\bibinfo{author}{Goshen E}, \bibinfo{author}{Beilin L}, \bibinfo{author}{Stern
  E}, \bibinfo{author}{Kenig T}, \bibinfo{author}{Goldkorn R},
  \bibinfo{author}{Ben-Haim S}.
\newblock Feasibility study of a novel general purpose CZT-based digital SPECT
  camera: initial clinical results.
\newblock EJNMMI Physics  \bibinfo{year}{2018}; \bibinfo{volume}{5}.
\newblock https://doi.org/10.1186/s40658-018-0205-z.
%Type = Article
\bibitem[{Morelle et~al.(2020)Morelle, Bellevre, Hossein-Foucher, Manrique \&
  Bailliez}]{Morelle_2019}
\bibinfo{author}{Morelle M}, \bibinfo{author}{Bellevre D},
  \bibinfo{author}{Hossein-Foucher C}, \bibinfo{author}{Manrique A},
  \bibinfo{author}{Bailliez A}.
\newblock First comparison of performances between the new whole-body
  cadmium–zinc–telluride SPECT-CT camera and a dedicated cardiac CZT camera
  for myocardial perfusion imaging: Analysis of phantom and patients.
\newblock J Nucl Cardiol  \bibinfo{year}{2020}; p.
  \bibinfo{pages}{1261–1269}.
\newblock https://doi.org/10.1007/s12350-019-01702-2.
%Type = Article
\bibitem[{Desmonts et~al.(2020)Desmonts, Bouthiba, Enilorac, Nganoa, Agostini
  \& Aide}]{Desmonts_2020}
\bibinfo{author}{Desmonts C}, \bibinfo{author}{Bouthiba M},
  \bibinfo{author}{Enilorac B}, \bibinfo{author}{Nganoa C},
  \bibinfo{author}{Agostini D}, \bibinfo{author}{Aide N}.
\newblock Evaluation of a new multipurpose whole-body CZT-based camera:
  comparison with a dual-head Anger camera and first clinical images.
\newblock EJNMMI Phys  \bibinfo{year}{2020}; \bibinfo{volume}{7}.
\newblock https://doi.org/10.1186/s40658-020-0284-5.
%Type = Article
\bibitem[{Nudi et~al.(2017)Nudi, Iskandrian, Schillaci, Peruzzi, Frati \&
  Biondi-Zoccai}]{NUDI_2017}
\bibinfo{author}{Nudi F}, \bibinfo{author}{Iskandrian A.~E},
  \bibinfo{author}{Schillaci O}, \bibinfo{author}{Peruzzi M},
  \bibinfo{author}{Frati G}, \bibinfo{author}{Biondi-Zoccai G}.
\newblock Diagnostic Accuracy of Myocardial Perfusion Imaging With CZT
  Technology: Systemic Review and Meta-Analysis of Comparison With Invasive
  Coronary Angiography.
\newblock JACC: Cardiovascular Imaging  \bibinfo{year}{2017};
  \bibinfo{volume}{10}:\bibinfo{pages}{787--94}.
\newblock https://doi.org/10.1016/j.jcmg.2016.10.023.
%Type = Article
\bibitem[{{Hutton} et~al.(2018){Hutton}, {Occhipinti}, {Kuehne}, {Máthé},
  {Kovács}, Waiczies, Erlandsson, Salvado, Carminati, Montagnani, Short,
  Ottobrini, van Mullekom, Piemonte, Bukki, Nyitrai, Papp, Nagy, Niendorf,
  de~Francesco, Fiorini \& consortium}]{INSERT_2018}
\bibinfo{author}{{Hutton} B}, \bibinfo{author}{{Occhipinti} M},
  \bibinfo{author}{{Kuehne} A}, \bibinfo{author}{{Máthé} D},
  \bibinfo{author}{{Kovács} N}, \bibinfo{author}{Waiczies H} et~al.
\newblock Development of clinical simultaneous SPECT/MRI.
\newblock Br J Radiol  \bibinfo{year}{2018};
  \bibinfo{volume}{91}:\bibinfo{pages}{1081}.
\newblock https://doi.org/10.1259/bjr.20160690.
%Type = Article
\bibitem[{Bouckaert et~al.(2014)Bouckaert, Vandenberghe \&
  Van~Holen}]{Bouckaert_2014}
\bibinfo{author}{Bouckaert C}, \bibinfo{author}{Vandenberghe S},
  \bibinfo{author}{Van~Holen R}.
\newblock Evaluation of a compact, high-resolution SPECT detector based on
  digital silicon photomultipliers.
\newblock \PMB  \bibinfo{year}{2014};
  \bibinfo{volume}{59}:\bibinfo{pages}{7521--39}.
\newblock https://doi.org/10.1088/0031-9155/59/23/7521.
%Type = Article
\bibitem[{{Popovic} et~al.(2014){Popovic}, {McKisson}, {Kross}, {Lee},
  {McKisson}, {Weisenberger}, {Proffitt}, {Stolin}, {Majewski} \&
  {Williams}}]{Popovic_2014}
\bibinfo{author}{{Popovic} K}, \bibinfo{author}{{McKisson} J.~E},
  \bibinfo{author}{{Kross} B}, \bibinfo{author}{{Lee} S},
  \bibinfo{author}{{McKisson} J}, \bibinfo{author}{{Weisenberger} A.~G} et~al.
\newblock Development and Characterization of a Round Hand-Held Silicon
  Photomultiplier Based Gamma Camera for Intraoperative Imaging.
\newblock IEEE Transactions on Nuclear Science  \bibinfo{year}{2014};
  \bibinfo{volume}{61}:\bibinfo{pages}{1084--1091}.
\newblock https://doi.org/10.1109/TNS.2014.2308284.
%Type = Inproceedings
\bibitem[{{Rando} et~al.(2016){Rando}, {Corti}, {Dazzi}, {De Angelis},
  {Dettlaff}, {Dorner}, {Fink}, {Fouque}, {Grundner}, {Haberer}, {Hahn},
  {Hermel}, {Korpar}, {Kukec Mezek}, {Maier}, {Manea}, {Mariotti}, {Mazin},
  {Mehrez}, {Mirzoyan}, {Podkladkin}, {Reichardt}, {Rhode}, {Rosier},
  {Schultz}, {Stella}, {Teshima}, {Wetteskind}, {Zavrtanik} \& {CTA
  Consortium}}]{SiPM-LST}
\bibinfo{author}{{Rando} R}, \bibinfo{author}{{Corti} D},
  \bibinfo{author}{{Dazzi} F}, \bibinfo{author}{{De Angelis} A},
  \bibinfo{author}{{Dettlaff} A}, \bibinfo{author}{{Dorner} D} et~al. {Silicon
  Photomultiplier Research and Development Studies for the Large Size Telescope
  of the Cherenkov Telescope Array }.
\newblock \bibinfo{booktitle}{Proc of The 34th International Cosmic Ray
  Conference {\textemdash} PoS(ICRC2015)}  \bibinfo{year}{2016}; p.
  \bibinfo{pages}{940}.
\newblock https://doi.org/10.22323/1.236.0940.
%Type = Article
\bibitem[{Hahn et~al.(2018)Hahn, Dettlaff, Fink, Mazin, Mirzoyan \&
  Teshima}]{HAHN_2018}
\bibinfo{author}{Hahn A}, \bibinfo{author}{Dettlaff A}, \bibinfo{author}{Fink
  D}, \bibinfo{author}{Mazin D}, \bibinfo{author}{Mirzoyan R},
  \bibinfo{author}{Teshima M}.
\newblock Development of three silicon photomultiplier detector modules for the
  MAGIC telescopes for a performance comparison to PMTs.
\newblock NIM-A  \bibinfo{year}{2018};
  \bibinfo{volume}{912}:\bibinfo{pages}{259 -- 63}.
\newblock https://doi.org/10.1016/j.nima.2017.11.071.
\newblock \bibinfo{note}{New Developments In Photodetection 2017}.
%Type = Article
\bibitem[{Guberman et~al.(2019)Guberman, Cortina, Ward, {Do Souto Espiñera},
  Hahn \& Mazin}]{Guberman_2019}
\bibinfo{author}{Guberman D}, \bibinfo{author}{Cortina J},
  \bibinfo{author}{Ward J}, \bibinfo{author}{{Do Souto Espiñera} E},
  \bibinfo{author}{Hahn A}, \bibinfo{author}{Mazin D}.
\newblock The Light-Trap: A novel concept for a large SiPM-based pixel for Very
  High Energy gamma-ray astronomy and beyond.
\newblock NIM-A  \bibinfo{year}{2019}; \bibinfo{volume}{923}:\bibinfo{pages}{19
  -- 25}.
\newblock https://doi.org/10.1016/j.nima.2019.01.052.
%Type = Article
\bibitem[{{Nagai} et~al.(2019){Nagai}, {Alispach}, {Barbano}, {Coco}, {Della
  Volpe}, {Heller}, {Montaruli}, {Ekoume}, {Troyano-Pujadas} \&
  {Renier}}]{Nagai_2019}
\bibinfo{author}{{Nagai} A}, \bibinfo{author}{{Alispach} C},
  \bibinfo{author}{{Barbano} A}, \bibinfo{author}{{Coco} V},
  \bibinfo{author}{{Della Volpe} D}, \bibinfo{author}{{Heller} M} et~al.
\newblock {Characterization of a large area silicon photomultiplier}.
\newblock NIM-A  \bibinfo{year}{2019};
  \bibinfo{volume}{948}:\bibinfo{pages}{162796}.
\newblock https://doi.org/10.1016/j.nima.2019.162796.
%Type = Article
\bibitem[{{Mallamaci} et~al.(2019){Mallamaci}, {Baibussinov}, {Busetto},
  {Corti}, {De Angelis}, {Di Pierro}, {Doro}, {Lessio}, {Mariotti}, {Rando},
  {Prandini}, {Vallania}, {Vigorito} \& {CTA LST Project}}]{Mallamaci_2019}
\bibinfo{author}{{Mallamaci} M}, \bibinfo{author}{{Baibussinov} B},
  \bibinfo{author}{{Busetto} G}, \bibinfo{author}{{Corti} D},
  \bibinfo{author}{{De Angelis} A}, \bibinfo{author}{{Di Pierro} F} et~al.
\newblock {Design of a SiPM-based cluster for the Large-Sized Telescope camera
  of the Cherenkov Telescope Array}.
\newblock NIM-A  \bibinfo{year}{2019};
  \bibinfo{volume}{936}:\bibinfo{pages}{231--2}.
\newblock https://doi.org/10.1016/j.nima.2018.09.141.
%Type = Inproceedings
\bibitem[{Gómez et~al.(2016)Gómez, Gascón, Fernández, Sanuy, Mauricio,
  Graciani \& Sanchez}]{Gomez_2016}
\bibinfo{author}{Gómez S}, \bibinfo{author}{Gascón D},
  \bibinfo{author}{Fernández G}, \bibinfo{author}{Sanuy A},
  \bibinfo{author}{Mauricio J}, \bibinfo{author}{Graciani R} et~al. {MUSIC: An
  8 channel readout ASIC for SiPM arrays}.
\newblock \bibinfo{booktitle}{Proc SPIE Optical Sensing and Detection IV}
  \bibinfo{year}{2016}; p. \bibinfo{pages}{85--94}.
\newblock https://doi.org/10.1117/12.2231095.
%Type = Inproceedings
\bibitem[{Taylor et~al.(2019)Taylor, Adams, Ambrosi, Ambrosio, Aramo, Benbow,
  Bertucci, Bissaldi, Bitossi, Boiano, Bonavolontà, Bose, Brill, Buckley,
  Caprai, Di~Venere, Feng, Fiandrini, Giglietto, Giordano, Hervet, Hughes,
  Humensky, Ionica, Jin, Kaaret, Kieda, Kim, Licciulli, Loporchio, Masone,
  Meures, Mukherjee, Pantaleo, Okumura, Petrashyk, Powell, Paoletti, Ribeiro,
  Rugliancich, Santander, Shang, Stevenson, Stiaccini, Tosti, Vagelli,
  Valentino, Vandenbroucke, Vassiliev, Wilcox \& Williams}]{leslie}
\bibinfo{author}{Taylor L}, \bibinfo{author}{Adams C}, \bibinfo{author}{Ambrosi
  G}, \bibinfo{author}{Ambrosio M}, \bibinfo{author}{Aramo C},
  \bibinfo{author}{Benbow W} et~al. {Camera Design and Performance of the
  Prototype Schwarzschild-Couder Telescope for the Cherenkov Telescope Array}.
\newblock \bibinfo{booktitle}{Proc of 36th International Cosmic Ray Conference
  {\textemdash} PoS(ICRC2019)}  \bibinfo{year}{2019}; p. \bibinfo{pages}{807}.
\newblock https://doi.org/10.22323/1.358.0807.
%Type = Inproceedings
\bibitem[{Adams et~al.(2019)Adams, Ambrosi, Ambrosio, Aramo, Benbow, Bertucci,
  Bissaldi, Bitossi, Boiano, Bonavolontà, Bose, Brill, Buckley, Caprai,
  Covault, Venere, Feng, Fiandrini, Gent, Giglietto, Giordano, Halliday,
  Hervet, Hughes, Humensky, Ionica, Jin, Kaaret, Kieda, Kim, Licciulli,
  Loporchio, Masone, Meures, Mode, Mukherjee, Okumura, Otte, Pantaleo,
  Paoletti, Petrashyk, Powell, Powell, Ribeiro, Rousselle, Rugliancich,
  Santander, Shang, Stevenson, Stiaccini, Taylor, Tosti, Vagelli, Valentino,
  Vandenbroucke, Vassiliev, Wilcox \& Williams}]{leonardoSPIE_2019}
\bibinfo{author}{Adams C}, \bibinfo{author}{Ambrosi G},
  \bibinfo{author}{Ambrosio M}, \bibinfo{author}{Aramo C},
  \bibinfo{author}{Benbow W}, \bibinfo{author}{Bertucci B} et~al.
  {Characterization and assembly of near-ultraviolet SiPMs for the
  Schwarzschild-Couder medium-size telescope proposed for the CTA Observatory}.
\newblock \bibinfo{booktitle}{Proc. SPIE Hard X-Ray, Gamma-Ray, and Neutron
  Detector Physics XXI}  \bibinfo{year}{2019}; p. \bibinfo{pages}{52--60}.
\newblock https://doi.org/10.1117/12.2530617.
%Type = Article
\bibitem[{Simmons(1988)}]{Simmons_1988}
\bibinfo{author}{Simmons G.~H}.
\newblock On-line Corrections for Factors That Affect Uniformity and Linearity.
\newblock J. Nucl. Med. Technol.  \bibinfo{year}{1988};
  \bibinfo{volume}{16}:\bibinfo{pages}{82--9}.
%Type = Article
\bibitem[{Passeri et~al.(1993)Passeri, Formiconi \& Meldolesi}]{Passeri_1993}
\bibinfo{author}{Passeri A}, \bibinfo{author}{Formiconi A.~R},
  \bibinfo{author}{Meldolesi U}.
\newblock Physical modelling (geometrical system response, Compton scattering
  and attenuation) in brain {SPECT} using the conjugate gradients
  reconstruction method.
\newblock \PMB  \bibinfo{year}{1993};
  \bibinfo{volume}{38}:\bibinfo{pages}{1727--44}.
\newblock https://doi.org/10.1088/0031-9155/38/12/004.
%Type = Article
\bibitem[{Li et~al.(2010)Li, Wedrowski, Bruyndonckx \& Vandersteen}]{Li_2010}
\bibinfo{author}{Li Z}, \bibinfo{author}{Wedrowski M},
  \bibinfo{author}{Bruyndonckx P}, \bibinfo{author}{Vandersteen G}.
\newblock Nonlinear least-squares modeling of 3D interaction position in a
  monolithic scintillator block.
\newblock \PMB  \bibinfo{year}{2010};
  \bibinfo{volume}{55}:\bibinfo{pages}{6515--32}.
\newblock https://doi.org/10.1088/0031-9155/55/21/012.
%Type = Article
\bibitem[{Morozov et~al.(2015)Morozov, Solovov, Alves, Domingos, Martins, Neves
  \& Chepel}]{Morozov_2015}
\bibinfo{author}{Morozov A}, \bibinfo{author}{Solovov V},
  \bibinfo{author}{Alves F}, \bibinfo{author}{Domingos V},
  \bibinfo{author}{Martins R}, \bibinfo{author}{Neves F} et~al.
\newblock Iterative reconstruction of detector response of an Anger gamma
  camera.
\newblock \PMB  \bibinfo{year}{2015};
  \bibinfo{volume}{60}:\bibinfo{pages}{4169--84}.
\newblock https://doi.org/10.1088/0031-9155/60/10/4169.
%Type = Article
\bibitem[{{Allison} et~al.(2006){Allison}, {Amako}, {Apostolakis}, {Araujo},
  {Dubois}, {Asai}, {Barrand}, {Capra}, {Chauvie}, {Chytracek}, {Cirrone},
  {Cooperman}, {Cosmo}, {Cuttone}, {Daquino}, {Donszelmann}, {Dressel},
  {Folger}, {Foppiano}, {Generowicz}, {Grichine}, {Guatelli}, {Gumplinger},
  {Heikkinen}, {Hrivnacova}, {Howard}, {Incerti}, {Ivanchenko}, {Johnson},
  {Jones}, {Koi}, {Kokoulin}, {Kossov}, {Kurashige}, {Lara}, {Larsson}, {Lei},
  {Link}, {Longo}, {Maire}, {Mantero}, {Mascialino}, {McLaren}, {Lorenzo},
  {Minamimoto}, {Murakami}, {Nieminen}, {Pandola}, {Parlati}, {Peralta},
  {Perl}, {Pfeiffer}, {Pia}, {Ribon}, {Rodrigues}, {Russo}, {Sadilov},
  {Santin}, {Sasaki}, {Smith}, {Starkov}, {Tanaka}, {Tcherniaev}, {Tome},
  {Trindade}, {Truscott}, {Urban}, {Verderi}, {Walkden}, {Wellisch},
  {Williams}, {Wright} \& {Yoshida}}]{Allison_2006}
\bibinfo{author}{{Allison} J}, \bibinfo{author}{{Amako} K},
  \bibinfo{author}{{Apostolakis} J}, \bibinfo{author}{{Araujo} H},
  \bibinfo{author}{{Dubois} P.~A}, \bibinfo{author}{{Asai} M} et~al.
\newblock {Geant4 developments and applications}.
\newblock IEEE Transactions on Nuclear Science  \bibinfo{year}{2006};
  \bibinfo{volume}{53}:\bibinfo{pages}{270--8}.
\newblock https://doi.org/10.1109/TNS.2006.869826.
%Type = Article
\bibitem[{Stockhoff et~al.(2017)Stockhoff, Jan, Dubois, Cherry \&
  Roncali}]{Stockhoff_2017}
\bibinfo{author}{Stockhoff M}, \bibinfo{author}{Jan S}, \bibinfo{author}{Dubois
  A}, \bibinfo{author}{Cherry S.~R}, \bibinfo{author}{Roncali E}.
\newblock Advanced optical simulation of scintillation detectors in {GATE}
  V8.0: first implementation of a reflectance model based on measured data.
\newblock \PMB  \bibinfo{year}{2017};
  \bibinfo{volume}{62}:\bibinfo{pages}{L1--L8}.
\newblock https://doi.org/10.1088\%2F1361-6560\%2Faa7007.
%Type = Article
\bibitem[{Gundacker et~al.(2019)Gundacker, Martinez~Turtos, Kratochwil, Pots,
  Paganoni, Lecoq \& Auffray}]{Gundacker_2019}
\bibinfo{author}{Gundacker S}, \bibinfo{author}{Martinez~Turtos R},
  \bibinfo{author}{Kratochwil N}, \bibinfo{author}{Pots R},
  \bibinfo{author}{Paganoni M}, \bibinfo{author}{Lecoq P} et~al.
\newblock Experimental time resolution limits of modern SiPMs and TOF-PET
  detectors exploring different scintillators and Cherenkov emission.
\newblock \PMB  \bibinfo{year}{2019};
  \bibinfo{volume}{65}:\bibinfo{pages}{025001}.
\newblock https://doi.org/10.1088/1361-6560/ab63b4.
%Type = Masterthesis
\bibitem[{{Occhipinti}(2015)}]{Occhipinti_2015}
\bibinfo{author}{{Occhipinti} M}.
\newblock {Development of a gamma-ray detection module for multimodal SPECT/MR
  imaging}.
\newblock Master's thesis.
\newblock  Politecnico di Milano; \bibinfo{year}{2015}. \URLprefix
  \url{http://hdl.handle.net/10589/114190}.
%Type = Article
\bibitem[{{Di Francesco} et~al.(2016){Di Francesco}, {Bugalho}, {Oliveira},
  {Pacher}, {Rivetti}, {Rolo}, {Silva}, {Silva} \& {Varela}}]{TOFPET2_2016}
\bibinfo{author}{{Di Francesco} A}, \bibinfo{author}{{Bugalho} R},
  \bibinfo{author}{{Oliveira} L}, \bibinfo{author}{{Pacher} L},
  \bibinfo{author}{{Rivetti} A}, \bibinfo{author}{{Rolo} M} et~al.
\newblock {TOFPET2: a high-performance ASIC for time and amplitude measurements
  of SiPM signals in time-of-flight applications}.
\newblock Journal of Instrumentation  \bibinfo{year}{2016};
  \bibinfo{volume}{11}:\bibinfo{pages}{C03042}.
\newblock https://doi.org/10.1088/1748-0221/11/03/C03042.
%Type = Inproceedings
\bibitem[{{Gómez} et~al.(2019){Gómez}, {Sánchez}, {Gascón}, {Cela},
  {Freixas}, {Graciani}, {Manera}, {Marín}, {Mauricio}, {Navarrete}, {Oller},
  {Pérez}, {Rato Mendes}, {Sanmukh} \& {Vela}}]{HRFlexToT}
\bibinfo{author}{{Gómez} S}, \bibinfo{author}{{Sánchez} D},
  \bibinfo{author}{{Gascón} D}, \bibinfo{author}{{Cela} J.~M},
  \bibinfo{author}{{Freixas} L}, \bibinfo{author}{{Graciani} R} et~al. A High
  Dynamic Range ASIC for Time of Flight PET with pixelated and monolithic
  crystals.
\newblock \bibinfo{booktitle}{Proc 2019 IEEE Nuclear Science Symposium and
  Medical Imaging Conference (NSS/MIC)}  \bibinfo{year}{2019}; p.
  \bibinfo{pages}{1--3}.
\newblock 10.1109/NSS/MIC42101.2019.9059762.
%Type = Article
\bibitem[{Calva-Coraza et~al.(2017)Calva-Coraza, Alva-Sánchez,
  Murrieta-Rodríguez, Martínez-Dávalos \&
  Rodríguez-Villafuerte}]{CALVACORAZA_2017}
\bibinfo{author}{Calva-Coraza E}, \bibinfo{author}{Alva-Sánchez H},
  \bibinfo{author}{Murrieta-Rodríguez T}, \bibinfo{author}{Martínez-Dávalos
  A}, \bibinfo{author}{Rodríguez-Villafuerte M}.
\newblock Optimization of a large-area detector-block based on SiPM and
  pixelated LYSO crystal arrays.
\newblock Physica Medica  \bibinfo{year}{2017};
  \bibinfo{volume}{42}:\bibinfo{pages}{19--27}.
\newblock https://doi.org/10.1016/j.ejmp.2017.08.003.
%Type = Article
\bibitem[{Park \& Lee(2019)}]{PARK_2019}
\bibinfo{author}{Park H}, \bibinfo{author}{Lee J.~S}.
\newblock Highly multiplexed SiPM signal readout for brain-dedicated TOF-DOI
  PET detectors.
\newblock Physica Medica  \bibinfo{year}{2019};
  \bibinfo{volume}{68}:\bibinfo{pages}{117--123}.
\newblock https://doi.org/10.1016/j.ejmp.2019.11.016.
%Type = Article
\bibitem[{Mottaghian et~al.(2010)Mottaghian, Koohi-Fayegh, Ghal-Eh \&
  Etaati}]{Mottaghian_2010}
\bibinfo{author}{Mottaghian M}, \bibinfo{author}{Koohi-Fayegh R},
  \bibinfo{author}{Ghal-Eh N}, \bibinfo{author}{Etaati G}.
\newblock Photocathode non-uniformity contribution to the energy resolution of
  scintillators.
\newblock Radiation protection dosimetry  \bibinfo{year}{2010};
  \bibinfo{volume}{140}:\bibinfo{pages}{16--24}.
\newblock https://doi.org/10.1093/rpd/ncq041.

\end{thebibliography}

\end{document}